\documentclass[useAMS,usenatbib,twocolumn]{aastex63}
\usepackage{times,graphicx,amsmath,amsfonts,amssymb}

\usepackage{hyperref}
\usepackage{hyperref}
  \hypersetup{bookmarksnumbered = true}
 \usepackage{subfigure}
\usepackage{epsfig} 
\usepackage{enumitem}
\usepackage{tabularx}
\usepackage{booktabs}
\usepackage{multirow}
\usepackage{tikz}

\usepackage{array}
\newcolumntype{L}[1]{>{\raggedright\let\newline\\\arraybackslash\hspace{0pt}}m{#1}}
\newcolumntype{C}[1]{>{\centering\let\newline\\\arraybackslash\hspace{0pt}}m{#1}}
\newcolumntype{R}[1]{>{\raggedleft\let\newline\\\arraybackslash\hspace{0pt}}m{#1}}

\newcommand{\tikzcircle}[2][red,fill=red]{\tikz[baseline=-0.5ex]\draw[#1,radius=#2] (0,0) circle ;}%
	\definecolor{candypink}{rgb}{0.89, 0.44, 0.48}
	\definecolor{chestnut}{rgb}{0.8, 0.36, 0.36}

\def\ltsim{\lower.5ex\hbox{$\; \buildrel < \over \sim \;$}}
\def\gtsim{\lower.5ex\hbox{$\; \buildrel > \over \sim \;$}}
\def\ltsim{\lower.5ex\hbox{$\; \buildrel < \over \sim \;$}}
\def\gtsim{\lower.5ex\hbox{$\; \buildrel > \over \sim \;$}}
\def\lobs{L_\textrm{obs}}
\def\lobsl{L_{\textrm{obs},l}}

\def\lint{L_\textrm{int}}
\def\lintl{L_{\textrm{int},l}}
\def\ngal{N_\textrm{gal}}
\def\nuint{\nu_\textrm{int}}
\def\IO{I^0_{\nu_0}}
\def\da{d_\mathrm{A}}
\def\tda{{\tilde d}_\mathrm{A}}

\def\be{\begin{equation}}
\def\ee{\end{equation}}
\def\ba{\begin{eqnarray}}
\def\ea{\end{eqnarray}} 

\mathchardef\mhyphen="2D 
\def\zc{z_\mathrm{c}}
\def\hmpc{\; \mathrm{h}^{-1}\,{\rm Mpc}}
\def\nunits{\mathrm{h}^{3}\,{\rm Mpc}^{-3} }
\def\hgpc{h^{-1}\,{\rm Gpc}}
\def\kms{\, {\rm km }\, {\rm s}^{-1}}

\def\dd{\mathrm{d}}

\def\dl{d_\mathrm{L}}
\def\dc{d_\mathrm{C}}
\def\lsfr{\log  \textrm{SFR}}
\def\lsfrobs{\log \textrm{SFR}_{\textrm{obs}}}
\def\vs{V_\mathrm{s}}
\def\vsh{\mathcal{V}}

\def\myr{\; \mathrm{M}_\odot \; \textrm{yr}^{-1}}

\def\dgal{\delta_\textrm{gal}}
\def\vgal{V_\textrm{gal}}
\def\dg{\delta_\textrm{gal}}
\def\dgs{\delta^s_{\textrm{gal}}}
\def\ds{\delta^{s}}
\def\sigs{\sigma_{s}}
\def\siggs{\sigma_{s,\textrm{gal}}}
\def\sigeps{\sigma_{\epsilon^s}}
\def\lcdm{\Lambda{\rm CDM}}

\def\ln{{\rm ln}\,}

\newcommand{\av}[1]{\langle{#1}\rangle}
\newcommand{\eq}[1]{Eq.~\ref{#1}}
\newcommand{\eqs}[2]{Eqs.~\ref{#1} \& \ref{#2}}
\newcommand{\figr}[1]{Fig.~\ref{#1}}
\newcommand{\figrs}[2]{Figs.~\ref{#1} \& \ref{#2}}

\def\br{\mathbf{r}}
\def\grad{\mathbf{\nabla}}
\def\bv{\mathbf{v}}

\def\tsfr{\textrm{SFR}}

\def\rs{R_\mathrm{s}}

\def\sfr{\textrm{SFR}}
\def\vpr{V_\textrm{rec}}
\def\vg{V_\textrm{gal}}

\def\vmp{V_{\mathrm{M}_{\textrm{p}}}}
\def\apjl{ApJL}

\newcommand{\tum}{\textsc{UM}}
\newcommand{\sag}{\textsc{SAG}}
\newcommand{\meth}{\textsc{SfLM}}
\newcommand{\sage}{\textsc{SAGE}}
\newcommand{\galc}{\textsc{Galacticus}}

\newcommand{\mdg}{\textsc{MultiDark-Galaxies}}

\newcommand{\Ha}{H$\alpha$\ }
\shorttitle{SFR}
\shortauthors{ Nusser, Yepes \& Branchini}

\begin{document}

\title{Biasing relation, environmental dependencies and estimation of the growth rate from   star forming  galaxies}
\author{Adi Nusser}
\email{adi@physics.technion.ac.il} 
\affil{Department of Physics and the Asher Space Research Institute, Israel 
Institute of Technology Technion, Haifa 32000, Israel\\  }
\author{Gustavo Yepes}
\affil{Departamento de F\'{\i}sica Te\'orica M-8, Universidad Aut\'onoma de Madrid, Cantoblanco, E-28049 Madrid, Spain\\  }
\affil{Centro de Investigaci\'{o}n Avanzada en F\'{\i}sica Fundamental (CIAFF), Facultad de Ciencias, Universidad Aut\'{o}noma de Madrid, E-28049 Madrid, Spain}
\author{Enzo Branchini}
\affil{Dipartimento di Fisica, Universit\`a di Roma, 3, via della Vasca Navale 84, I-00146 Roma, Italy\\  }
\affil{INFN - Sezione di Roma Tre, via della Vasca Navale 84, I-00146 Roma, Italy\\}
\affil{INAF - Osservatorio Astronomico di Roma, via Frascati 33, I-00040 Monte Porzio Catone (RM), Italy\\}


\begin{abstract}

The connection between galaxy star formation rate (SFR) and  dark matter (DM) is of paramount importance for the extraction of cosmological information from next generation spectroscopic surveys that will target  emission line star forming   galaxies. 
Using  publicly available mock galaxy catalogs obtained 
from  various  semi-analytic models (SAMs)  we explore the 
SFR-DM  connection  in relation to   the {\em speed-from-light} method \citep{Feix2016} for inferring the growth rate, $f$, from luminosity/SFR shifts.  Emphasis is given to the dependence of the SFR distribution on the environment density on scales of 10s-100s Mpc.
We show that the application of the {\em  speed-from-light} method to an  Euclid-like survey is not biased by environmental effects. In all models, the  precision on the measured 
$\beta=f/b$ parameter is $\sigma_\beta\ltsim 0.17$ at $z=1$. This  translates  into errors of $\sigma_f \sim 0.22$ and  $\sigma_{(f\sigma_8)}\sim   0.1$, 
without invoking  assumptions on the mass power spectrum. These errors are in the same ballpark as  recent analyses of the redshift space distortions in galaxy clustering.
In agreement with  previous studies,
the bias factor, $b$ is roughly a scale-independent, constant function of the SFR for star forming galaxies. Its value at $z=1$ ranges from $1.2$ to $1.5$  depending on the SAM recipe.
Although in all SAMs denser environments host galaxies with higher stellar masses, the dependence of the SFR  on the  environment  is more involved.
In  most  models the SFR probability distribution is skewed to larger values in denser regions.  One model exhibits an \textit{inverted } trend where high SFR is suppressed  in  dense environment.

\end{abstract}

\section{Introduction}
\label{sec:intro}

The connection between the dark matter and galaxies is essential for understanding the processes that 
regulate the formation and the evolution of galaxies and, consequently, to derive cosmological parameters  from
the analysis of galaxy redshift surveys. 
In particular, estimation of cosmological parameters  from the clustering pattern  is inherently dependent on this connection.

In the  standard paradigm galaxies form by the condensation and cooling of gas inside DM dominated  halos (virialized objects)   \citep{Binney1977,Rees1977,Silk1977,White1978a}. The process is hierarchical \citep{Peeb80} with early forming galaxies 
collapse into and merge with other   galaxies and is greatly affected by  energy released from  supernovae \citep{Larson74,DS86} and 
AGN activities \citep{Silk1998}. 

Semi-analytic galaxy formation models (SAMs)  \citep{White1991,Kauffmann1993,Lacey1993,Somerville1999} have been  extensively employed  in an attempt to understand the vast amount of observational properties of galaxies and the link to the formation of supermassive blackholes. SAMs approximates complex  interconnected processes of star formation, energetic feedback and hydrodynamics in terms of 
simple forms involving a large number of free parameters. The importance of each process can then by assessed by 
tuning the relevant parameters to match certain observational data. 

State-of-the-art cosmological simulations can now follow the hydrodynamics in conjunction with 
elaborate (albeit poorly known) sub-grid physics of galaxy formation over large dynamical scales \citep{Gene2014,Khandai2015,Schaye2015,Dolag2016,Dubois2016}.   
Unfortunately, the box size of this type of simulations  remains  insufficient to describe  the structure on the 
large scales probed by large redshift surveys.  
As an example,  {Illustris TNG300} \citep{Springel2018}  has a box size of $205\hmpc$ on a side, equivalent to  only 25\% the volume probed by  the  low redshift Two Mass Redshift Survey \citep[2MRS,][]{2mrs2012} and substantially smaller  than  the volume coverage planned by future large surveys.

DM-only simulations on the other hand have been done for large simulation boxes of several Gpcs.
\cite{kn97} pioneered the approach of incorporating SAMs in DM only simulations. The simulation used in that work was only of a   box size of $128 $ Mpc. 
 The same approach was later applied by many 
 workers in the field using much larger simulations and more elaborate SAMs  \cite[e.g.][]{Kauffmann1999,Benson2000,Guo2011,Angulo2014,Baugh2019}. 
 
Numerical and semi-analytic methods have been extensively used to study the  \textit{biasing} relation between the distribution of  galaxies and the underlying distribution of mass (DM-dominated) as a function of the stellar mass. 
f{However, next-generation spectroscopic galaxy redshift surveys like the Dark Energy Spectroscopic Instrument (DESI) survey  \citep{DESICollaboration2016b}, 
the Euclid  space mission \citep{EuclidCollaboration2019} and the Roman space mission \citep{wfirst2019} will  mainly select objects based on SFR indicators like the  \Ha\ 
and the [OII]  emission lines. Therefore the biasing relation for galaxies selected based on star formation rates is becoming of particular interest \citep{Angulo2014}.}

In general, the biasing relation enters in any analysis based on the clustering of galaxies. Its knowledge is essential for a precise and accurate estimation of cosmological parameters. It is at the heart of methods relying on the anisotropy of clustering in redshift space (the so-called redshift space distortions, hereafter RSDs) \citep{Sargent1977} and,  to a lesser degree, in analysing signatures of Baryonic Acoustic Oscillations (BAO). The standard and most convenient assumption is that of \textit{linear biasing}. If  $1+\dgal$ and $1+\delta$ are the galaxy number density and the total mass density in units of their respective mean values, linear biasing dictates 
\begin{equation}
\label{eq:linbias}
\delta_\textrm{gal}(t,\br)=b(t)  \delta(t,\br) +\varepsilon\; ,
\end{equation} 
where $b$ is independent of $\delta$ and position $\br$, but is a function of time  as implied by continuity considerations \citep{ND94,Tegmark1998}. The term 
$\varepsilon$ represents stochastic (random) scatter around the mean relation  \citep[e.g.][]{dh99}.  Both density contrasts, $\dgal$ and 
$\delta$, are  assumed to have been  filtered with the same smoothing window. 
For gaussian random fields, the relation is valid on sufficiently large scales \citep{BBKS}.
The term $\varepsilon$ arises from an intrinsic scatter in the biasing relation as well as the Poisson fluctuations (shot noise) 
associated with the finite number of galaxies. The intrinsic scatter can be attributed to several factors that are not captured 
solely by the local mass density  field at a given time. 
 The  assembly and star formation  history, details of the feedback process and external gravitational tidal field that affects the galaxy rotation, 
 all can impact the galaxy properties.
Linear biasing has been demonstrated to hold on sufficiently large scales and its dependence on the stellar mass
and the host halo mass has been studied using simulations as well as analytic models.

{Modern spectroscopic redshift surveys are designed  to provide tight constraints on the growth rate, $f$,   of linear density fluctuations at high redshift. }
The growth rate is related to the growing density mode $D$ via
\citep{Peeb80}
 \begin{equation}
 f=\frac{\dd \ln D}{\dd \ln a}\approx \Omega^{\gamma} 
 \end{equation}
with  $\gamma\approx 0.55+ 0.05(1+w)$ for a dark energy model with an equation of state parameter $w$ \citep{Lind05}. Therefore, constraining  $f$ at different cosmic epochs could in principle yield important insight into the 
dark energy models {responsible} for  the accelerating expansion of the Universe. 
The aforementioned RSDs resulting from placing galaxies at their redshifts rather than actual distances are a traditional probe of $f$ via the combination
\begin{equation}
\label{eq:beta}
\beta\equiv \frac{f(\Omega)}{b} \; .
\end{equation}

But, placing galaxies at their redshift positions rather than actual distances, does not only result in RSDs. 
It also shifts the estimates  of the galaxy intrinsic  luminosities from  their true values obtained using  actual distances. 
To first order, the redshift position differs from the distance by the line of sight (los) peculiar velocity. 
Therefore, coherent large scale  luminosity variations  in space, can be used to constrain the peculiar velocity field. 
This  idea dates back
to the work of \cite{TYS} who correlated the magnitudes of nearby galaxies with their redshifts to constrain the velocity of the Virgo cluster relative to the Local Group. 

There are two techniques to probe the velocity field using luminosity variations. 
In the first one,  direct constraints on the velocity { are derived}  using the observed variations. This technique has been applied to the 2MRS at $z\ltsim 0.03$ \citep{Nusser2011a,abate2012a,Branchini2012}  and the Digital Sloan Sky Survey  (SDSS) at $z\sim 0.1$ by \cite{Feix2015,Feix2016}, and led to interesting constraints on the amplitude of the  bulk flow. This technique is susceptible to environmental dependence of the luminosity distribution. Indeed,
the dependence of the luminosity function on the large scale density field could mimic variations due to peculiar velocities. For nearby surveys like the 2MRS, the luminosity shift from  peculiar velocities  {dominates over environmental effects}. However, at $z\approx 1$, relevant to next generation surveys this is not the case anymore, making the application of this technique less attractive or even irrelevant in comparison to low redshift surveys. 

The second technique  relies on a  simultaneous estimation of the luminosity fluctuation and the peculiar velocity field from the actual galaxy distribution in redshift space \citep{Nusser2012}. The derived  velocity field depends on $\beta$.  The true galaxy luminosities are then estimated using the  distances derived from the redshifts by subtracting the los peculiar velocities. 
The parameter $\beta$ is  derived   by minimizing the large scale spatial luminosity variations. Assuming the environmental dependence of the luminosity function are mostly via the 
large scale density, this method will yield an unbiased estimate of $\beta$ thanks to the lack of correlation between the  density and the peculiar velocity at a given point in space.
In general, Galilean invariance can be invoked to conclude that the 
peculiar motion of a galaxy cannot affect any of its internal properties. Therefore, any mechanism that affects the luminosity distribution to a given point in space must be uncorrelated with the peculiar velocity at the same point. 
Thus environmental dependencies will only affect the statistical uncertainty in the derived estimates of $\beta$.
 We gave the name Speed-from-Light (hereafter, \meth ) \citep{Feix2016} to this second method.
 
The third goal of this paper is  to provide  an assessment of the applicability of this method for constraining $\beta$ from next generation spectroscopic redshift survey and to verify its sensitivity to environmental effects.
We wish to confirm that environmental dependencies do not bias the $\beta$ estimate from the \meth\ and to estimate 
 how they affect the random error assigned to $\beta$.

 We will rely on mock galaxy catalogs obtained by applying different SAM recipes to large DM-only simulations. Mock catalogs are likely inaccurate at describing galaxy properties at high redshift. We therefore 
compare predictions from different models to achieve different goals. First, we aim at identifying common features that can be used as a predictions for planned surveys. Second, the discrepancies among the models will serve to appreciate the scatter in current theoretical predictions. And third  we will assess how environmental dependencies, which can be quantified in future observations, can provide useful constraints on models of galaxy formation


{The outline of the paper is as follows. In \S\ref{sec:simulations}, we present the simulations and the corresponding mock galaxy catalogs that we use in this work and their relation to next generations spectroscopic catalogs.  In \S\ref{sec:sfr} we use mock catalogs to study how stellar masses and star formation rate depend on the environment.
In \S\ref{sec:biasing}
we focus on galaxy biasing and its dependence on the SFR and stellar mass. The impact of the large scale environment on the estimate of $\beta$ from the \meth\ method is studied in \S\ref{sec:growth}. We discuss our results and offer our conclusions in Sections \ref{sec:discussion} and \ref{sec:conclusions}, respectively.
}


\section{Mocks and simulations}
\label{sec:simulations}

We use publicly available mock galaxy catalogs extracted from three DM-only
simulations ( SMDPL, MDPL2, BigMDPL)
of the \textsc{MultiDark} suites.   {The relevant parameters of the  simulations  are listed in Table \ref{tab:sim}. }
Mock galaxy catalogs from the \sag\  \citep{Cora2018} , \sage\  \citep{Croton2016} and \galc\ \citep{Benson2012}  semi-analytic galaxy formation models (SAMs), are 
publicly available only for the MDPL2 simulation. 
These  mocks, referred to as  \mdg\    \citep{Knebe2018} have been downloaded from the \textsc{CosmoSim} data base\footnote{\href{http://www.cosmosim.org}{https://www.cosmosim.org}}.
Mock galaxies extracted from the SMDPL simulation 
are available only for the \textsc{UniverseMachine} \citep[hereafter \tum ,][]{Behroozi2019} { a self-consistent empirical} galaxy formation model and have been kindly provided by Peter Behroozi.   The \tum\ model offers a simple recipe for assigning SFR to halos.
 We use that recipe to populate the MDPL2 and BigMDPL  simulations with galaxies based on the \tum\  mocks in the SMDPL.

\begin{table}[]
\centering
\caption{Relevant parameters of the three MDPL simulations used in this work. All simulations correspond to 
a $\Lambda$CDM cosmology with parameters $h=0.6777$, $\Omega_\Lambda=0.692885$, $\Omega_\textrm{m}=0.307115$ and 
$\sigma_8=0.8228$}
\bigskip
\begin{tabular}{l|l|l|l}
\label{tab:sim}
                        & SMDPL          & MDPL2  & BigMDPL \\ \hline
$L \; [\hmpc ] $        & 400            & 1000   & 2500    \\ \hline
$m_p \; [10^8 M_\odot]$ & $1.42 $ & $22.3$ & $348$   \\ \hline
\end{tabular}
\end{table}

 All  SAMs incorporate  the same basic processes of galaxy formation, gas cooling,
  star formation according to the amount of cold gas, stellar winds  and AGN feedback.  They include recipes for tracing the mass in the main galaxy components: disks, bulges, black holes. 
  The models have a large number of free parameters that are fixed by matching observations of the galaxy population. 
  {Despite their similarities, the three SAMs differ in the way baryonic physical processes are implemented.}
For example,   all of them basically follow the gas cooling treatment presented 
 in \cite{White1991}, but differ in the details of metal cooling.
 \galc\  follows the SFR recipe in \cite{Krumholz2009}, while 
 \sag\  initiates star formation  only once the   cold gas mass in the forming disk exceeds a certain  value. 
In addition to the radio mode AGN feedback employed in \sag\ and \galc ,  \sage\ also includes a quasar wind mode. 
Another  difference between \sage\  and the two other models is the treatment of galaxies which no longer have  identifiable parent subhalos in the simulation. \sage\ disperses the stellar content of orphan galaxies into the 
main halo,  while \galc\ and \sag\ maintain them as separate entities, 
 However, since we will focus on  galaxies with high SFR, we do not expect our results to depend on this aspect of the models.

 Comparison between \textsc{SAG}, \textsc{SAGE} and \textsc{Galacticus}  applied to MDPL2 is provided in \cite{Knebe2018}. 
All models have been calibrated using  
 low redshift, $z\approx 0$,  key observational data such as the black hole - bulge mass relation and the stellar mass functions, but 
 not always to the same 
data compilation \citep[c.f. Table 1 in][]{Knebe2018}. 

We are mostly interested in results at $z=1$ to match the typical redshift of next generation surveys. Therefore we shall focus on $z=1$ simulated catalogs and only consider the $z=0$ case to explore the biasing of star forming galaxies at lower redshift.

\subsection{\textsc{UniverseMachine} Mock galaxies for {MDPL2} and {BigMDPL}}
\label{sec:tum}
The \tum\ models  available for us are only for the SMDPL simulation of insufficient volume to probe the dependence of the SFR and luminosity distributions on  the  galaxy environment. 
Therefore, we wish to generate \tum\ mock galaxies for the larger boxes as well. The main usage for \tum\ in these larger boxes will be {to estimate the } cosmic variance. 
Fortunately,  in the   \tum\ models, the  SFR in a halo at redshift $z$ depends mainly on $\vmp$, the maximum of the rotation curve measured at the 
peak mass through the history of the halo until $z$ \citep{Behroozi2019}. {This is not necessarily true for satellite galaxies that do not play a significant role in our analysis since we are interested in large star forming galaxies.}   Also, since we are mainly  interested in the SFR rather than stellar mass, we avoid running the whole \tum\ machinery by   assigning  SFRs  to  MDPL2 and {BigMDPL} halos
from a  random re-sampling of the SFR  conditional  probability distribution given  $\vmp$,  taken  from  the full \tum\ in  the SMDPL  simulation
\footnote{The quantity $\vmp$ is available from the Rockstar catalogues  for 
all halos in all simulations.}.
{Instead, for} each halo in the larger simulations, we select a value for SFR from the distribution in the $\vmp-\textrm{SFR}$ in SMDPL. 
The re-sampling will  capture  environmental dependencies 
present in the distribution of  $\vmp$, but will  miss those associated with other parameters that can have an effect on the SFR. 
Nonetheless, these re-sampled catalogs will  mainly serve for the 
  statistical assessments of the \meth . An added value for random re-sampling is that several  realisations of SFRs can be generated for the same halo. This allows an assessment of randomness (stochasticity) in the SFR  per halo assignment, at least in the \tum\ models.

\begin{figure}
\vskip 0.2in
  \includegraphics[width=0.48\textwidth]{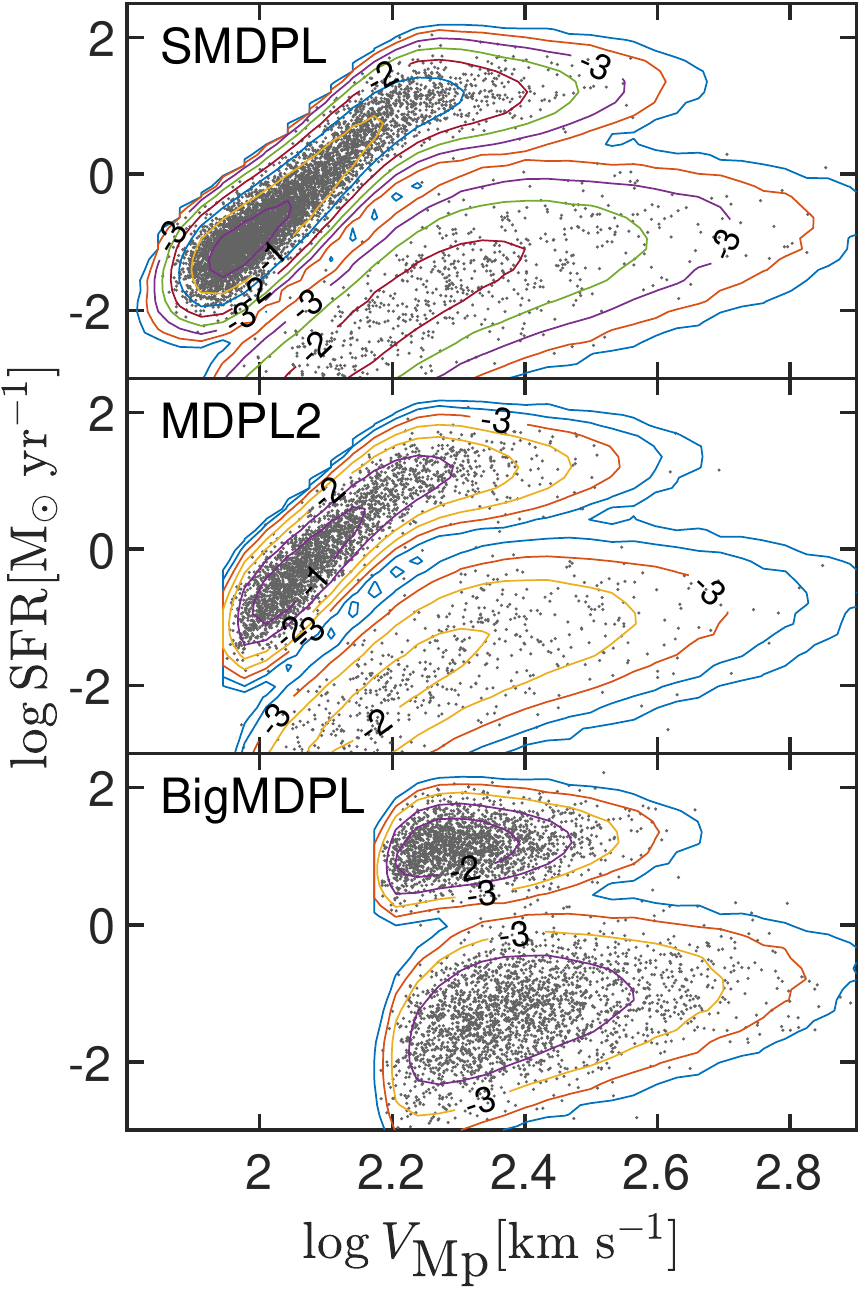}  
  \vskip 0.0in
 \caption{ The distribution of \tum\ galaxies in the plane  $\log \textrm{SFR}$ and 
 $\log V_{{\textrm{M}}_\textrm{p}} $. Top: galaxies in the \tum\  catalog extracted from the SMDPL  simulation.
 In  middle (MDPL2) and bottom (BigMDPL), SFRs were assigned to halos with a given $\vmp$ by re-sampling of the distribution in the top panel. Contours designate certain values of the logarithm of  2D probability distribution function $P(\log \textrm{SFR}, \log V_{{\textrm{M}}_\textrm{p}}$).}
\label{fig:VMP}
\end{figure}

The top panel in \figr{fig:VMP} shows the distribution of a randomly selected fraction of mock galaxies
 in the  $\textrm{SFR} - \vmp$ plane from   the full \tum\  model in SMDPL.  Also plotted are contours of the  two dimensional (2D) probability distribution function (PDF) of 
$\log\tsfr$ and $\log \vmp$. The two branches of star forming and quenched (low star formation) galaxy population are clearly visible. 
As we shall see in the next Section, next generation surveys like Euclid will observe galaxies with relatively high $\tsfr>\; 10 \myr$, sampling  the tip of the  PDF.

The middle and bottom panels show, respectively,  the same distributions for MDPL2 and BigMDPL obtained by re-sampling SFR at a given $\vmp$ from the distribution represented in the top panel.  
In practice,  we partition  the SMDPL galaxies  in  100 bins in $\log \vmp$ and 
match the bins to   $\vmp$ of halos in the larger simulations. Then the distribution of \tum\ SFR values 
in
each  bin  is randomly sampled to assign SFR values to the respective halos in the larger simulations. 
 Due to the larger particle mass in the larger simulations, they do not resolve low  $\vmp$   as  the  SMDPL does, yielding lower number densities of galaxies relative to the SMDPL.  
 Table \ref{tnumgal} lists  the  number densities  of \tum\ galaxies in SMDPL and in the re-sampled \tum\ in the two larger simulations, which  we label \tum\ MDPL and \tum\ BigMDPL. This table refers to redshift $z=1$ and two different $M_*$ cuts, as indicated. We emphasize that the re-sampling does not provide $M_* $  for the larger simulations, only SFR. 
 The number densities in SMDPL and MDPL2
 are comparable, with the latter having only a 11\% lower value, while 
 \tum\ BigMDPL is significantly more dilute.  Still, the number densities of all models in the Table are consistent with each other  within a factor of two.
 In addition, the BigMDPL box represents a sizable  fraction of  the volume that will be probed by the Euclid's spectroscopic survey and contains a comparable total number  of galaxies \citep{EuclidCollaboration2019}.
 
\begin{figure}
\vskip 0.2in
  \includegraphics[width=0.48\textwidth]{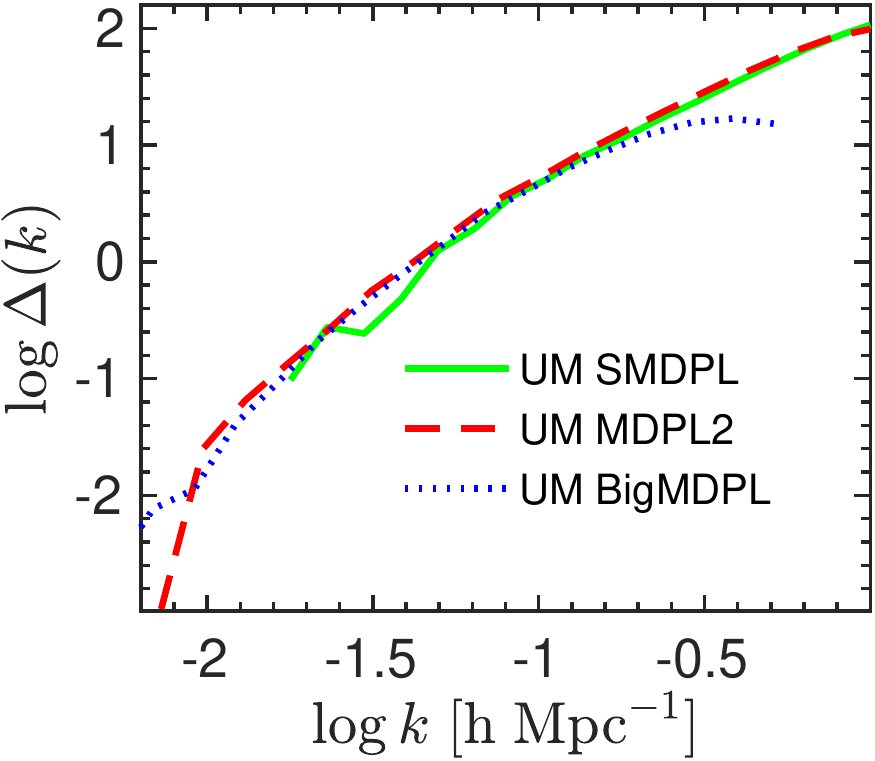} 
  \vskip 0.0in
 \caption{ Consistency of power spectra of the mock galaxies in the the full \tum\ in SMDPL and in the \tum\ MDPL and \tum\ BigMDPL samples.}
\label{fig:umPk}
\end{figure}

An important  check of the re-sampling procedure  is whether it yields consistent  clustering properties among the three \tum\ catalogs. 
  In \figr{fig:umPk} we plot  the quantity $\Delta(k)\propto k^3 P_\textrm{gal}(k)$ where  $P_\textrm{gal}(k)$ is the galaxy power spectrum as a function of the wave number $k$. The details of computing $P_\textrm{gal}$ are  described in \S\ref{sec:biasing}. 
  The figure indeed demonstrates a very good agreement between the three power spectra.

\subsection{Volumes and Galaxy Densities: connection to observations}
\label{sec:volnum}
The results of the present study are relevant for any spectroscopic surveys that, like DESI, Euclid and Roman-WFIRST, will target emission line galaxies over large sky areas at intermediate redshifts. Here we will focus on the Euclid case and consider it as representative of a typical next generation spectroscopic surveys.



 Euclid's survey will detect \Ha\ galaxies with line flux larger than $ 2\times 10^{-16} \; \textrm{erg}\; \textrm{s}^{-1}\; \textrm{cm}^{-2}$, 
 over 15000 $\textrm{deg}^2$  of the sky in the  redshift range $0.9<z<1.8$
 corresponding to comoving volume of $\sim 43\;  h^3 \;\textrm{Gpc}^{-3}  $. 
 Based on models calibrated on available observations of \Ha\  emitters \citep{Pozzetti2016}, 
the Euclid collaboration has recently provided their forecast for the number density of 
 \Ha\  galaxies: one expects
 $n\sim 6.9 \times 10^{-4} \nunits$ at $z\sim  1$, gradually  decreasing to  
 $n\sim 4.2 \times 10^{-4} \nunits$ at $z\sim 1.4$ and dropping to $n\sim 2.6 \times 10^{-4} \nunits $ at $z\sim 1.7$. 
 
Galaxy formation models  provide the  SFR rather than the  luminosity, $L_{H\alpha}$, of the \Ha\ line.  
To link these quantities we adopt the transformation  \citep[e.g.][]{Dom2012}
\begin{equation}
\log \textrm{SFR}[\textrm{M}_\odot \; \textrm{yr}^{-1}]=\log L_{H\alpha}[\textrm{erg}\; \textrm{s}^{-1}]-41.1\; .
\end{equation}
The Euclid  \Ha\ flux cut corresponds then to a lower SFR threshold $\sfr_\textrm{lim}=10\myr$
at $z=1$.
For mock galaxies with $M_*>5\times 10^9 \textrm{M}_\odot$, the  number densities, listed  in Table \ref{tnumgal} are higher than the official Euclid forecast. 
However, they are comparable to those of  Model 1 of \cite{Pozzetti2016} since they expect a number density of  $2.6\times 10^{-3} h^{3} {\rm Mpc}^{-3}$ 
galaxies above the Euclid \Ha\ flux threshold.  This is actually  close to the number densities 
of galaxies with $\sfr >10\myr $ and $M_*>5\times 10^8\textrm{M}_\odot$ in the mock catalogs, 
 as seen in Table \ref{tnumgal}. 
This is perhaps not surprising since the Euclid forecast account for instrumental effects and observational biases that are not included in the simulations.

The MDPL2 box is $\sim 2.3\%$  the total volume probed by the full Euclid survey and $\sim 12\%$   the volume in the redshift bin $[0.9-1.1]$.  In this redshift range the \cite{EuclidCollaboration2019}  expects to observe more than $(10^6)$ galaxies, i.e. roughly the number of objects required for a successful application of the \meth\ method. 
The mocks, however, approximately contain this number of galaxies in the MDPL2 volume alone. Therefore, it is fortunate that the number density in the simulations is higher that the expectations of  \citep{EuclidCollaboration2019}, as this will allow us to test the \meth\ already with the mocks we have. The shortcoming is  that the smaller simulated volume prevents a proper assessment of cosmic variance.

The larger simulation BigMDPL is $\sim 36\% $ of the full spectroscopic survey,  still small to perform cosmic variance estimation. However,  it is  helpful in constraining cosmic variance on smaller scales, which  can point  towards its magnitude  for the whole  Euclid survey.

\begin{table*}[]
\centering
\caption{Number of galaxies with $\textrm{SFR}>10 \myr$ in the mock catalogs at $z=1$. Top two entries list the total number of galaxies in the simulation boxes for two different stellar mass thresholds.  The bottom two entries list the galaxy number densities $n=N_\mathrm{gal}/L^3$ in the same boxes.   }
\bigskip
\begin{tabular}{l|l|l|l|l|l|l}
\label{tnumgal}
             Mock      catalog                        & SAG   & SAGE  & Galac & UM    & UM MDPL2 & UM BigMDPL \\ \hline
Simulation                              & MDPL2 & MDPL2 & MDPL2 & SMDPL & MDPL2    & BigMDPL    \\ \hline
$N_\textrm{gal} $, $M_*>5\times 10^9 \textrm{M}_\odot  $      &                 $   3.29\times 10^6   $    &  $  4.48 \times 10^6  $  &  $  2.07\times 10^6  $   &  $  3\times 10^5  $  &  $  4.15 \times 10^6  $     & $3.4\times 10^7  $     \\ \hline
$N_\textrm{gal} $, $M_*>5\times 10^{10} \textrm{M}_\odot$      &               $   7.43\times 10^5   $    &  $  1.73 \times 10^6  $  &  $  9.5\times 10^4 $   &  $  5.88\times 10^4  $  &  $ - $     & $-  $     \\ \hline
$n \; [ \mathrm{h}^{3} {\rm Mpc}^{-3}]$, $M_*>5\times 10^9 \textrm{M}_\odot    $ &    $ 3.29 \times 10^{-3}$ & $ 4.48 \times 10^{-3}$  & $ 2.07\times 10^{-3}$   &$  4.67 \times 10^{-3}$  & $ 4.15 \times 10^{-3}$    & $ 2.20\times 10^{-3}  $      \\ \hline
$n \; [ \mathrm{h}^{3} {\rm Mpc}^{-3}]$,$M_*>5\times 10^{10} \textrm{M}_\odot  $  & $ 7.43 \times 10^{-4}$ & $ 1.73 \times 10^{-3}$  & $ 9.5\times 10^{-5}$   &$    9.12\times 10^{-4}$  & $ -$    & $ -  $      \\ \hline
\end{tabular}
\end{table*}


\vskip 2cm

\section{Stellar mass and star formation rate} 
\label{sec:sfr}
Properties of  the \mdg\ have been studied extensively by \cite{Knebe2018}. Nonetheless,  for completeness and as a basic check on our analysis, we compute  the stellar mass and SFR distribution  functions from the downloaded data and  compare  our results with \cite{Knebe2018} whenever relevant. For this validation test we will consider two epoch, $z=0$ and $z=1$.

\begin{figure}
\vskip 0.2in
  \includegraphics[width=0.48\textwidth]{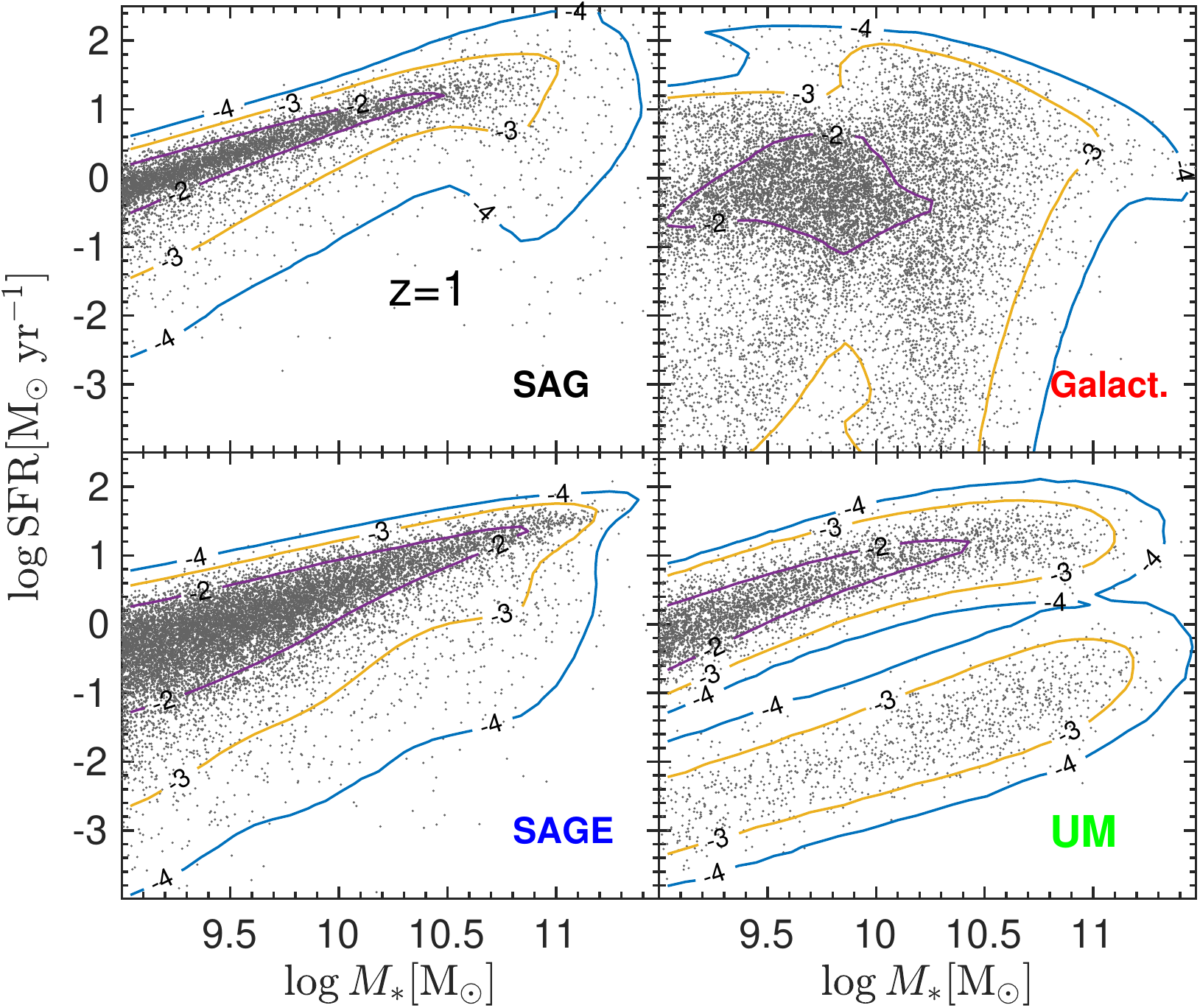} 
  \vskip 0.0in
 \caption{The distribution of galaxies in the $M_*-\textrm{SFR}$ plane at $z=1$ in different mock catalogs. Contours designate certain values of the 2D PDF of $\log \textrm{SFR}$ and $\log M_*$.}
\label{fig:PMstSFR_Z1}
\end{figure}

In \figrs{fig:PMstSFR_Z1}{fig:PMstSFR_Z0}, we plot the  distribution of a randomly selected fraction of galaxies in the plane $\log \textrm{SFR}-\log M_*$ at $z=1$ and $z=0$, respectively. 
Instead of the SFR, \cite{Knebe2018} plot the specific SFR defined as the SFR per unit stellar mass. Since we are interested in galaxies selected according to the SFR, it is more instructive 
for our purposes to explore the distribution $\textrm{SFR}-M_*$ plane. A bimodal structure is recognizable at both redshifts for the \tum\ (SMDPL) galaxies. 
This is not surprising since these models impose a division into a quenched and active galaxies. 
At $z=1$, all models except \galc\ produce  a tight ``main sequence" of star forming galaxies with similar slope and normalization, although it is broader in the SFR direction for \sage\ as can be seen from the plotted contours. A main sequence  can be identified in \galc\ at $z=0$,  as the  point encompassed by  purple contours, but the overall distribution is much more diffuse than the other models. 
This figure demonstrates the complexity of the relation between the SFR and stellar mass. The large scatter and the shape of the distribution make it hard to associate a 
well-defined  stellar mass to a given SFR. 

 The left and right panels of  \figr{fig:phi}  show  the 1D  distribution functions for SFR  and the  stellar mass, respectively. 
 Apart from the \mdg\ mocks,  the figure also shows results from the \tum\ (SMDPL) simulation.  
Models generally agree with the measured PDF of the SFR except for \sage\ that under-predicts the counts in the high-SFR tail.
The stellar mass functions at $z=0$ (right bottom panel) for \sag , \sage\ and \galc\  are in agreement with the corresponding curves at $z=0.1$ in figure 1 of \cite{Knebe2018}. All curves in each panel are roughly in the same ballpark, but the deviations  are significant even at $z=0$. This is not surprising due to the  differences in the modelling  and calibration to observations.
As pointed out by \cite{Knebe2018}, \sage\ produces the best match to the observed stellar mass distribution  at $z=0$ as reported by \cite{Moustakas2013} (bottom right). Also, since
\cite{Moustakas2013} found  little evolution of the observed stellar mass distribution since $z=1$ we can take  observations  at $z=0$  as representative to those at at $z\approx 1$ and see
 that only \sag\ over-produces galaxies at the high mass end. 
 At $z=0$, both \sag\ and \galc\ curve are above the observations at the high mass end. 
\cite{Knebe2018} attribute this $z\approx 0$ excess to 
less efficient AGN suppression of star formation compared \sage .

\begin{figure}
\vskip 0.2in
  \includegraphics[width=0.48\textwidth]{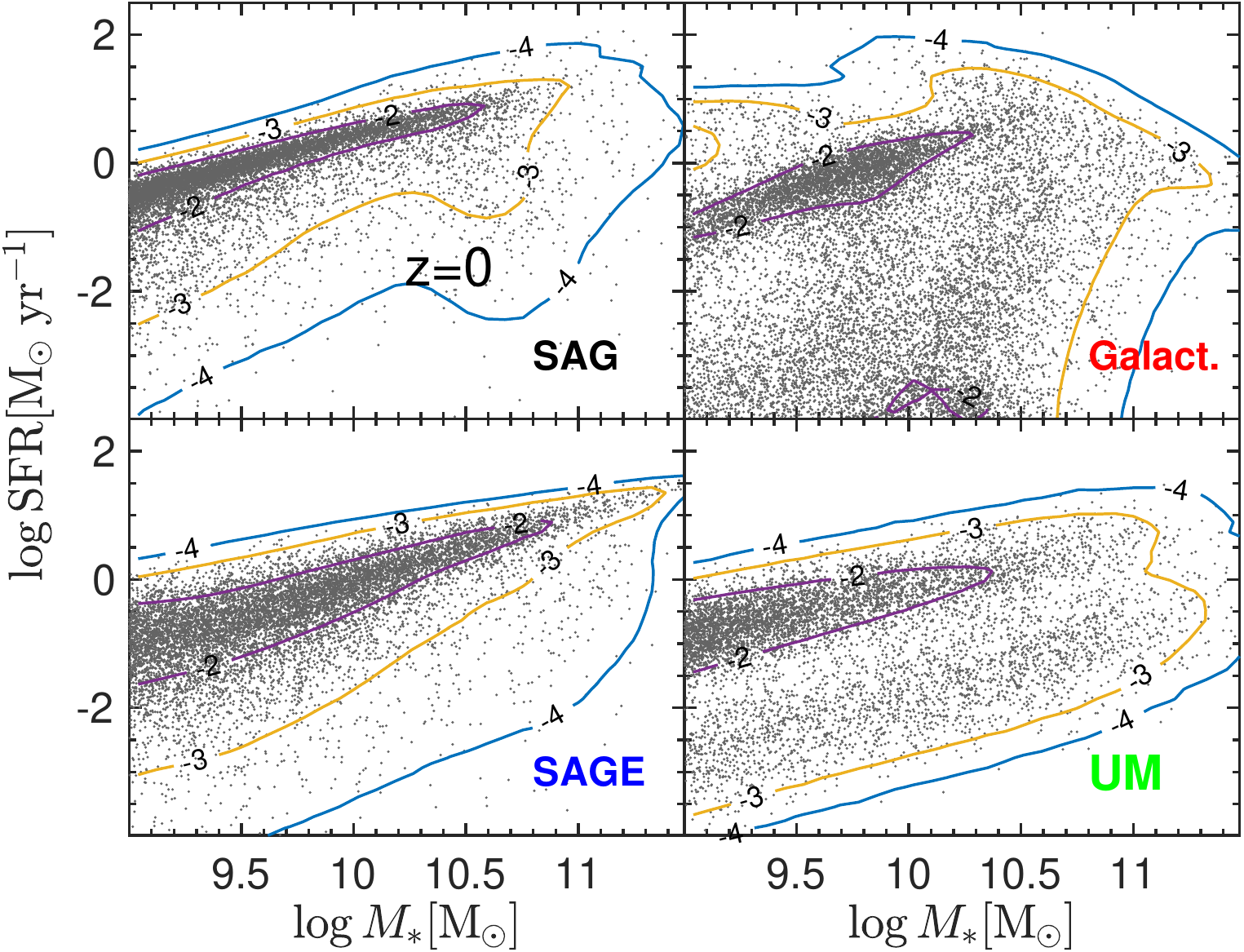} 
  \vskip 0.0in
 \caption{Same as the previous figure, but  for $z=0$.}
\label{fig:PMstSFR_Z0}
\end{figure}

\begin{figure}
\vskip 0.2in
  \includegraphics[width=0.48\textwidth]{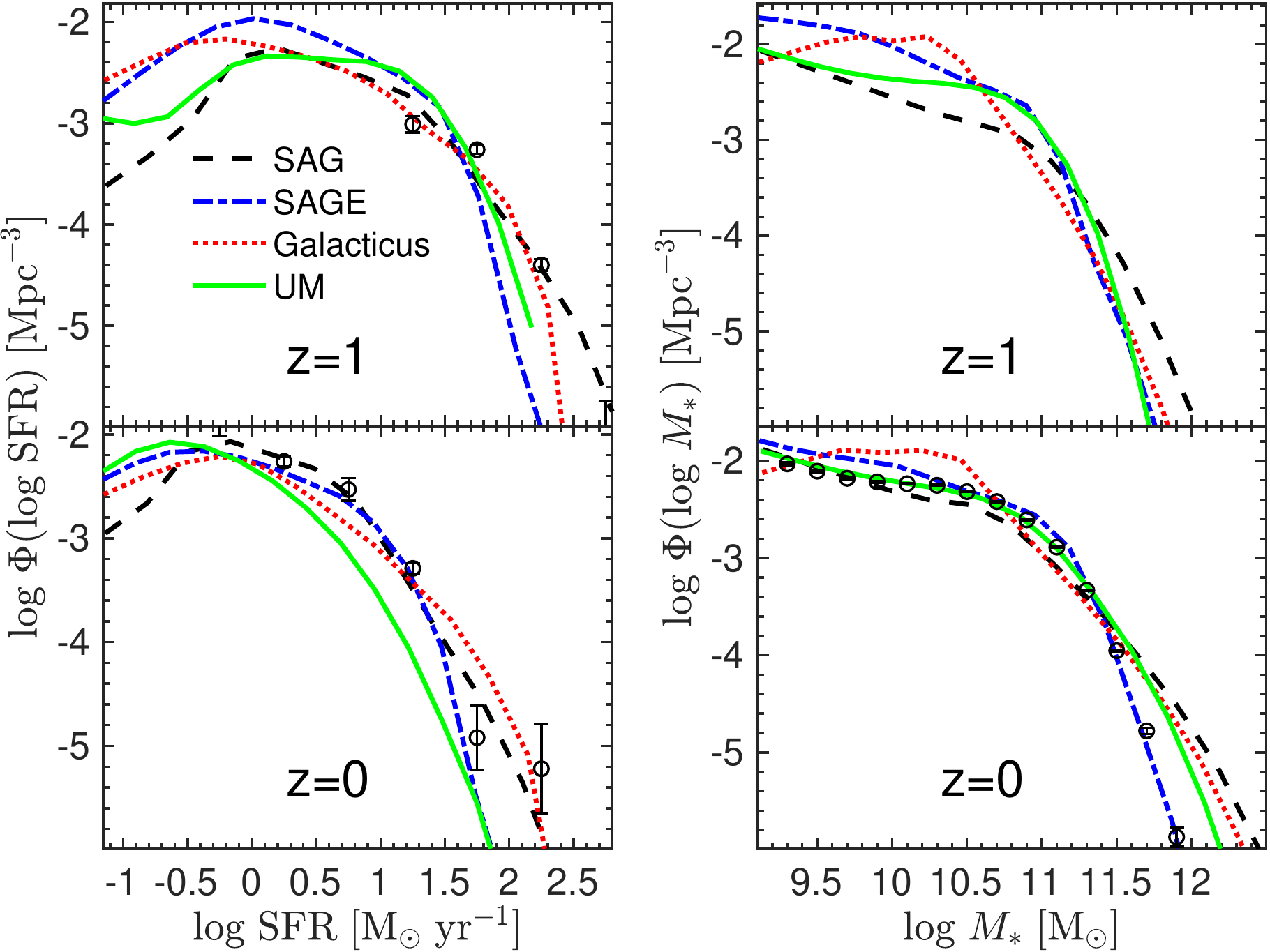}  
  \vskip 0.0in
 \caption{  The number density of  SAM galaxies as a function of the SFR (panels to the left) and the stellar mass (to the right right)  for redshift $z=0$ (bottom) and $z=1$ (top). 
 Different model predictions are represented by curves with different linestyles, specified in the labels.
 The open circles are data with errorbars from \cite{Gruppioni2015} (left) and  \cite{Moustakas2013} (right).}
\label{fig:phi}
\end{figure}


\subsection{Environmental dependencies}
The large scale environment  can play a role in shaping the properties of galaxies  \citep[e.g.][]{Xu2020}. 
Minor differences in  assembly history of halos of the same mass  \citep[e.g][]{Gottlober2001,Sheth2002,Gao2007} can lead to significant differences in their SFR evolution and the final 
stellar mass. 
Here we are interested in the modulation of  the SFR and stellar mass distributions,  as a function of the large scale density smoothed on scales of tens to hundreds of Mpc. 

In all mocks, the mass density of the DM is provided on a cubic grid. 
We use Fast Fourier Transform (FFT) to smooth the mass density with a Top-Hat (TH)  window of width $\rs=20 $ and $100 \hmpc$, respectively.    
Densities, $\delta_{i}$, at  the galaxy positions are obtained by linear interpolation of the smoothed density fields on the grid.  

We compare SFRs and $M_*$-values in low vs. high density environment by comparing 
the PDFs of $\lsfr$  and $\log M_*$ estimated for galaxies with 
the  lowest versus  highest 20\% values of $\delta_{i}$.

\begin{figure}
\vskip 0.2in
  \includegraphics[width=0.48\textwidth]{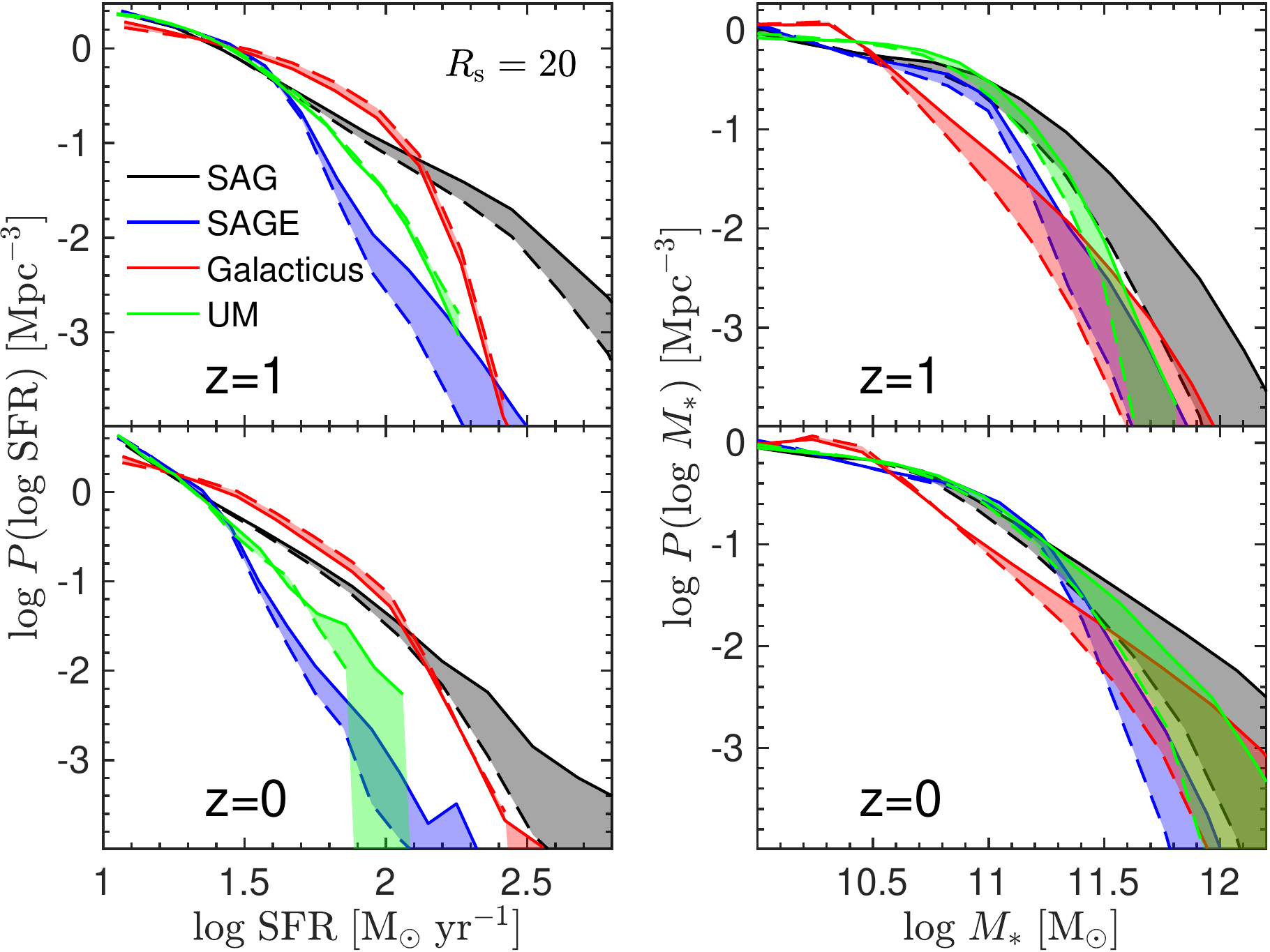} 
  \vskip 0.0in
 \caption{  The PDF of  $\lsfr$ (left) and $\log M_*$  (right) as a function of the 
 DM density smoothed with a TH window of width $R_s=20  \hmpc$, at $z=1$  (top) and $z=0$ (bottom).
 Dashed and solid curves correspond to least  and most  20\% dense regions. 
The area between these curves for each SAM is color marked, as indicated in the figure. 
The figure refers to galaxies satisfying our Euclid cut of SFR mass greater than $ 10\;  \myr$ and stellar mass $5\times 10^9\textrm{M}_\odot$, respectively.}
\label{fig:P20}
\end{figure}

\begin{figure}
\vskip 0.2in
  \includegraphics[width=0.48\textwidth]{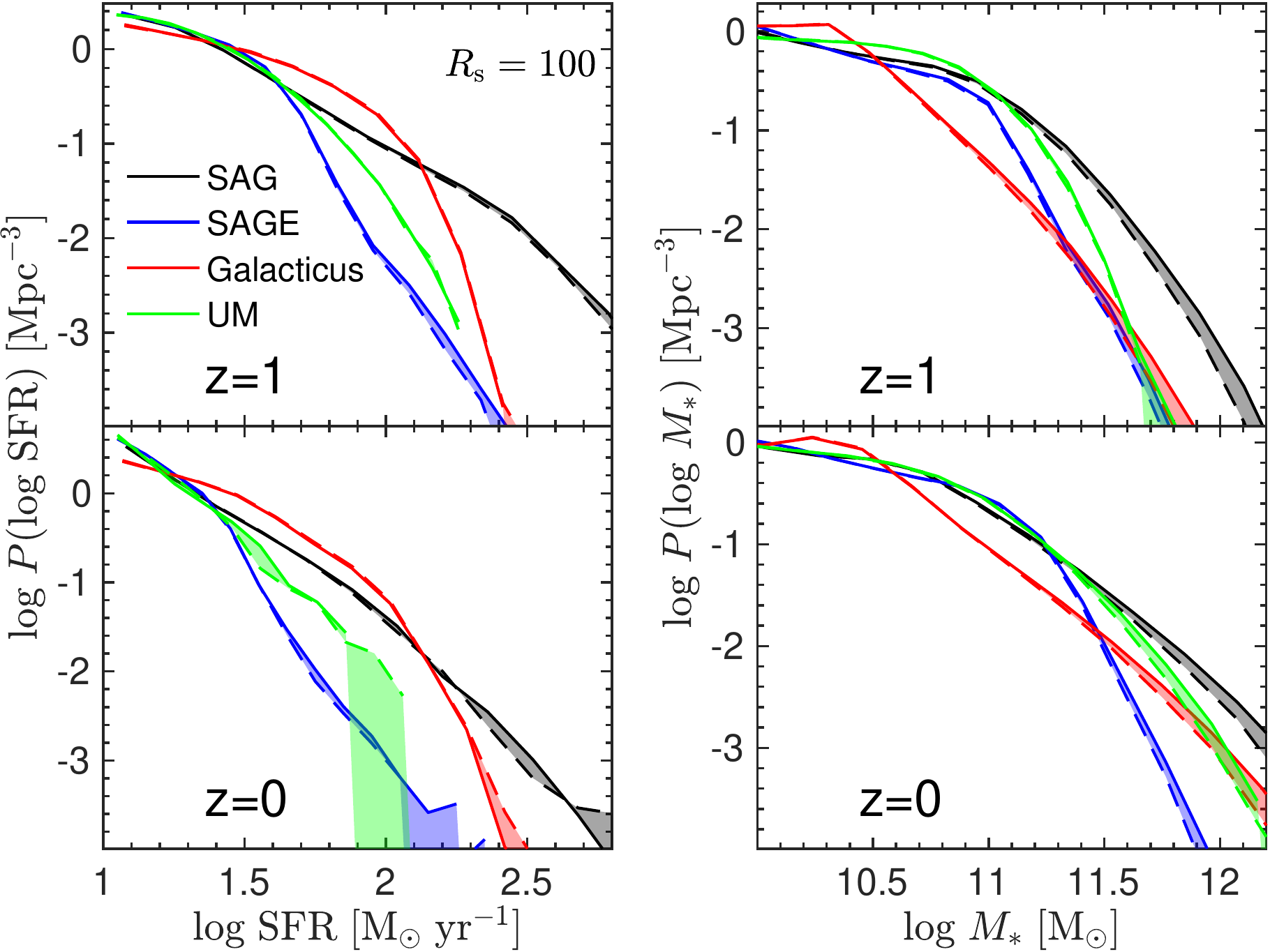} 
  \vskip 0.0in
 \caption{
 Same as Fig \ref{fig:P20},  but for $R_s = 100 \hmpc$. }
\label{fig:P100}
\end{figure}

The PDF of $\lsfr$ is computed for galaxies with $\sfr > 10 \myr$ to match the cut of the Euclid survey
and $M_*>5\times 10^9 \; \textrm{M}_\odot$.
Conversely, no cut in $\lsfr$ is imposed in the 
the PDF of $\log M_* $. The results are plotted  in \figr{fig:P20} and \figr{fig:P100}, referring  to two different environment scales of $\rs=20$ and $100\hmpc$, respectively.  
{ In all models} there is a clear dependence on the environment density, which is more pronounced at the high end of either $\lsfr$ or  $\log M_*$. 
A  reduction in the abundance of high $M_*$ galaxies in  low density environments, is  evident  in the panels to the right,  where  at high $M_*$  the  dashed curve (low $\delta_{i}$) is below the solid (high $\delta_{i}$) for all models.

The SFR relation to the environment is  more involved. 
Except  \galc, high density environments are associated with higher SFRs.
 \galc\ exhibits an intriguing  ``inverted dependence" on $\delta_{i}$; the PDF is skewed toward higher SFR  for galaxies in a low rather than high density environment.
 This implies a relatively more active star formation in   galaxies in  low  than high density environments . We can compromise  this behavior   in \galc\ with the trend of  increased fraction of high  $M_{*}$  at high densities, 
 if  the  star formation in dense regions  is preferentially intensified  well earlier than  $z=1$.

The curves are closer to each other for the larger $\rs=100\hmpc$ smoothing. The reason for that is mainly  the narrower density range  
 in the larger smoothing. Note the density ranking is not preserved between the
two smoothed density fields, otherwise the two figures would be identical. 

\begin{figure}
\vskip 0.2in
  \includegraphics[width=0.48\textwidth]{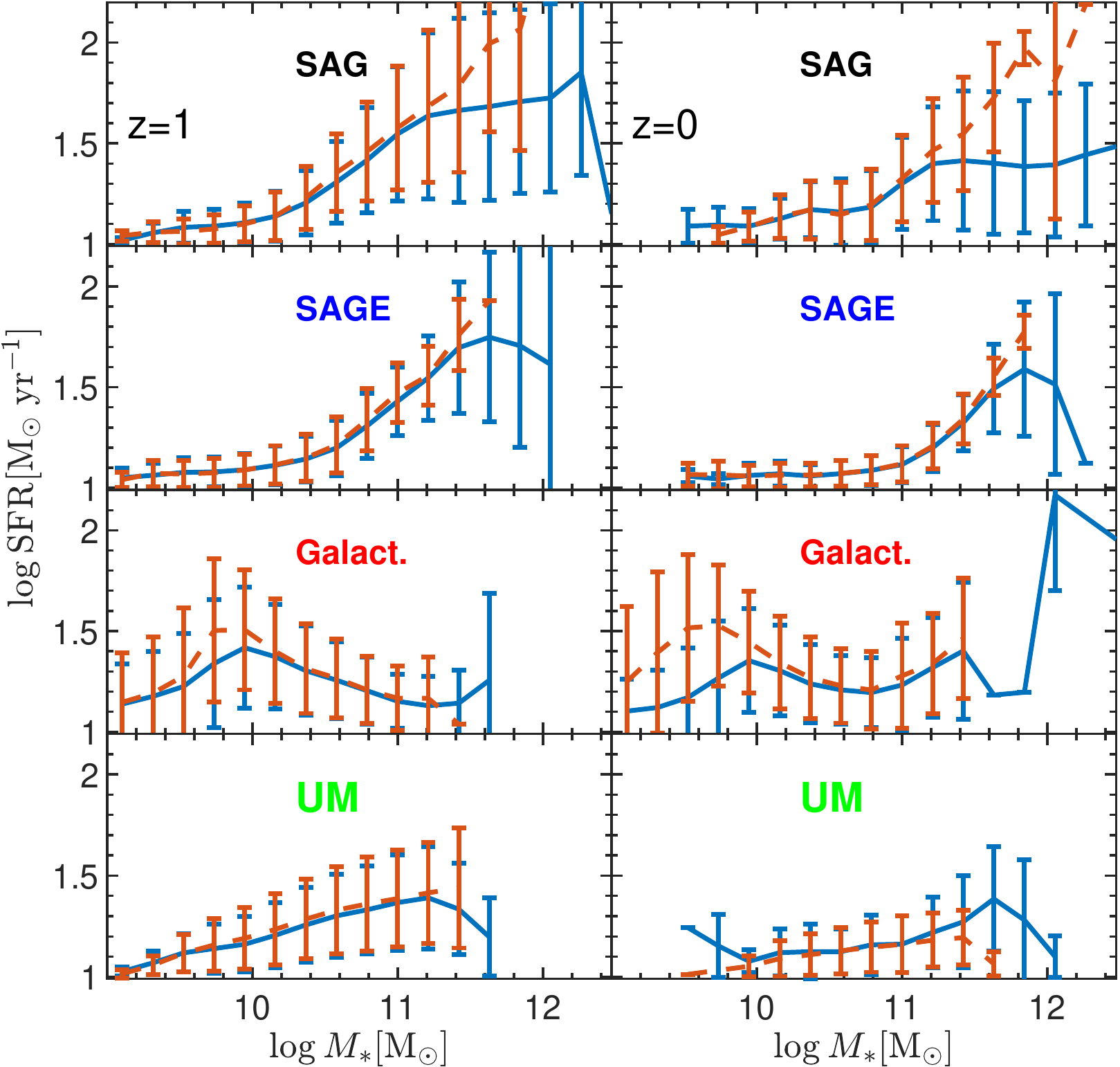} 
  \vskip 0.0in
 \caption{The mean $\log \textrm{SFR}$ at a given $M_*$ for galaxies with $\textrm{SFR}>10\myr$. Red  dashed and blue solid lines correspond to galaxies in the 20\%   least and most dense regions. Error bars represent scatter around the mean.
 The TH smoothing width for this plot only is $\rs=8\hmpc$}.
\label{fig:MstSFRlh}
\end{figure}

It is also interesting to examine the mean $\lsfr$ at a given $\log M_*$ versus density.  This is plotted in \figr{fig:MstSFRlh}   where the red dashed and blue solid lines, respectively, 
correspond to  galaxies with  lowest  20\%  and highest  20\% densities. For this plot only, the density is smoothed on a scale $\rs=8\hmpc$. The error bars represent the rms of the scatter of individual galaxies around the mean curves. The red curves do not reach as   high $M_{*}$ as the blue, simply because of the reduction of galaxies with this high $M_{*}$ in low density environments.  The \tum\ and \sage\ galaxies exhibit very little   dependence on the density of the environment. 
\sag\   galaxies follow similar curves in low and dense environments for  $\log M_*  \ltsim 11$ at both redshifts, even for this small $\rs=8\hmpc$.  For \galc ,   the only signature of the environment is a boost in the  SFR  in low densities for 
  $\log M_* \ltsim 10$.


\subsubsection{Parametrisation of Environmental Dependencies} 
We focus here on the SFR as 
it  is relevant for the \meth\ applied to emission line surveys. We parameterize the dependence  of  the  SFR on the smoothed mass density, $\delta$, as 
 \begin{eqnarray}
 \label{eq:fitone}
 \lsfr_i & =& c_1 + c_2 \delta_i+\mathcal{R}_i\; ,\\
 \label{eq:fittow}
{ \mathcal{R}}^2_i & =& d_1 + d_2 \delta_i+\mathcal{E} \; .
 \end{eqnarray}
 where the index $i$ refers to a galaxy lying at a point $\br_i$ with smoothed density $\delta_i$. The residuals, $\mathcal{R}$, and $\mathcal{E}$ are
   random numbers with zero mean values and the parameters $c_2$ and $d_{2}$ describe how the mean and variance of $\lsfr$ vary with  the environment density.
 For galaxies  with densities $\delta_i$ close to a certain $\delta_0$, the mean and variance are    
 \begin{equation}
  \overline \lsfr |_{\delta_0} =c_1+c_2 \delta_0\; ,
  \end{equation}
   and 
   \begin{equation}
   \sigma^2_{\lsfr}|_{\delta_0}=d_1+d_2\delta_0 \; .
   \end{equation}
The mean of $\lsfr$ over all galaxies is, however, 
  \begin{equation}
  \overline \lsfr=c_1+ c_2\overline {\delta_i}\; . 
  \end{equation}
 The average, $\overline {\delta}=\sum_i \delta_i/\ngal $ of $\delta_i$ over \textit{all galaxies} is close to zero, but not strictly so. To evaluate it we 
work in the  continuous limit 
 $\sum_i (\cdot ) \rightarrow \int \dd^3 r n(\br) (\cdot)$ where $n(\br)=\bar n (1+\dgal)$ with  $\bar n $  the mean number density of galaxies.  We find, 
 \begin{equation}
 \overline {\delta}=\av{\delta \dgal}_\textrm{vol} \; , 
 \end{equation}
 where  the volume average on the r.h.s is over the product of the smoothed mass density times the 
un-smoothed  density inferred from the galaxy distribution, and   we have  made use of the vanishing 
volume average of $\delta$. Therefore,
 \begin{equation}
 \label{eq:cone}
 c_1=  \overline \lsfr-c_2 \av{\delta \dgal} \approx  \overline \lsfr\; .
 \end{equation}
where in the last step  it is assumed that the density rms is $\sigma_\delta\ll 1$ and that $c_2$ is sufficiently small.   Similarly, 
 \begin{equation}
 d_1\approx \sigma^2_\mathcal{R} \approx \sigma^2_{\lsfr} \; .
 \end{equation}
 
In practice, { we first estimate   $c_1$ and $c_2$  from the mocks} by an ordinary least square fitting to  $\lsfr_i$. Then the two parameters are used to  compute the residual  
 $\mathcal{R}_i$ for every galaxy. 
Finally, the parameters  $d_1$ and $d_2$ are  derived by least square fitting to  $\mathcal{R}^2_i$ given from the previous step. The results are summarized in Table \ref{tab:env} for all mock galaxies at $z=1$ and with $\sfr>10 \myr$.
The entry  for each mock  lists the inferred parameters for the  smoothing widths $\rs=20 $ and $100\hmpc$, respectively,  in the top and bottom lines. 
 Because of the relatively small simulation box,  results with only  $\rs=20\hmpc$ are listed for \tum\ mocks in the SMDPL. 
  
For each model, the parameter $c_1$  varies very little with $\rs$ and is very close to the corresponding  $\overline {\lsfr}$ (second column), consistent with \eq{eq:cone}. 
The parameter  $c_2$ which is indicator for the modulation of the mean of $\lsfr$ versus $\delta $, exhibits some dependence on $\rs$  and, as expected, has a small amplitude. Also  
$d_1 \approx  \sigma^2_{\lsfr}$, as expected. 

For 
\sag\ and \galc,  $c_2$ has similar values  for the two values of $\rs$.   The remaining two models, \sage\ and \tum , 
yield stronger dependence on $\rs$ with difference of more than more than 50\% in $c_2$. 

 The \galc\ mock stands out in two respects, it has the strongest sensitivity to the  environment
 (largest $|c_2|$), and,   in accordance with  \figrs{fig:P100}{fig:P20},  exhibits an inverted dependence on $\delta$ (negative $c_2$).
The results are a function of the SFR threshold, but we find similar numbers for $\sfr>12\myr$. For example, the inverted dependence in  $\galc$ persists to $\sfr>12 \myr$ with 
$(c_2,d_2)=(-0.0371\pm 0.0007, -0.005\pm 0.0002 )$  and $(-0.036\pm  0.0004,-0.0031\pm 0.0015)$, respectively for  $\rs=20$ and $100\hmpc$.  This model shows the strongest change with the SFR threshold. Parameters in the other mocks 
change at $<15\%$.

\begin{table*} 
\centering
\caption{The parameters of the fitting formulae in \eqs{eq:fitone}{eq:fittow} for a threshold  $\textrm{SFR}>10\myr$  }
\begin{tabular}{c|c |c  | c |  c | c | c| }
\label{tab:env}
          & $ \overline {\lsfr} $ &  $c_1$ & $c_2$     & $ \sigma^2_{\lsfr}$ &      $ d_1$ & $ d_2$ \\
\hline
\multirow{2}{*}{SAG}   & \multirow{2}{*}{1.2926}  &        $ 1.2899 \pm 0.0002  $ & $0.0213 \pm 0.0005 $ &       \multirow{2}{*}{0.0634}       &     $ 0.0614 \pm 0.0001 $ &  $0.0151 \pm 0.0003$ \\ 
                                          &         & $ 1.2925 \pm 0.0001  $ & $ 0.0263 \pm 0.0033 $  &                                             &  $ 0.0633 \pm 0.0001  $ &  $0.0175 \pm 0.0019$ \\                            
                                                           \hline
\multirow{2}{*}{SAGE}     &  \multirow{2}{*}{1.2566}     & $ 1.2553 \pm 0.0001 $ &  $ 0.0132 \pm 0.0003 $ & 	\multirow{2}{*}{0.0343}	& $ 0.0340 \pm 0.0001 $ &  $ 0.0030 \pm 0.0001$ \\ 
                                              & &  $ 1.2566 \pm 0.0001 $ & $ 0.0174 \pm 0.0021 $  & 						& $ 0.0343 \pm 0.0001 $ & $ 0.0033 \pm 0.0005$ \\                                      \hline
\multirow{2}{*}{Galacticus} & \multirow{2}{*}{1.3598} & $ 1.3644 \pm 0.0002  $ & $  -0.0428 \pm 0.0007 $ & 	\multirow{2}{*}{0.0699}	& $ 0.0705 \pm 0.0001 $ & $  -0.0066 \pm 0.0002 $\\ 
                                                &  &  $ 1.3599 \pm 0.0002 $ & $ -0.0444 \pm 0.0044 $ & 						& $ 0.0699 \pm 0.0001$ & $ -0.0045 \pm 0.0015 $\\ 
                                 \hline
 \multirow{2}{*}{UM SMDPL}   & \multirow{2}{*}{1.2724}        & $ 1.2724 \pm 0.0004 $ & $ -0.0005 \pm 0.0015 $  & \multirow{2}{*}{0.0416} & $ 0.0416 \pm 0.0001 $ & $ 0.0002 \pm 0.0005$ \\ 
						&  &  $ - $ & $ - $ &         											      &  $ -$ & $-$ \\ 
                                     \hline
 \multirow{2}{*}{UM MDPL2}  & \multirow{2}{*}{1.2676} & $ 1.2665 \pm 0.0001 $ & $ 0.0086 \pm 0.0004 $ &          \multirow{2}{*}{0.0414}&   $ 0.0411 \pm 0.0001 $ & $ 0.0014 \pm 0.0001$ \\ 
						&   & $ 1.2676 \pm 0.0001 $ & $ 0.0049 \pm 0.0024 $ &                                            & $ 0.0413 \pm 0.0000 $ & $ 0.0003 \pm 0.0008 $\\ 
                                     \hline                                 
\end{tabular}
\end{table*}


\section{Galaxy biasing as a function of the SFR } 
\label{sec:biasing}

Assume we have a (volume limited) sample of $\ngal$ galaxies with positions $\br_i$ in a large volume $V$. Theoretically,  the number density  contrast $\dgal$ is expressed in terms of a sum over Dirac delta functions
\begin{equation}
\dgal=\frac{V}{\ngal}\sum_{i=1}^{\ngal} \delta^\textrm{D}(\br -\br_i) -1 \; .
\end{equation}
This form, although of little practical use, stresses the importance of shot noise resulting from the discrete nature of the distribution of galaxies. 
Practically, we  
 generate a galaxy density field from  each simulation output, by interpolating  the galaxy distribution on a cubic grid using the Cloud-in-Cell
(CIC) scheme. The grid size is   and $256^3$ for SMDPL, $512^3$  for MDPL2 and $470^3$ for BigMDPL. Here also we use  galaxies with  $M_*>5\times 10^9 \textrm{M}_\odot$  and $\sfr >10 \myr$. 

\begin{figure*}
\vskip 0.2in
  \includegraphics[width=0.98\textwidth]{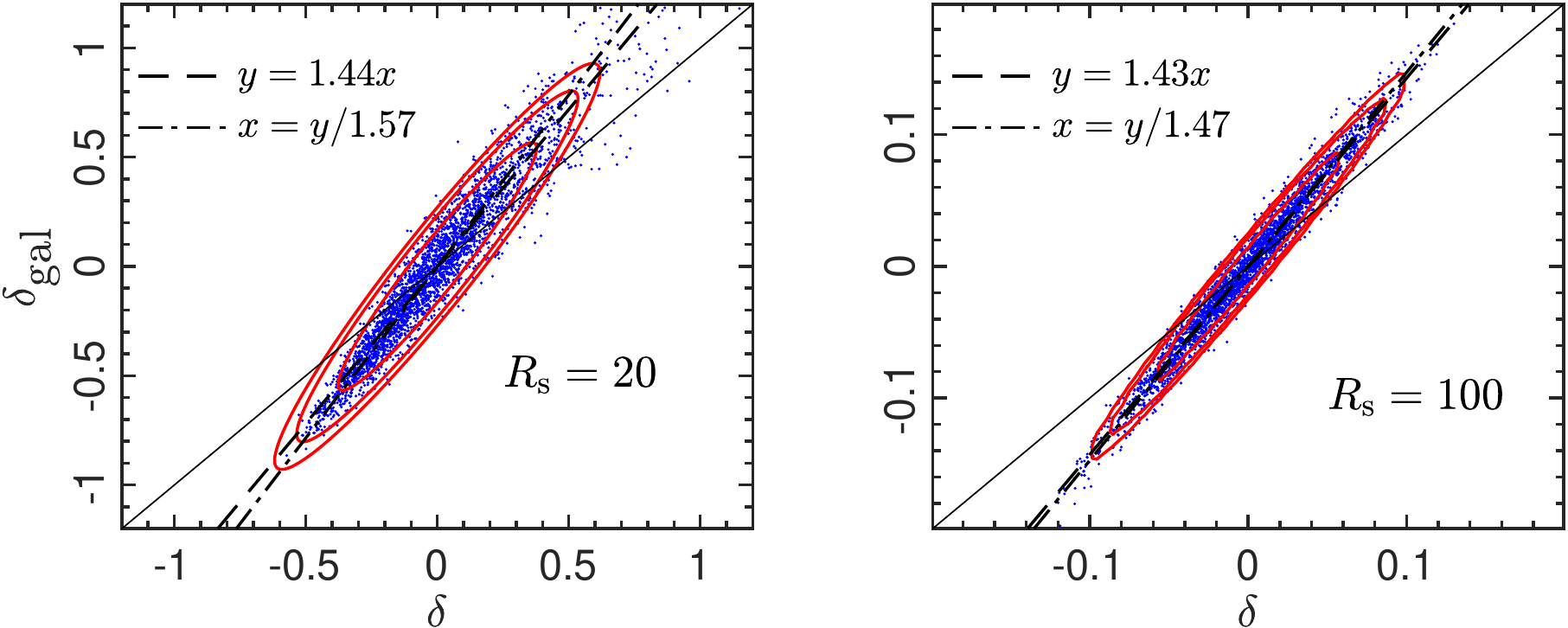} 
  \vskip 0.0in
 \caption{ Scatter plot of the smoothed galaxy versus DM density fields for the $z=1$ \sag\ galaxies. Left and right panels correspond, respectively to   TH smoothing of   widths $R_s=20$ and 100 $\hmpc$.  Slopes obtained via linear regression of  $\dgal$ on $\delta$ and vice versa are plotted as the dashed and dash-dotted lines, respectively, and indicated in the labels. To guide the eye, the diagonal line $\dgal=\delta$ is plotted in thin solid }
\label{fig:scattbiasSAG}
\end{figure*}

We make various comparisons between   galaxy and mass density fields  on the grid. 
A visual impression of the biasing relation is offered in terms of a scatter plot of $\dgal$ vs $\delta$ in  Fig.~\ref{fig:scattbiasSAG}. 
For clarity only a small fraction of the densities on the  grid are plotted as the blue points. 
The contour lines  mark the boundaries containing 68\%. 90\% and 95.4\% of the points. The contours were  computed by fitting a 2D   gaussian normal PDF to the distribution of points in the plane $\delta-\dgal$. 
The dashed and dash-dotted lines represent, respectively, the linear regression of $\dgal$ on $\delta$ and vice versa. 
Using the expressions in \S\ref{sec:slopes}, in which we discuss the details of the regression procedure,
the slope of the simple linear regression of $\dgal $ on $\delta$ is given by
\begin{equation}
p=\frac{\sum\delta_\alpha \delta_{\textrm{gal},\alpha}}{\sum \delta_\alpha}\; , 
\end{equation} 
where the subscript $\alpha$ refers to grid points.  
If the biasing relation is indeed well described  by \eq{eq:linbias}, then the ensemble average of this expression is approximated as 
$\av{p}=b$ thanks to  $\av{\delta \varepsilon}=0$. Therefore, the slope of the dashed lines in the figure should serve as a statistically unbiased estimate of $b$. The statistical $1\sigma$ uncertainty on the slope is given by 
\begin{equation}
\label{eq:sigma}
\sigma^2_p=\frac{\sum(\delta_{\textrm{gal},\alpha}-p \delta_\alpha)^2}{N_1\sum \delta_\alpha^2}\; ,
\end{equation}
where $N_1$ is the number of independent grid points. 
Since the densities are smoothed on a grid, we need to consider that  only a fraction of the points are statistically independent. We estimate the number of independent grid points as $N_1\sim (3/4\pi)(L/\rs)^3$ and apply the expression in \eq{eq:sigma} using density values at $N_1$ randomly selected  grid points. 
This gives $\sigma_p\approx 0.006$ and $0.015$, respectively, for the smaller and larger $\rs$.
Thus, the bias factors, $b=1.44$ and 1.43, estimated as the slopes of the dashed lines in the two panels of  Fig.~\ref{fig:scattbiasSAG} are consistent within the $1\sigma$ statistical error.

Given the inferred slopes, the variance of the stochastic term $\varepsilon$ in Eq. \ref{eq:linbias} is estimated  as $\sigma_\varepsilon^2=\textrm{Var}(\dgal -p\delta)$. 
For $\rs =100\hmpc $,  we find $ \sigma^2_\epsilon=1.1\times 10^{-4}$, which includes shot noise and intrinsic scatter in the bias relation. 
Following  \S\ref{sec:SN},  the shot noise contribution in this case is $\sigma^2_\textrm{SN}=7.3\times 10^{-5}$. Since 
$\sigma^2_\delta=1.75 \times 10^{-3}$ and  $ \sigma^2_{\dg}=3.7\times 10^{-3}$ are much larger than $  \sigma^2_\epsilon$, the inverse regression of $\delta$ on $\dgal $ leads to a similar slope.   
The same conclusion applies  to  $\rs=20\hmpc$ where  $\sigma^2_\epsilon =1.18\times 10^{-2}$, 
$\sigma^2_\textrm{SN}=9.3\times 10^{-3}$, $\sigma^2_\delta=6.2\times 10^{-2}$ and $\sigma^2_{\dg}=1.4\times 10^{-1}$.

The relation in Eq.~\ref{eq:linbias} is assumed to hold  between between the density field on any scale as long as it is large enough. Decomposing the fields in Fourier modes, the relation yields 
\begin{equation}
\label{eq:linbiask}
\delta_{\textrm{gal},k}=b \delta_k +\epsilon_k\; . 
\end{equation} 
We examine now  the power spectra $P_\textrm{g}(k)=\av{|\delta_{\textrm{gal},k}|^2}$ of the galaxy distribution  and 
$P_\textrm{DM}(k)=\av{|\delta_k|^2} $ of the corresponding dark matter density field. 
We FFT the un-smoothed $\dgal$ and $\delta$  on the grid into Fourier space and compute the respective  power spectra. 
We remove the contribution, $\ngal^{-1}$, of the shot noise from the  galaxy power spectrum  $P_\textrm{g}$ \citep{Peeb80}. 
Thanks to the large number of DM particles in the 
simulation, shot noise is negligible in $P_\textrm{DM}$.

\begin{figure}
\vskip 0.2in
  \includegraphics[width=0.4\textwidth]{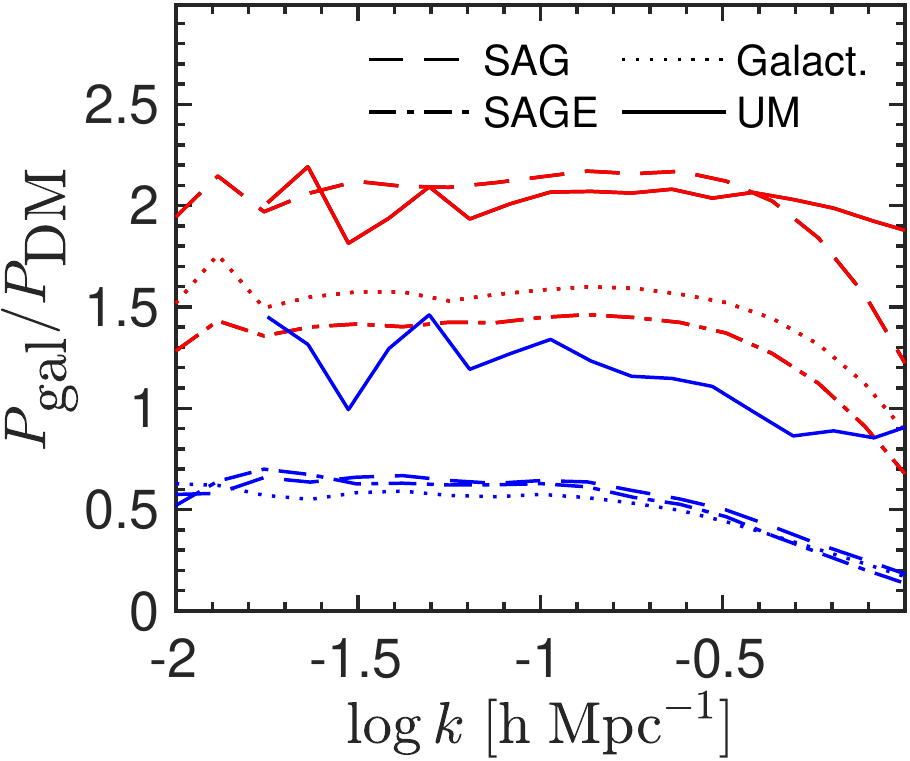} 
  \vskip 0.0in
 \caption{The ratio of the galaxy to the DM power spectrum  at $z=1$ (red curves) and $z=0$ (blue). Only the full \tum\ in SMDPL is plotted.}
\label{fig:SFRbias}
\end{figure}

In Fig~.\ref{fig:SFRbias}, we plot the ratio of the galaxy to the DM power spectra 
for the various mocks  at 
$z=0$ (blue curves) and $z=1$ (red).  
The Nyquist frequency $k_\textrm{N}=\pi N /L=1.6 \; {h}\;  \mathrm{Mpc}^{-1} $  and $2.0\;  {h}\;   \mathrm{Mpc}^{-1}$ for MDPL2 and MDPL2, respectively. The \tum\ curves are noisier than the others due to to the significantly smaller number of \tum\ mock galaxies (c.f. Table \ref{tnumgal}).
The decline of the ratio at $\log k \gtsim -0.5$  for mocks from the MDPL2 simulation (i.e. all curves except  \tum ) is due to aliases of the CIC interpolation. Since we are interested 
in the large scale regime,  we have not made any special effort to correct for these aliases  \citep[e.g.][]{Jing2005}.
The power spectrum ratio versus the wavenumber, $k$,   is an indication to the dependence of the bias factor on scale. 
For all models, the figure clearly demonstrates a  very weak dependence on $k$  in the range  $-2 < \log k <-1$, strongly motivating linear biasing. 
At $z=0$ (blue) the ratio is larger than unity only for   \tum . In the remaining models  at this redshift,  the ratio is less than unity, meaning that
the galaxy distribution is less clustered than the dark matter.

\begin{figure}
\vskip 0.2in
  \includegraphics[width=0.4\textwidth]{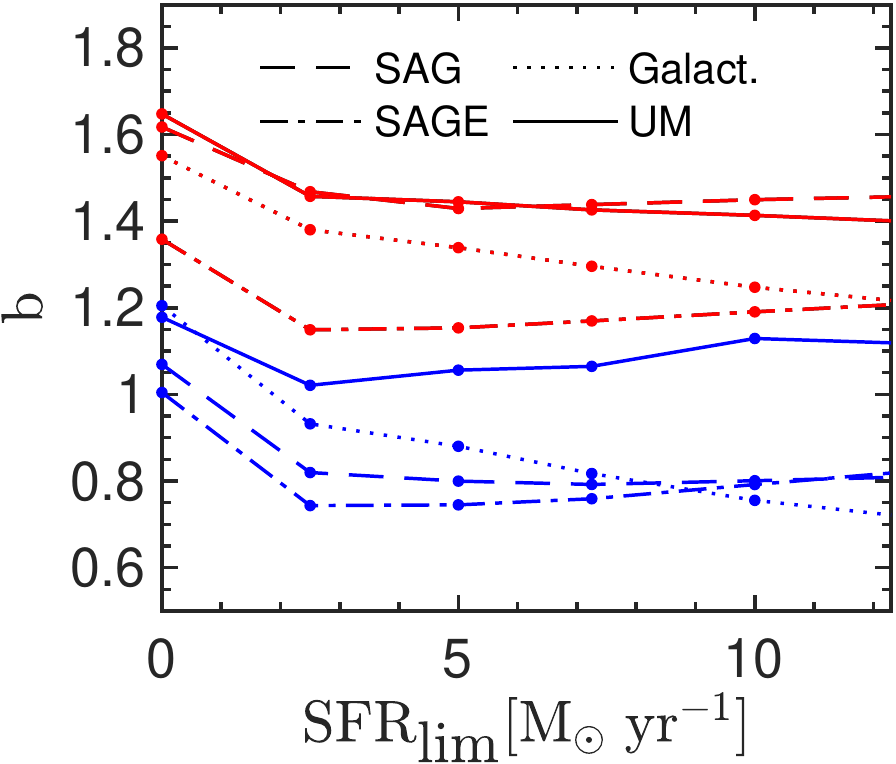} 
  \vskip 0.0in
 \caption{The bias factor computed using Eq. \ref{eq:bPk} of galaxies with SFR above a limiting value. 
 Red and blue curves correspond to $z=1$ and $z=0$, respectively.   }
\label{fig:biasSFR}
\end{figure}

It is interesting to examine the biasing relation as a function of the threshold imposed on SFR. 
We compute  the bias factor according 
\begin{equation}
\label{eq:bPk}
b^2=\frac{\overline{P_\textrm{gal}}}{\overline{P_\textrm{DM}}}
\end{equation}
where the overline refers to the average of the power spectrum over  the range $-2 < \log k <-1$. 
\figr{fig:biasSFR}  plots  $b$ as a function of a lower threshold imposed on the SFR.  
The bias factor of galaxies with very low SFR (quenched star formation) is highest.  
 These galaxies tend  to live in massive halos
and thus are  strongly biased (clustered). Actively star forming galaxies are associated with less massive halos with lower $b$, as seen in 
\figr{fig:biasSFR} for all models. As soon as star formation is active, the  the bias factor is nearly constant versus the SFR threshold
with \galc\ showing the strongest dependence.
{This is in agreement with \cite{Angulo2014} who analysed the biasing of SFR selected galaxies extracted 
from  the Millennium-XXL simulation \cite{Angulo2012}  using the L-Galaxies SAM \cite{Springel2005}.  
Based on the  galaxy and DM  correlation functions at separations $60-70\hmpc$, they find that the bias factor depends weakly on the SFR (expressed in terms of number density in their case) with $b\approx 1.1$ and $.7$, respectively,  at $z=1$ and $z=0$. In \figr{fig:biasSFR},  these values are best matched by the \sage\ galaxies, while the other SAMS predict larger $b$ values.}
 
The mean stellar mass for the \tum\ mock galaxies with $\sfr >10 \myr$ at $z=0$ 
is $M_*=7.9\times 10^{10}\textrm{M}_\odot$ and we find a similar value $M_*=5\times 10^{10}\textrm{M}_\odot$
for \sag.  For these values of $M_*$
the corresponding halo mass in the models is $\sim 2\times 10^{12} \textrm{M}_\odot$  
with a factor of 2 scatter 
 \citep[cf. Fig.~8 in][]{Knebe2018}. 
According to \cite{Comparat2017}  who analyzed halo bias in the \textsc{MultiDark} simulations, the relevant bias for this halo mass range is around unity with some scatter. Taking into account the range of the halo mass and the scatter in the biasing relation, we find the difference in  $b$,  measured from the slopes in Fig.\ref{fig:scattbiasSAG}  
and $P_\textrm{gal}/P_\textrm{DM}$ between \tum\ and the other models is completely reasonable.

\section{ The Growth rate from the Speed-from-Light Method (\meth )}
\label{sec:growth}
Redshift surveys provide ``observed redshifts'', $z$, of galaxies.   LOS peculiar velocities  introduce a shift between $z$ and the cosmological redshifts $\zc$ according to \citep{SW}, 
\begin{equation}
\label{eq:zzct}
\frac{z-\zc}{1+z}\approx \frac{V}{c}\; , 
\end{equation}
where $V$ is the physical peculiar velocity of a galaxy,   the speed of light is $c$ and we have neglected terms related to the gravitational potential and higher order in $V/c$. Further, we will not consider the important effect of magnification by gravitational lensing in this paper (cf.  \S\ref{sec:dl} for some considerations of this effect).

Cosmological redshifts can only be derived from actual distances  and, therefore,   are impossibles to measure for most galaxies in redshift surveys. Therefore, the (luminosity) distances, $\dl$ computed at  $z$ rather than $\zc$, are  used to derive  SFRs and  stellar masses of    galaxies   from  the observed fluxes. 
This obviously holds for the derivation of any  intrinsic luminosity of galaxies, but we will phrase the  relevant relations in terms of the SFR.

Given the measured flux, $F$,  the observed SFR is given by
\begin{equation}
\sfr_\textrm{obs}=4\pi \dl^{2}(z) F\; .
\end{equation}
The true intrinsic  SFR is instead (cf. \S\ref{sec:dl})
\begin{equation}
\sfr=4\pi \da^2(\zc)(1+z)^4 F \; ,
\end{equation} 
where $\da(\zc)=\dl(\zc)/(1+\zc)^2 $ and similarily  for $\da(z)$,  is the angular diameter distance. 
Collecting up the terms, we write
\begin{equation}
 \label{eq:logSFRobs}
 \log \textrm{SFR}_\textrm{obs}=\log \textrm{SFR}+ \vsh  \; ,
 \end{equation}
 with
  \begin{equation}
 \label{eq:NU}
\vsh=2\log \left[\frac{\da(z)}{\da(\zc)} \right]\; .
 \end{equation}
In \S\ref{sec:dl} we show  that to first order in $V$,  
\begin{equation}
 \label{eq:vsha}
 \vsh\approx  \frac{V}{c}  \mathcal{D}(z)\; ,
\end{equation}
where
 \begin{equation}
 \label{eq:Dz}
 \mathcal{D}(z)=0.868
 \left[\frac{c}{H(z) \da(z)}-1 \right]\; .
 \end{equation}
  The different signs for the  two terms in the square brackets reflect the fact that the corrections for 
  relativistic beaming  and 
 the shift in $\dl(z)$ are in opposite directions. 
 For a given $V$, the first term  is $\sim V/cz$ for $z\ll 1$, while the second term is independent of redshift.
 At $z=1$, this gives $\vsh=1.4\times 10^{-4} (V/100\kms)$. Lets compare this to the modulation of $\lsfr$ 
 due to environmental density dependence.  Taking the parameter $c_2=0.026$ from Table \ref{tab:env} for \sag\ as an example. The density rms  for $\rs=100\hmpc$ is $\sigma_\delta=0.042$. Thus the modulation in the mean $\lsfr$ versus the environment density is typically 
 $c_2\sigma_\delta=10^{-3}$. This is almost an order of magnitude larger than $\vsh$. 
   Thus, as expected, at $z\sim 1$, in the $\lcdm$ framework the \meth \ cannot be used as a direct probe of  $V$ as was done  at $z\ll 1$ for 2MRS and SDSS by \cite{Branchini2012,Feix2015,Feix2016}.  Of course, it can still be useful for constraining non-standard models predicting unusually large amplitude of 
   the velocity field at high redshift. 

 But now let us focus on the \meth  { method}.
The distribution of galaxies in the redshift survey,  allows a reconstruction of the peculiar velocity field. 
 Linear  theory of gravitational instability relates
the 3D physical  velocity field, $\bv$ to the mass density field in \textit{real (comoving distance) space }as 
\begin{eqnarray}
\label{eq:lin}
{\nabla^2\phi}&=&a H f(\Omega)   \delta 	  \\
\label{eq:linbeta}
&\approx & aH \beta  \dgal \; ,
\end{eqnarray}
where  we assume a potential flow, $\bv=-\grad \phi$,  and  adopted  a linear biasing relation \eq{eq:linbias} with negligible scatter to arrive at the second line.  
The solution to this Poisson equation  is obtained from the galaxy distribution 
as a function of a single parameter\footnote{The parameter $H$ disappears from the equation 
if distances  are expressed in $\kms$.} $\beta$ in Eq. \ref{eq:beta}
The spatial derivative in the relations \eqs{eq:lin}{eq:linbeta} is with respect to the comoving distance coordinate, 
$\br$, and  the density fields  are assumed in real  space.
In the application to redshift surveys, \eq{eq:linbeta} needs to be modified to account for the difference in the comoving distance being derived from  the observed redshift $z$ rather than the cosmological redshift $\zc$.  
Denoting the galaxy density in \textit{redshift space} by $\dgal^\textrm{red}$, the equation becomes \citep[e.g.][]{ND94}, 
\begin{equation}
{\nabla^2\phi}+ \frac{\beta}{r^2}\frac{\partial}{\partial r}\left( r^2\frac{\partial \phi}{\partial r}   \right) =a H \beta \dgal^\textrm{red} \; . 
\end{equation}
We see that the degeneracy between  $f$ and $b$ is maintained and the solution remains  only a function of  $\beta$. There is an important difference though; the solution is not linearly proportional to $\beta$, and the equation needs to be solved for every value of $\beta$ 
of interest \citep[cf.][for details]{ND94}. 
Non-linear effects are important on small scales, especially in redshift space where incoherent motions and  finger-of-god effects smear out structure in the los direction. 
However, we are interested in tens  of Mpc scales where linear theory is completely satisfactory for the purpose of recovering the peculiar velocity \citep{KN16}. 
The tests we perform below are tailored to the  expected uncertainties from the finite number of particles and environmental  
density effect. 
 Thus, for simplicity  of presentation and clarity of the results, we choose to work with real space density fields and thus we only use velocity reconstructed from \eq{eq:linbeta}.  
 
The \meth\ seeks $\beta$ as the value that renders a minimum in  the function 
 \begin{equation}
\label{eq:chis}
 \tilde  \chi^2(\beta)=\sum_{i \in \textrm{gal}} \left[\lsfr_{i,\textrm{obs}}-\vsh_i(\beta)\right]^2\; .
 \end{equation}
 
 We will assume the approximate expression $\vsh=V \mathcal{D}/c$  (see \eq{eq:vsha}) and that the velocity model is linear in $\beta$. We also assume that we are given the velocity field $V_1$ corresponding to a solution  \eq{eq:linbeta}  with a fixed value for $\beta=\beta_1$. Thus, 
 $\vsh(\beta)=\beta \vsh_1/\beta_1$ and the problem reduces to a simple linear regression as described in \S\ref{sec:slopes}. The minimum  condition   $\partial \tilde \chi^2/\partial \beta=0$, gives
 \begin{equation}
 \label{eq:slope}
 \frac{\beta}{\beta_1}=\frac{\sum_i \Delta \vsh_{i,1} \; \Delta \lsfr_{i,\mathrm{obs}}}{\sum_i (\Delta \vsh_{i,1})^2}\; .
 \end{equation} 
 We have defined 
 \begin{equation}
\Delta \vsh_{i,1}= \vsh_{i,1}-\overline{\vsh}_1
\end{equation}
where an over-line indicates a mean over galaxies   in the sample and the same definition applies to $\lsfr_\textrm{obs}$. The $1\sigma$ error on $\beta $ is given by 
\begin{equation}
\label{eq:rigerror}
\frac{\sigma^2_\beta}{\beta_1^2}=\frac{\sum_i \left(\Delta \lsfr_{i,\mathrm{obs}} -\frac{\beta}{\beta_1}\Delta \vsh_{i,1}\right)^2 }{N_\textrm{gal}\sum_i (\Delta \vsh_{i,1})^2}\; ,
\end{equation}
 where $N_\textrm{gal}$ is the number of galaxies.

\subsection{A rough estimate of the error}
Before we present full results for the expected errors on  $\beta$ we give here a 
rough estimate. 
This is most easily achieved if  we take 
  $\lsfr_\textrm{obs}=\lsfr$, so that minimization of the function $ \tilde \chi^2$  should yield $\beta$ consistent with zero for a model of the form $\vsh(\beta)=\beta\vsh_0$, where $\vsh_0$ is computed from the true peculiar velocities (cf. \eq{eq:NU}). Of course, in reality $\vsh$ will be obtained from the velocities recovered from $\dgal$, but we are 
 only  interested in a rough estimate in this section.

 Galilean invariance implies the absence of correlation between the properties of galaxies and their velocities. Therefore, 
the ensemble average of \eq{eq:slope}  over 
 all possible realizations 
 of $\lsfr_i$ yields $\langle \beta \rangle=0$..
 This is valid even if the star formation depends on the underlying mass density since the velocity and density are also uncorrelated.
An estimate of the variance of the  scatter around  $\langle \beta \rangle=0$, is
 \begin{equation}
\sigma_\beta^2\equiv \langle \beta^2 \rangle=\frac{1}{N}\frac{\sigma^2_{\lsfr}}{\sigma^2_\vsh}\; . 
 \end{equation}
 where $\sigma^2_\vsh=\overline {(\vsh-\overline \vsh)^2}$ and $\sigma^2_{\lsfr} =\av{ (\lsfr -\av{\lsfr})^2}$. In the limit of  a large number of objects  $\sigma^2_{\lsfr}=\overline {(\lsfr-\overline \lsfr)^2}$.
 At $z=1$, \eq{eq:NU} gives $\vsh= 4.2\times 10^{-4} (V/300)$. For the \sag\ mock galaxy catalogs, $\sigma_{\lsfr}\approx 0.25$ for $\sfr>10$ $\textrm{M}_\odot \textrm{yr}^{-1}$. Thus 
\begin{equation}
 \label{eq:rougherror}
\sigma_\beta={0.13}\left(\frac{2\times 10^7}{N}\right)^{0.5}\left(\frac{300\kms}{\sigma_V}\right)
\end{equation} 
At $z=1$, the simulations give $\sigma_V= 300 \kms$ for the 1D  rms of unfiltered galaxy velocities.


\begin{figure*}
\vskip 0.2in
  \includegraphics[width=0.9\textwidth]{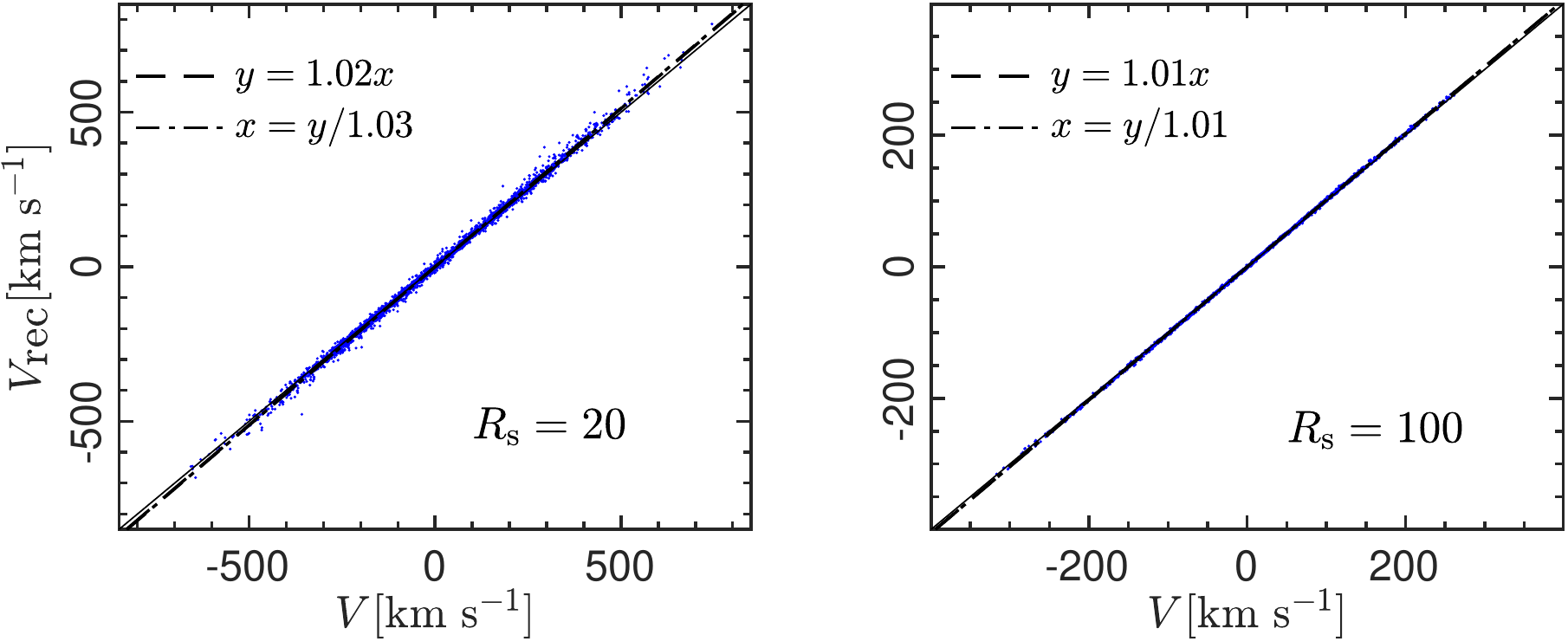} 
  \vskip 0.0in
 \caption{
 Comparison  between the velocity predicted from the DM density field
 using the linear relation \eq{eq:lin}.
 Both velocity fields  are given on a grid and  smoothed with a tophat window of  width $20\hmpc$ (left) and $100\hmpc$ (right), as indicated in the figure. For clarity, only a small fraction of the grid points are shown as the blue dots. 
 Dashed and dash dotted lines represent linear regression of $V_\textrm{pred }$ on $V$ and its  inverse. The slopes are indicated in the plots. The diagonal  (thin solid) is plotted to guide the eye.
}
\label{fig:gvDMDM}
\end{figure*}

\begin{figure*}
\vskip 0.2in
  \includegraphics[width=0.9\textwidth]{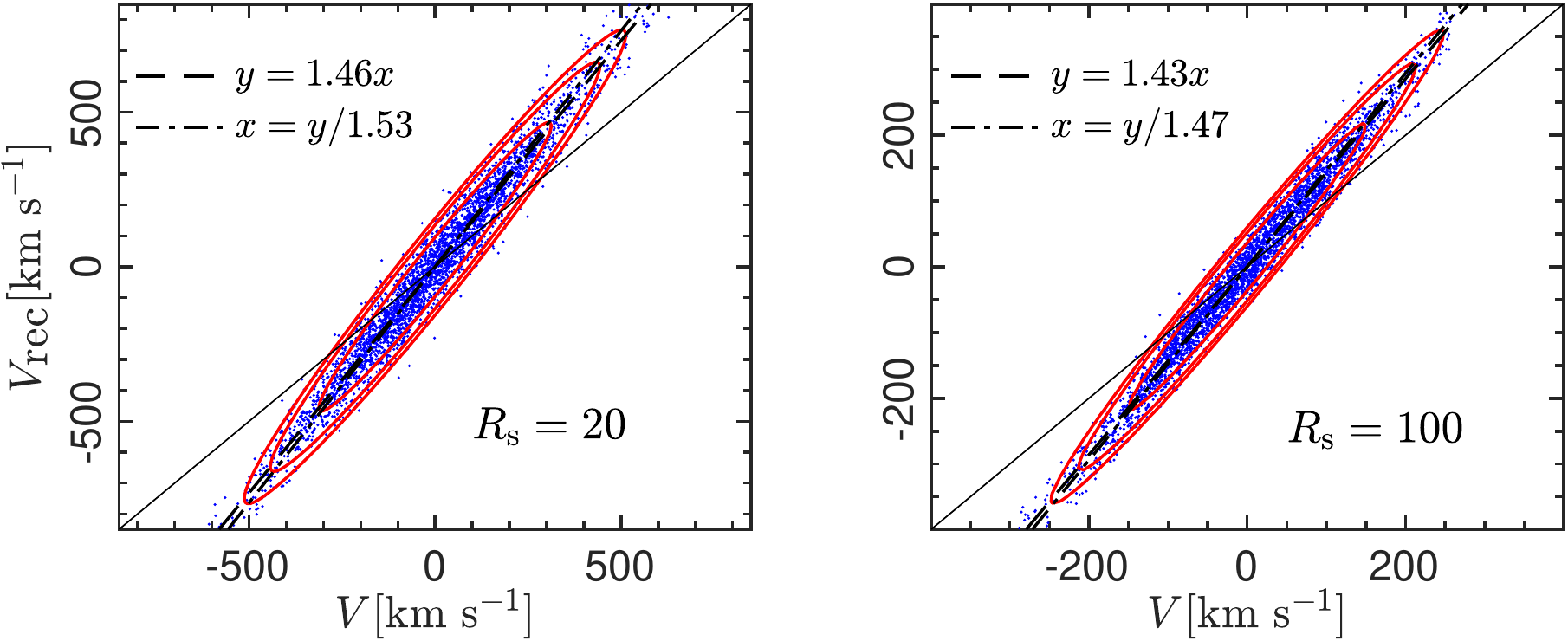} 
  \vskip 0.0in
 \caption{ 
 Same as the previous figure but for  $V_\textrm{pred}$ from the galaxy density field in the SAG simulations.
Red contours enclose   \%68, \%90 and \%95  of the points. 
The slope of the dashed line
is consistent with the bias factor for SAG galaxies with $SFR_\textrm{lim}\approx 10$ in Fig.~\ref{fig:biasSFR} (dashed curve).}
\label{fig:gvsmooth}
\end{figure*}

\begin{figure*}
\vskip 0.2in
  \includegraphics[width=0.9\textwidth]{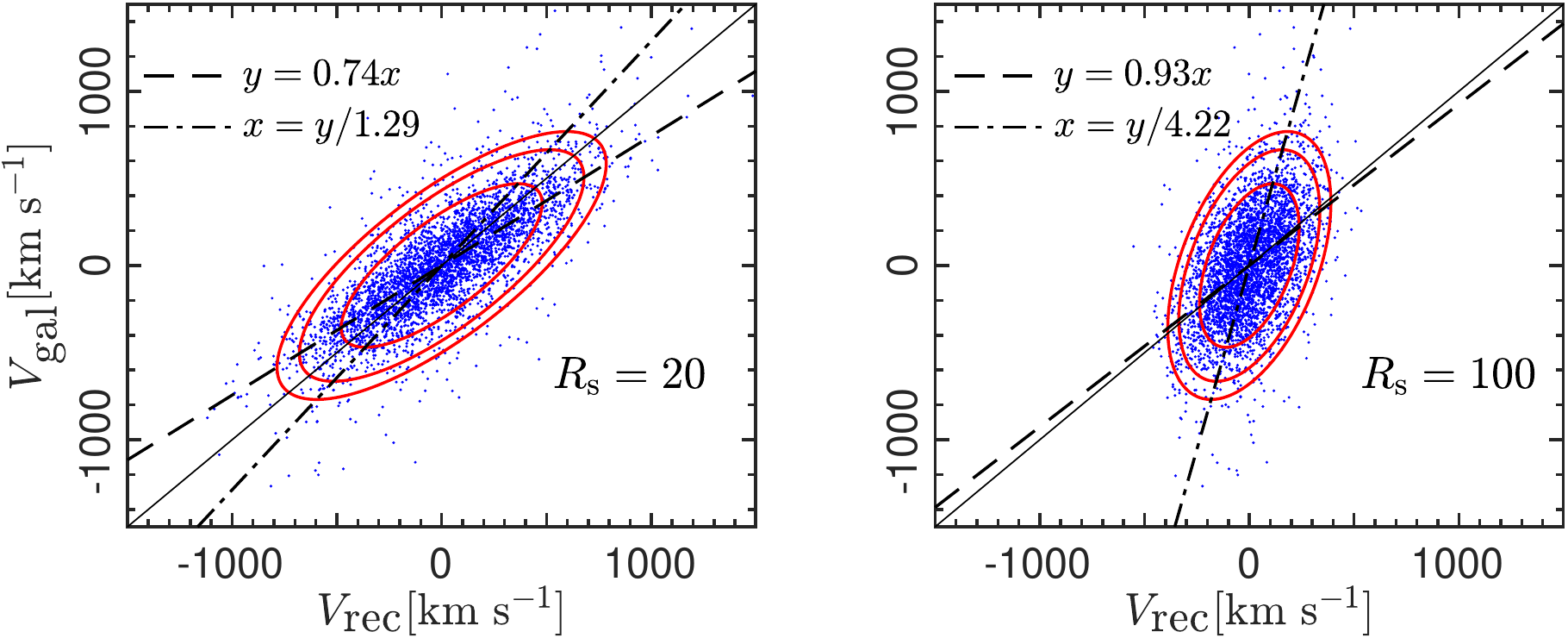} 
  \vskip 0.0in
 \caption{The actual galaxy velocities versus the prediction from the relation using the galaxy density field smoothed on $R_\mathrm{s}=20$ and $100\hmpc$. In contrast to Fig~.\ref{fig:gvsmooth} the galaxy velocities here are 
 the raw un-smoothed galaxy velocities.
 The three velocity components are compared for 1/15000 of the SAG mock galaxies.}
\label{fig:gvPART}
\end{figure*}

\subsection{Recovered vs. true velocities: smoothing matters}
\label{sec:recv}

The rough error estimate presented above is based on the true los velocities, while in a realistic application only the reconstructed
los velocities, $\vpr$,  from the galaxy distribution  smoothed on a large scale are available.  
Therefore, the $\beta$ estimate from the  \meth\  is  basically the slope of the  regression of the true 
$\vsh=\mathcal{D}V/c $ in \eq{eq:logSFRobs}
on the reconstructed  $\vsh_\textrm{rec}=\mathcal{D}\vpr/c $, where  $\vsh$ is perturbed by the random spread in $\lsfr$. 

Smoothing alone causes a statistical bias in the estimate of the  slope ($\beta$), i.e. 
even if  the smoothed true velocities, $V_\mathrm{s}$ where
used instead of $\vpr$, we expect a slope of the regression of $V$ on $\vs$ to differ from unity due to 
the correlation between $V-V_\mathrm{s}$ and $V_\mathrm{s}$ for the TH smoothing. 

In this section we explore the expected statistical  bias in the estimates of the slope.
We begin with a basic assessment of  ability of the linear theory relation, \eq{eq:lin}, at reconstructing the 
peculiar velocity from the DM density field, $\delta$.
In \figr{fig:gvDMDM} Cartesian velocity components $\vpr$ reconstructed from $\delta$ with the true value of the parameter $f$ are plotted 
against the true velocities, $V$. Both $\vpr$ and $V$ are provided on a grid and have been smoothed with a TH window of width $\rs=20$ (left panel) and $100\hmpc$ (right panel). For clarity, only a small randomly selected 
fraction of the grid points is plotted.  This figure refers to the $z=1$ output from the MDPL2 simulation.
Linear theory in this case performs extremely well. The slopes  of of the regression of $\vpr$ on $V$ as well as that of the inverse regression, 
are clearly very close to unity. Further, 
the scatter  in $\vpr$ versus $V$ is negligible even for the smaller smoothing.  

Next we turn to reconstruction from the galaxy density field, using \eq{eq:linbeta} with $\beta=f$. 
The results are shown in \figr{fig:gvsmooth} for the $z=1$ \sag\ galaxies 
selected above the Euclid cut $\sfr >10 \myr$ 
 (cf. Table \ref{tnumgal}).
The contours contain  68\%. 90\% and 95.4\% of the points, obtained by fitting a 2D  normal   PDF to the distribution of points in the $ \vpr-V$ plane.
The results are quite different from the previous figure.  The regression slopes   clearly deviate from  unity in this case. 
Since $ \vpr$ is derived for $\beta=f$, the slope of the regression of $\vpr$ on $V$ should yield  the galaxy bias factor,  $b$.
Indeed, the slope  is very close  to the values obtained from the density scatter plot in \figr{fig:scattbiasSAG}  and from the ratio of the power spectra  as seen in \figr{fig:SFRbias} \&  \figr{fig:biasSFR} for \sag .

We know already from \figr{fig:gvDMDM} that inaccuracies associated with  linear theory reconstruction on the scales considered are negligible. That implies that the origin of the scatter in \figr{fig:gvsmooth} is   stochasticity in the biasing relation and the effect of shot noise
on $\vpr$. 
The rms of the residual between $\vpr $ and the best fit lines (dashed curves) in \figr{fig:gvsmooth} is 
$63.4\kms$ and $26.8\kms$ for $\rs=20$ and $100\hmpc$, respectively. 
To quantify the contribution of the shot noise we resort to 
appendix    \S\ref{sec:SN} where we derive the following  expression for the variance  of the shot noise effect on $\vpr$  \citep[see also][]{Strauss1992}
\begin{equation}
\label{eq:VSN}
\sigma^2_{V,\textrm{SN}}=\frac{(a H \beta )^2}{10 \pi \bar n\rs}\; .
\end{equation}
For \sag\ at $z=1$ and $\beta=f$ ,  we find  $\sigma_{V,\textrm{SN}}=36.4\kms$ and $16.3\kms$ for the smaller and larger $\rs$, respectively.  
This makes the contribution of biasing stochasticity  $51.9\kms$ and $21.2\kms$, respectively, for the two smoothing widths. 
Thus, the intrinsic stochasticity in the biasing relation  is the dominant contribution to the scatter.

So far we have made comparisons between $\vpr$ and $V$ smoothed on the same scale. In contrast, the model velocity in the \meth\  is smoothed, while the data $\lsfr_\textrm{obs}$  involves the true galaxy velocities. 
In \figr{fig:gvPART} we plot the true un-smoothed \sag\ galaxy velocities, $\vgal$,  vs $\vpr$ from the galaxy density field. Note that in this plot $\vpr$ is plotted on the $x-$axis.

Only   a sharp $k-$cutoff filtering, yields a vanishing correlation between the residual  $\vgal-\vpr$ and
  $\vpr$.  However,  an idealised sharp $k-$cutoff  filtering is impossible to apply to real data due to complicated
  observational window functions. 
For the more practical  TH smoothing the correlation  affects the regression slopes (cf. \S\ref{sec:slopes}). In addition, as  we have seen in \figr{fig:gvsmooth}, stochastic biasing and shot noise will 
add scatter to $\vpr$ as recovered from the galaxy density field. 

The scatter will also affect the slope but as we have seen the effect is small since both the forward and inverse regression  in \figr{fig:gvsmooth} have similar slopes. 
Therefore, it is not surprising that the regression slopes in \figr{fig:gvPART} are quite different from those in
\figr{fig:gvsmooth} and from each other. The relevant slope for the \meth\ is that of the regression of $\vgal$ on $\vpr$. For both smoothing the slopes are  closer to unity than the obtained by a regression of the identical smoothing case of $V$  on $\vpr$ in \figr{fig:gvsmooth}. 
Just for comparison,  the  slope of $\vpr$ 
on $\vgal$ is  4.22 for $\rs=100\hmpc$  and is 1.43 for the identical smoothing case in the right panel of 
\figr{fig:gvsmooth}.

Therefore, in order to infer a statistically unbiased  $\beta$ estimate, one should 
carefully calibrate the result to account for the inherently different smoothing between the models and the data. 
The same point  also has been repeatedly emphasised  in regard to  the velocity-velocity comparison local surveys by \cite{Nusser1995,DN10,DNW96}.

\begin{figure*}
\vskip 0.2in
  \includegraphics[width=0.9\textwidth]{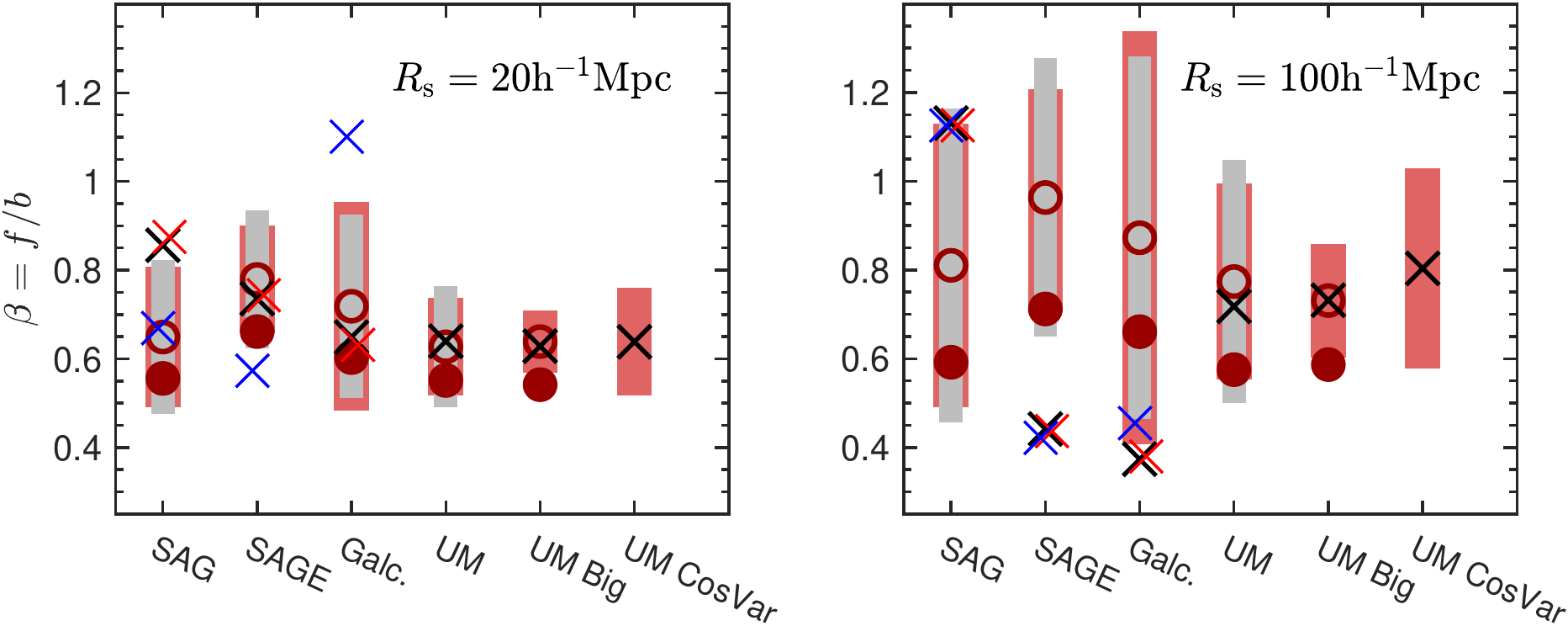} 
  \vskip 0.0in
 \caption{The slope of the regression of $\lsfr_\textrm{obs}$ on the peculiar velocity estimated  using 
 Eq~.\ref{eq:lin} 
 from the galaxy distribution is shown as the crosses for various mocks. Except for the last point to the right in each panel.  The error bars represent the $rms$ uncertainty due to the scatter in the SFR distribution. The last point, UM CosVar, also includes cosmic variance expected for a  volume of $1\hgpc$. 
 Left and right panels correspond to velocities recovered from the galaxy density field smoothed on $20$ and 
 $100\hmpc$ TH smoothing, as indicated. }
\label{fig:slope}
\end{figure*}

\subsection{Tests of \meth\ with mock catalogs}
\label{sec:testmock}

 We are now in a position to present  the results of our tests of  the \meth\ applied to mock catalogs and its sensitivity to SFR  and environment.
This exercise is to be viewed as an intermediate phase testing of the \meth\ towards detailed forecast and full application to realistic surveys.
Therefore, we do not make any attempt to run the tested on mock catalogs generated in light cones 
{matching the expected} number  of galaxies vs redshift. Instead, we simply place the simulation  box at $z=1$ and apply \meth\ to the simulated sample\footnote{For The MDPL2 volume of (comoving) $1(\hgpc)^{3}$ is equivalent   
to the volume of a \%5  of the sky between $z=1$ and 1.2.  For reference, the sky coverage of Euclid is \%35 of the sky } (c.f.    \S\ref{sec:volnum}).
\def\bmeth{\beta_{\meth}}
For the sake of conciseness  we define $ \bmeth$ as the $\beta$ estimate obtained from the application of the \meth . 

Following the discussion in \S\ref{sec:volnum},  since the  MDPL2 mocks have substantially smaller volume and less objects than  the typical corresponding sample obtained form next-generation surveys, our tests will provide upper bounds on the expected errors on $\beta$ from the \meth .  Furthermore, mocks  taken from the same MDPL2 parent simulations cannot be used to quantify cosmic variance. For this purpose we resort to   \tum\ BigMDPL  to quantify  the   uncertainty in  $\bmeth$ due to cosmic variance on the scale of the MDPL2. This is done by partitioning the BigMDPL  box into cubic sub-volumes each matching  the size of the MDPL2  box. The \meth\ is then  applied to the mock galaxies in each of these sub-volumes individually.  The scatter in the $\bmeth$ values from all the sub-volumes serves as an indication of the cosmic variance on the scale of the MDPL2 box and should 
serve as  an upper bound on the cosmic variance on the scale of real, upcoming surveys.

The ``observed" quantities  $\lsfrobs$  are  obtained by shifting the true $\lsfr$  by $\vsh$, as in  \eq{eq:logSFRobs}. 
We work with the  approximate expression in  \eq{eq:vsha} for $\vsh$ and write  the model $\vsh$ as
\begin{equation}
\vsh(\beta)=\frac{\beta}{f} \vsh_1  ,
\end{equation}
where 
\begin{equation}
\vsh_1=\frac{V_{\textrm{rec},1}}{c}\mathcal{D}\; \, 
\end{equation}
with  $V_\textrm{rec,1}$ being the  reconstructed  peculiar velocity $\vpr$  obtained from the galaxy density field  $\dgal$, as a solution to \eq{eq:linbeta} for $\beta=f(\Omega)$. 
Solutions are   provided for a TH smoothing of $\dgal$ with    widths of  $\rs=20$ and $100\hmpc$, respectively. 
In all cases we impose $\sfr >10 \myr$ 
 and  consider the three Cartesian components of the  true raw velocities of galaxies, $\bv_\textrm{gal}$,  to obtain three sets of $\lsfrobs $, respectively. Thus for each mock, we have  three values  of $\bmeth$ corresponding to  the three Cartesian components. 
In the results referring to MDPL2 mocks  we consider the mean of the three values for each mock.  Since the three components of the velocity are statistically independent, we effectively are assessing  
the \meth\ with a number of objects which is triple what we actually have in each mock. 
The   increase in the statistical sampling  should yield a more reliable error estimate to be used as an  indicator to precision that can be achieved with  $\mathcal{O}(10^7)$
as in the full Euclid's spectroscopic survey.

The $\bmeth$  values obtained from each set of mocks are summarized in \figr{fig:slope} for $\rs=20$ (left panel) and $100\hmpc$ (right panel), as indicated in the figure.
The labels \sag , \sage , \galc\ on the $x$-axis, correspond to the MDPL2 mocks generated using the full galaxy formation machinery in each model. 
 UM refers   \tum\ mocks in MDPL2, obtained by re-sampling the \tum\ SMDPL galaxies, as described in \S{\ref{sec:tum}}.  UM Big is the same as UM but for the BigMDPL simulation.
UM CosVar  show the results obtained from the 
sub-volumes of UM BigMDPL to assess cosmic variance on the scale of MDPL2. 

 The different symbols in the plots indicate $\bmeth$ values and their scatter obtained with different procedures:
 \begin{itemize}
 \item \textit{Empty  red circles}(\tikzcircle[purple, fill=white, ultra thick]{3pt}):   These are the $\beta$ values obtained by a regression of the true galaxy velocities on the model $V_1$. Therefore, these values represent 
what the \meth\ would yield theoretically in the limit of an infinite number of objects. and should be regarded as the reference value for the $\beta$ parameter.
 
 \item \textit{Black crosses } ($\pmb {\textcolor{black}{\times}}$):  { These are the $\bmeth$ value obtained from the \meth\ method in each mock, averaging over the three Cartesian velocity components } For the \tum\ mocks, they correspond to the average $\bmeth$ obtained from  20 random re-samplings of the full \tum\ in SMDPL.
 
\item \textit{Filled red circles}  (\tikzcircle[purple, fill=purple]{3pt}): {These symbols represent $\beta=f/b$ using the true $f$  and  $b$ obtained via a regression of the  galaxy density on the  DM density (both smoothed with the same window, e.g. \figr{fig:scattbiasSAG})}.
Very similar  values are obtained for $b$ estimated from the regression of the smoothed 
 $V_\textrm{rec,1}$ on the smoothed true velocities (e.g. \figr{fig:gvsmooth}).  
 
\item  \textit{Blue  crosses}  ($\pmb {\textcolor{blue}{\times}}$): \meth\ $\beta$ estimates including environmental dependencies using
 $\delta$  in \eq{eq:chisdel} (see \S\ref{sec:envmeth}). They are provided for the  \sag, \sage\ and \galc { cases} only.

\item  \textit{Red crosses} ($\pmb {\textcolor{red}{\times}}$):  These $\beta$ values are the same as the blue crosses but  using $\dgal$ in \eq{eq:chisdel} (see \S\ref{sec:envmeth})

\item \textit{Red vertical bands}  ( \tikz[baseline=-0.5ex]\draw[chestnut,fill=chestnut] (0pt,-4pt) rectangle ++(.2,.3); ): For all cases except UM CosVar. They represent the $1\sigma$ error  on $\bmeth$, computed according  \eq{eq:rigerror}. As stated above, this is the error on the mean of the three values obtained for the three Cartesian components or equivalently from 
 a single $\beta$ estimated from triple the number of galaxies in each mock. The red bands are centered on the empty red circles. 
 
 \item \textit{Red vertical bands }  ( \tikz[baseline=-0.5ex]\draw[chestnut,fill=chestnut] (0pt,-4pt) rectangle ++(.2,.3);  ): For the UM CosVar case. The bands represent the rms of the $\bmeth$ values obtained from sub-volumes 
 of the simulation BigMDPL. This scatter accounts for the combined error from 
 cosmic variance and the scatter in $\lsfr$.
 

 \item \textit{Grey vertical bands} ( \tikz[baseline=-0.5ex]\draw[gray,fill=gray] (0pt,-4pt) rectangle ++(.15,.3); ): These bands provide a crude estimate of the error expected for a Euclid-like survey, obtained by re-scaling the errors shown as red bands.
 Their estimate account for a) the number density vs. $z$ given in Table 3 in \cite{EuclidCollaboration2019}, b) the monotonic decrease of $\mathcal{D}$ with $z$ in \eq{eq:Dz}, and c) the dependence of  $\sigma^2_{\lsfr}$ with $z$ as a result of change in the limiting SFR for a given observed flux limit.

 \end{itemize}

 The error bars (red and grey bands) for $\rs=100\hmpc$ (panel to the right) are  roughly a factor of two larger than the corresponding 
errors for $\rs=20\hmpc$.  The smaller smoothing clearly captures  more details of the structure  of the velocity field and is bound to yield a tighter  fit to the data.  Quantitatively, the relative difference in the errors can be understood from  \eq{eq:rougherror} taking $\sigma_V=210$  and 
$110\kms$  for the small and large $\rs$, respectively. The increase in the error is consistent with the ratio of $\sigma_V$.

 The reason for the differences between the empty and filled red circles is explained in  \S\ref{sec:recv}. The empty  red circles are regressions of raw  on smoothed velocities which leads to distinct slopes 
from regressions done with equally  smoothed velocities. This  accounts for  the larger difference  for $\rs=100\hmpc$ and also the fact that 
the empty red circles are always above the filled red ones. 

Given the red error bars, $\bmeth$ (back crosses)  for \sag, \sage\ and \galc\ 
are consistent with theoretical 
values that would be obtained from an infinite number of data points (empty red circles). 
The largest deviation is for \sage\ with $\rs=100\hmpc$. We could not identify  any specific 
cause   for this
$\sim 2 \sigma $ deviation. Recovered vs true velocities follow the same correlation as for other models and it seems that this is just a random statistical fluctuation. 

The black crosses and the empty red circles for the \tum\ mocks in the 
fourth and fifth  sets, are very similar. This is expected since the \meth\ beta values here are averages over many random re-sampling realisations of the original \tum\ in SMDPL. 
Furthermore, the rms of the scatter in $\beta$ around the average is consistent with the error 
computed with \eq{eq:rigerror}.
 
 {The red error bar on the set UM CosVar  is comparable to the red error bands  for the \tum\ set}. This is important since it implies that the contribution of the cosmic variance to the error budget is small. The main error is due to the finite number of data points.
 
 The  $1\sigma$ error on $\beta$ obtained by   averaging the grey bands  over \sag , \sage , \galc\ and \tum\   is 
 $\sigma_\beta=0.17$ and  $0.34$ for the small and large smoothing widths, respectively. 
Using the bias factors computed for each mock, the corresponding errors on $f$ are 
$\sigma_f=0.22$ and $0.45$. 

\subsection{Environment in the \meth\ }
\label{sec:envmeth}
As already pointed out,   environmental dependencies should not introduce systematic errors on $\bmeth$, only affect the random uncertainty.
To prove this statement, we perform a simultaneous  \meth\  fitting of   $\beta$ and the  parameters $c_{1}$ and $c_2$ characterizing environmental dependencies (c.f. \eq{eq:fitone}). 
Generalizing \eq{eq:chis}, we write
 \begin{equation}
\label{eq:chisdel}
 \tilde \chi^2(\beta,c_1,c_2)=\sum_{i \in \textrm{gal}} \left(\lsfr_{\textrm{obs},i}-\frac{\beta}{f}\vsh_{1,i}-c_1-c_2\delta\right)^2\; . 
 \end{equation}
 The parameters $d_{1}$ and $d_{2} $  introduce a dependence on the scatter  of $\lsfr$ and do not affect into the description here. They would, however, enter the analysis if  each data point  $\tilde \chi^{2}$ is assigned a weight  according to the expected scatter in $\lsfr$. Even then the variation of the weighting  would be mild  at the level of $\sigma_{\delta }d_{2}/\sigma^{2}_{\lsfr}< 10\%$, taking $\sigma_{\delta}=0.24$ and $d_{2}$ values from  Table \ref{tab:env}.
 
 The minimum is obtained for, 
\begin{eqnarray}
\label{eq:betac}
\beta & = & \frac{\sum (\Delta \vsh_{1,i} {\Delta \lsfrobs}_i- c_2 \Delta \vsh_{1,i} \Delta \delta_i )}{\sum \vsh_{1,i}^2} \; ,\\
\label{eq:ctwo}
c_2&=& \frac{\sum(\Delta  \delta_i {\Delta \lsfrobs}_i- \beta  \Delta \vsh_{1,i} \Delta \delta_i ) }{\sum \delta^2_i}\; .
\end{eqnarray} 
where the $\Delta$ symbol implies that the mean value (over the galaxies) have been subtracted.  
{ The parameter $c_{1}$ accounts for the fact that the mean of $\delta$ over galaxies is not strictly zero, and will not be considered further in our analysis}

 In realistic applications, only  the galaxy density $\dgal$  is accessible to observations, 
and therefore   we repeat the minimization procedure using  $\dgal$ instead of 
 $\delta$.
The blue  and red crosses in \figr{fig:slope} are  $\bmeth$ as a solution to \eq{eq:beta} obtained with  $\delta$ and $\dgal$, respectively. 
The results are shown  for  \sag, \sage\ and \galc\ only.
For the larger smoothing (right panel) the results   are very close to the black crosses corresponding to  minimizing
\eq{eq:chis}. The results with $\dgal$ (red crosses) are well within the statistical error  for the smaller smoothing. The same is true for the blues crosses  for the \sag\ and \sage .
However, for the small smoothing,  the \galc\   estimate obtained  with $\dgal$ is significantly higher than all other estimates. 
This is perhaps not surprising since, as mentioned above, the   \galc\ galaxies are peculiar, exhibiting inverted environmental dependencies.

Since we have tripled the number of fitting parameters relative to the estimates in the previous sections, we should naively expect that the error on $\beta$ is substantially increased.  Lets us examine this issue in detail. 
Since the ensemble average $\av{\Delta \delta \Delta \vsh}=0$, we find that  the ensemble averages 
$\av{\beta} $ and $\av{c_2}$ in \eqs{eq:betac}{eq:ctwo} is the same as in the previous section (c.f. \eq{eq:slope}). 
For the same reason, the ensemble average of the  error covariance matrix is  diagonal. 
 It, therefore, suffices to examine \eq{eq:betac} 
alone to conclude that the added error due to  the term with   $ c_2$  is $\mathcal{O}(1/N_\textrm{gal})$   compared to $\mathcal{O}(1/\sqrt{N_\textrm{gal}}) $ for  the first term. As a result, we find that the computed errors on $\beta$ estimated with and without the inclusion of $c_2$ are very close to each other.


\section{Discussion}
\label{sec:discussion}


The  effects of the large scale environment   on galaxy properties have been  the focus of 
numerous  observational studies
\citep[e.g.][]{Hoyle2005,Bernardi2005,Park2007,Disney2008,Blanton2009,Tempel2011a,Hahn2015}.
Although the  environment has a strong impact on distribution of a single property (e.g. stellar mass),  all evidence points that it plays a meagre role in shaping   intrinsic  relations between galaxy properties  such as the fundamental plane and the Tully-Fisher relations \citep{Bernardi2005, Disney2008,Nair2010}. 
Despite the importance of the  environmental dependence 
 on these scaling relations 
has not yet been invoked as a strong constrain on galaxy formation models \citep[but c.f.][]{moyang04}. Part of the reason is that these relations involve observations which are sensitive to the distribution and kinematics of stars which are not properly followed in the SAM-DM simulations combination.
Also,  hydrodynamical simulations of galaxy formation may not be large enough to { quantitatively }explore this dependence 
There are, however, certain intrinsic  relations that involve properties which can be studies with SAM. 
{Among these, the one we touched upon in this work is the mean SFR at a given
stellar mass, shown in \figr{fig:MstSFRlh} for galaxies with high SFR ($>10\myr$). 
Our analysis has revealed important differences among various galaxy formation models. } The \tum\ and \sage\ models showed little evidence for dependence on 
environment, while in \sag, the SFR is boosted in less dense environment only for
stellar masses above $10^{11}\textrm{M}_\odot$. \galc\ galaxies are  associated with higher SFR in less dense environments only for $M_*\ltsim 10^{10}\textrm{M}_\odot$. 

A central goal of this work is to a make an assessment of the \meth\ at constraining the parameter $\beta$ with mock galaxy catalogs that incorporate as much physical effects as possible {and focus on star forming, line emitting objects.}
The application of the method at $z\simeq1$ relies  on the availability of a large number of galaxies in the survey and therefore we have geared the tests toward the  next generation spectroscopic surveys, focusing on the Euclid satellite mission.

The \meth\ has been criticized  on the ground that it could be susceptible to environmental dependencies in the SFR/luminosity distribution. The concern was already raised and addressed in previous papers by two of the current authors \citep{Feix2014,Feix2015} who argued that environmental dependencies affects only  a direct identification  of luminosity spatial modulations with the large scale velocity field. They pointed out that the environment cannot lead to  any systematic effect on the estimation of $\beta$ in terms of fitting a  velocity field 
reconstructed  using  the spatial distribution of the same galaxies. This is a direct consequence of  Galilean invariance which guarantees a vanishing
correlation between  peculiar velocity and density field at the same point. 
We have demonstrated this point in \S\ref{sec:envmeth}. Environmental 
dependencies do not bias the $\beta$ estimate but
can potentially contribute to the errors.  Our tests demonstrate that this effect is small. Its magnitude can be inferred by comparing the black cross with the  empty red circles in \figr{fig:slope}. 

{Ideally,    \meth\ 
should be tested with  more realistic mock catalogs that mimic the footprint and the selection effects of a specific survey. In this work we have been focusing on the paradigmatic case of the Euclid survey. Unfortunately the  available mock catalogs, though highly valuable to identify and assess environmental dependencies, do not cover a sufficiently large volume needed to mimic the full Euclid survey. In absence of publicly available, realistic mock Euclid catalogs, a reasonable compromise is to   generate light cone mock data by stacking the available simulation outputs to cover a sufficiently large volume.  We leave this for future work.
In this paper we have estimated the errors by rescaling those obtained from the available mocks at $z=1$ (red bands in Fig. \ref{fig:slope}) to the expected number of galaxies in the Euclid survey (grey band errors, same figure).  We expect the real error to lies in between the grey and red bands.

It is worth emphasizing that the application of \meth\ to these spectroscopic surveys will be quite straightforward. Clustering analyses have been focusing on detecting and fitting the BAO peak to extract cosmological parameters. Most of these analyses include reconstruction procedures to enhance the signal-to-noise of the BAO peak by tracing galaxy orbits back to an epoch in which linear theory applies \cite{Padmanabhan+2012,Eisenstein2007a}. 
Among these, reconstruction methods based on the cosmological application of the least action principle \cite{peebles1989tracing,nusser2000least,Sarpa:2018ucb}  generate as "byproduct" a model peculiar velocity field at the epoch of observation that can be readily used as input to the \meth\ method.
}

In this paper we have opted to apply the simplest form of the method and avoid 
performing a full maximum likelihood  estimation (MLE) as in \cite{Feix2015}.  
Theoretically,  an MLE analysis would  exploit the details of the shape of the 
SFR/luminosity distribution and not only the second moment as is done here. 
We leave the application of the MLE to the future analysis of more realistic galaxy mocks. 

The amplitude of the velocity induced shift is a strong function of redshift. In fact, 
for a Planck cosmology the net effect vanished at $ z=1.6$ but then will pick up again with an opposite sign, dominated by the beaming effect (the $V/c$ term).  Therefore it can be relevant for all those redshift surveys like DESI and Roman-WFIRST that will target galaxies beyond this redshift.


Gravitational lensing  by the foreground mass distribution causes a magnification/de-magnification of galaxies  in the survey.
For a source at $z=1$, the mass distribution at $z<1$  causes a $2 \times 10^{-3}$    shift  in the mean $\lsfr$ in spheres of radius $20\hmpc$.  The amplitude of the effect is close to the environmental dependency estimated from the mocks.
It is rather easy to incorporate this effect in the analysis. Suppose the redshift survey 
 spans the redshift range from $z=1$ to $z=2$. Let us separate the lensing shift into a contribution from    mass distribution at $z<1$ and another from mass at  $z>1$. 
 The foreground matter at  $z<1$, will cause 
 shift  that  varies in a specific way with the distance of galaxies in the survey. 
 Lensing induced by matter  at $z>1$ is 
 easy to model using  observed galaxy distribution to infer the density field.  
 The spatial coherence of the total lensing contribution will be quite distinct from the shift due to velocities.
 We therefore expect little covariance between lensing and peculiar velocity-induced effects, which would make the two contributions easy to discern.

As demonstrated in \figr{fig:gvDMDM} , non-linear dynamical  effects are insignificant 
 and linear theory is  quite  satisfactory on the scales of interest. 
Larger information content is  captured by the recovered velocity field for smaller smoothing scales. We have seen that $\rs=20\hmpc$ is adequate for linear theory reconstruction and yields much smaller errors than $\rs=100\hmpc$.
The only possible concern is whether $\vpr$   smoothed on a $\sim 20 \hmpc$ scale is contaminated by  shot noise contribution when the
mean number density is of the order of the one expected in next-generation surveys, i.e.
$\bar n\sim 5\times 10^{-4} \nunits$  \citep{EuclidCollaboration2019}.
Substituting this value for $\bar n$  with $\rs=20\hmpc$ in  \eq{eq:VSN}, yields $\sigma^2_{V,\textrm{SN}}=(93\kms)^2$. 
This is larger than contribution of  biasing stochasticity,  $(51.9\kms)^2$, to the variance of $\vpr$   (see \S\ref{sec:recv}), but 
significantly smaller than the variance of the smoothed velocity at $z=1$, $(210\kms)^2 $.
Therefore, the expected shot noise contribution will be sub-dominant in the total error budget which is dominated by the scatter in the SFR.

 As we have seen, the \meth\ method provides an estimate of the $\beta$ parameter.
It is has become customary to express the results on $\beta$ via the combination $\sigma_{\textrm{gal},8}\beta=\sigma_8 f$, thus avoiding the appearance of neither $b$ nor $\beta$ in the  result. 
We find this to be  inadequate. In the case of RSD on large scales, for example, according to  \cite{k87}, only 
$\beta$ can be  directly inferred from the   ratio of    angular moments of the observed galaxy power spectrum \textit{independently} of any assumptions regarding $\sigma_8$ and the shape of the DM power spectrum. 
The combination  $b \sigma_8$ can also  be derived in this context, but only  by matching the amplitude of the moments to  an assumed shape  for the DM power spectrum. Once $b$  and $\beta$ are  known, $f\sigma_8$ can easily be derived. Therefore, since the inference of $\beta$ depends purely on the relative clustering anisotropy  and $b$ mainly on the clustering amplitude, it would be prudent to quote the values  separately. 
Although   $f\sigma_8$ is the amplitude of the velocity,  it is not the parameter responsible for RSD in the galaxy distribution. Quoting only $f\sigma_8$  creates the false impression that it is  the primary quantity which governs the RSD phenomenon.

It is instructive to compare the performance of  the  \meth\ in constraining $\beta$  to other probes.  One of those is by means of a comparison of direct observations of galaxy peculiar velocities versus the predicted velocities from the distribution of redshift surveys. This  is possible for local data (within distances $\sim 100 \textrm{Mpc}$ ) . 
 \cite{Davis2011} performed such a comparison using   the SFI++ velocity catalog and 2MRS gravity.  They  included a full error analysis  and found a $1\sigma$ uncertainty in $\beta$ at the level of $\delta \beta\approx 0.05$.  A similar uncertainty is found by \cite{p05} using different analysis technique and a different  compilation of data. 
 
{ The other probe is RSD in galaxy clustering. Let us focus on $z\sim1$ results, i.e. the same redshift we have been focusing on so far.
\cite{contreras13} have estimated $\beta$ from the anisotropic 2-point correlation function of about 34000 galaxies in the WiggleZ survey  at an effective redshft $z=0.76$ with an error 
$\sigma_{\beta}$ in the range $0.11-0.22$, depending on the model assumed for the real-space correlation function and on the minimum transverse separation considered in the analysis.

\cite{pezzotta17} performed an RSD analysis on the final release of the VIPERS galaxy catalog.
They quoted a result in terms of $f\sigma_8$ at $z\approx 0.85$ with an error 
of $\sim 0.11$. Using the results shown in Table 2, this implies an error $\sigma_{\beta} \sim 0.13$. 
 
 More recently, \cite{eboss}  presented the final results of the clustering analyses of various extra-galactic objects from  the  completed  Sloan  Digital  Sky  Survey  (SDSS).
 For our comparison we are interested in the RSD-only analysis of  about 170000
 emission line galaxies at the effective redshift $z=0.85$ \citep{Tamone2020}.
 They quote a consensus result of  $f\sigma_8=0.315\pm 0.095 $.  
 Taking $\sigma_8(z=0.85)=0.522$  from Planck $\lcdm$, we find that the error on $f$  $\sigma_f \sim 0.18$. This is actually comparable to the error $\sigma_f \sim 0.22$ that we predict by applying the \meth\ to upcoming spectroscopic samples at $z=1$. 
  

 Next generation surveys are designed to estimate the growth rate with higher precision.
 Focusing again on the Euclid case, 
\cite{amendola18} provide  forecasts for the growth rate from an RSD-only analysis based, however, on somewhat outdated, optimistic assumptions on the expected number of Euclid galaxies. Their Fisher analysis indicate a relative error on $f$ of about 1\%  at $z=1$
for a flat $\lcdm$ cosmology (see Table 4 of their paper). More recent forecasts based 
on updated predictions on Euclid galaxy density has been provided by the
\cite{EuclidCollaboration2019}. They, however, do not consider an RSD-only analysis. Instead, they studied the case of a full shape clustering analysis that include both RSD and BAO with no additional information from CMB, low redshift surveys and weak lensing.
They do not provide error on $f$ explicitly, but on the mass density parameter $\Omega_m$ and on the parameters 
of the dark energy equation of state.
From Table 16 in their paper, we conclude that the errors depend to a large extend on the assumed background cosmology.  The relative error on $f$ from the combined RSD+BAO range from 5\% for the standard flat $\lcdm$ to 21\% for non-flat $\lcdm$; smaller than the \meth\ case but very sensitive to the cosmological background model.
}


\section{Conclusions}
\label{sec:conclusions}

{ Motivated by the advent of large spectroscopic surveys designed to target line-emitting galaxies we have investigated the connection between the star formation rate and the underlying mass density using a suite of publicly available mock catalogs in which galaxy properties, including SFR and stellar masses, are predicted using different semi-analytic recipes for galaxy formation and evolution.
The main results of our analysis can be summarized as follows:}

\begin{itemize}
\item There are certain general properties which are qualitatively common to all mocks examined in this work: the validity of scale-independent linear bias on scales larger than $\sim 20 \ \hmpc$, insensitivity of the bias factor to the SFR for star forming galaxies with SFR larger than $\sim 3 \myr$, and the trend of large stellar masses and quiescent galaxies in denser environment. {This is in agreement with previous findings in the literature \citep[e.g.][]{Angulo2014}.}
\item Despite the above and the fact that the galaxy formation recipes incorporate the same physical processes, the corresponding mocks also exhibit distinct features that can be tested in next-generation surveys. In general, an important discriminator among models is the extent to which the large scale environment affects intrinsic relations between galaxy properties. 
The PDFs of the SFR and stellar mass, their redshift evolution and their variation with  environment depend on the SAM recipe, especially for objects  characterized by intense star formation activity and/or large stellar masses. The   linear bias parameter versus  SFR is model-dependent, especially at low redshifts. 
In particular the \galc\ model predicts, unlike all the others, a decreasing SFR as a function of the stellar mass; and for this reason should be the easiest to test through observations.

\item We have shown that the \meth\ is a viable method to infer the growth rate of density fluctuations. Firstly,  environmental effects do not bias the $\beta$ estimate obtained from the \meth. This point has already been made before \citep{Feix2014,Feix2015}. Here we have demonstrated it explicitly using mock catalogs for the specific case of {star} forming galaxies, implying that the \meth\ can be safely applied to  future surveys targeting emission line galaxies.
Environmental effect do contribute to random errors, though.  Secondly, a point in favor of the \meth\ is that these errors are small compared to shot noise and stochasticity. The former dominates the total error budget for the galaxy density expect for the Euclid survey at $z \simeq 1$. 
Thirdly, the uncertainties in the $\beta$ estimate from the \meth\ are comparable to those from RSD analyses of existing datasets at $z\sim 1$. 
Moreover, the SFR/luminosity shift on which the \meth\ method relies upon, is rather insensitive to  the parameters of the cosmological background.
For example, a  reasonable deviations from the Planck parameters represented by a
flat $\Lambda$ model with $\Omega_m=0.2$  changes the effect by about 6\%  at $z=1$. As a result,  unlike most of the RSD analyses, the \meth\ is rather insensitive to the 
assumed background cosmology and the underlying power spectrum.
{The \meth\ constraints on the growth rate are similar across 
the very 
different SAMs considered, thus we conclude that our results are robust and model independent.}
For all these reasons we believe that the \meth\ is an effective method to infer the growth rate.

\item The \meth\ method relies on SFR/luminosity shift induced by peculiar velocities $V$.
There are  two competing  effects that contribute to this shift:    1) relativistic beaming, reducing the amount of light emitted by a source with positive $V$; this effect is independent of redshift. 2) the familiar term which at low redshift results from taking the distance as $cz/H_0$; this term is redshift-dependent and increases the observed luminosity compared to the true one.  
For the Planck cosmological parameters, the two effects cancel out at $z\approx 1.6$ while the first one becomes dominant at higher redshifts.

\end{itemize}

\section*{Acknowledgements}
We thank the anonymous referee for useful comments.
We also thank Carmelita Carbone, Ben Granett and Lucia Pozzetti for useful comments and discussion. 
AN acknowledges the hospitality of the MIT Kavli Institute for Astrophysics and Space Research where part of this work has been done. This research was supported by the I-CORE Program of the Planning and Budgeting Committee,
THE ISRAEL SCIENCE FOUNDATION (grants No. 1829/12 and No. 936/18). 
GY acknowledges financial support from {\em Ministerio de Ciencia, Innovaci\'on y Universidades / Fondo  Europeo de DEsarrollo Regional,}  under research  grant  PGC2018-094975-C21.
EB is supported by MUIR/PRIN 2017 “From Darklight to Dark Matter: understanding the galaxy-matter connection to measure the Universe”,  ASI/INAF agreement n. 2018-23-HH.0 “Scientific activity for Euclid mission, Phase D”, ASI/INAF agreement  n. 2017-14-H.O “Unveiling Dark Matter and Missing Baryons in the high-energy sky”  and INFN project “INDARK”.
This research was supported by the Munich Institute for Astro- and Particle Physics (MIAPP) which is funded by the Deutsche Forschungsgemeinschaft (DFG, German Research Foundation) under Germany's Excellence Strategy – EXC-2094 – 390783311.
The CosmoSim database (www.cosmosim.org) used in this paper is a service provided by the Leibniz--Institute for Astrophysics Potsdam (AIP).

\bibliographystyle{aasjournal}
\bibliography{Library.bib}

\begin{thebibliography}{}
\expandafter\ifx\csname natexlab\endcsname\relax\def\natexlab#1{#1}\fi
\providecommand{\url}[1]{\href{#1}{#1}}
\providecommand{\dodoi}[1]{doi:~\href{http://doi.org/#1}{\nolinkurl{#1}}}
\providecommand{\doeprint}[1]{\href{http://ascl.net/#1}{\nolinkurl{http://ascl.net/#1}}}
\providecommand{\doarXiv}[1]{\href{https://arxiv.org/abs/#1}{\nolinkurl{https://arxiv.org/abs/#1}}}

\bibitem[{Abate \& Feldman(2012)}]{abate2012a}
Abate, A., \& Feldman, H.~A. 2012, MNRAS, 419, 3482,
  \dodoi{10.1111/j.1365-2966.2011.19988.x}

\bibitem[{{Akeson} {et~al.}(2019){Akeson}, {Armus}, {Bachelet}, {Bailey},
  {Bartusek}, {Bellini}, {Benford}, {Bennett}, {Bhattacharya}, {Bohlin},
  {Boyer}, {Bozza}, {Bryden}, {Calchi Novati}, {Carpenter}, {Casertano},
  {Choi}, {Content}, {Dayal}, {Dressler}, {Dor{\'e}}, {Fall}, {Fan}, {Fang},
  {Filippenko}, {Finkelstein}, {Foley}, {Furlanetto}, {Kalirai}, {Gaudi},
  {Gilbert}, {Girard}, {Grady}, {Greene}, {Guhathakurta}, {Heinrich},
  {Hemmati}, {Hendel}, {Henderson}, {Henning}, {Hirata}, {Ho}, {Huff},
  {Hutter}, {Jansen}, {Jha}, {Johnson}, {Jones}, {Kasdin}, {Kelly}, {Kirshner},
  {Koekemoer}, {Kruk}, {Lewis}, {Macintosh}, {Madau}, {Malhotra}, {Mand el},
  {Massara}, {Masters}, {McEnery}, {McQuinn}, {Melchior}, {Melton},
  {Mennesson}, {Peeples}, {Penny}, {Perlmutter}, {Pisani}, {Plazas}, {Poleski},
  {Postman}, {Ranc}, {Rauscher}, {Rest}, {Roberge}, {Robertson}, {Rodney},
  {Rhoads}, {Rhodes}, {Ryan}, {Sahu}, {Sand}, {Scolnic}, {Seth}, {Shvartzvald},
  {Siellez}, {Smith}, {Spergel}, {Stassun}, {Street}, {Strolger}, {Szalay},
  {Trauger}, {Troxel}, {Turnbull}, {van der Marel}, {von der Linden}, {Wang},
  {Weinberg}, {Williams}, {Windhorst}, {Wollack}, {Wu}, {Yee}, \&
  {Zimmerman}}]{wfirst2019}
{Akeson}, R., {Armus}, L., {Bachelet}, E., {et~al.} 2019, arXiv e-prints,
  arXiv:1902.05569.
\newblock \doarXiv{1902.05569}

\bibitem[{{Amendola} {et~al.}(2018){Amendola}, {Appleby}, {Avgoustidis},
  {Bacon}, {Baker}, {Baldi}, {Bartolo}, {Blanchard}, {Bonvin}, {Borgani},
  {Branchini}, {Burrage}, {Camera}, {Carbone}, {Casarini}, {Cropper}, {de
  Rham}, {Dietrich}, {Di Porto}, {Durrer}, {Ealet}, {Ferreira}, {Finelli},
  {Garc{\'\i}a-Bellido}, {Giannantonio}, {Guzzo}, {Heavens}, {Heisenberg},
  {Heymans}, {Hoekstra}, {Hollenstein}, {Holmes}, {Hwang}, {Jahnke},
  {Kitching}, {Koivisto}, {Kunz}, {La Vacca}, {Linder}, {March}, {Marra},
  {Martins}, {Majerotto}, {Markovic}, {Marsh}, {Marulli}, {Massey}, {Mellier},
  {Montanari}, {Mota}, {Nunes}, {Percival}, {Pettorino}, {Porciani},
  {Quercellini}, {Read}, {Rinaldi}, {Sapone}, {Sawicki}, {Scaramella},
  {Skordis}, {Simpson}, {Taylor}, {Thomas}, {Trotta}, {Verde}, {Vernizzi},
  {Vollmer}, {Wang}, {Weller}, \& {Zlosnik}}]{amendola18}
{Amendola}, L., {Appleby}, S., {Avgoustidis}, A., {et~al.} 2018, Living Reviews
  in Relativity, 21, 2, \dodoi{10.1007/s41114-017-0010-3}

\bibitem[{{Angulo} {et~al.}(2012){Angulo}, {Springel}, {White}, {Jenkins},
  {Baugh}, \& {Frenk}}]{Angulo2012}
{Angulo}, R.~E., {Springel}, V., {White}, S.~D.~M., {et~al.} 2012, \mnras, 426,
  2046, \dodoi{10.1111/j.1365-2966.2012.21830.x}

\bibitem[{{Angulo} {et~al.}(2014){Angulo}, {White}, {Springel}, \&
  {Henriques}}]{Angulo2014}
{Angulo}, R.~E., {White}, S.~D.~M., {Springel}, V., \& {Henriques}, B. 2014,
  \mnras, 442, 2131, \dodoi{10.1093/mnras/stu905}

\bibitem[{Bardeen {et~al.}(1986)Bardeen, Bond, Kaiser, \& Szalay}]{BBKS}
Bardeen, J.~M., Bond, J.~R., Kaiser, N., \& Szalay, A.~S. 1986, ApJ, 304, 15,
  \dodoi{10.1086/164143}

\bibitem[{Bartelmann \& Schneider(2001)}]{BS01}
Bartelmann, M., \& Schneider, P. 2001, Phys.Rep, 340, 291,
  \dodoi{10.1016/S0370-1573(00)00082-X}

\bibitem[{{Baugh} {et~al.}(2019){Baugh}, {Gonzalez-Perez}, {Lagos}, {Lacey},
  {Helly}, {Jenkins}, {Frenk}, {Benson}, {Bower}, \& {Cole}}]{Baugh2019}
{Baugh}, C.~M., {Gonzalez-Perez}, V., {Lagos}, C. d.~P., {et~al.} 2019, \mnras,
  483, 4922, \dodoi{10.1093/mnras/sty3427}

\bibitem[{Behroozi {et~al.}(2019)Behroozi, Wechsler, Hearin, \&
  Conroy}]{Behroozi2019}
Behroozi, P., Wechsler, R.~H., Hearin, A.~P., \& Conroy, C. 2019, MNRAS, 488,
  3143, \dodoi{10.1093/mnras/stz1182}

\bibitem[{Benson(2012)}]{Benson2012}
Benson, A.~J. 2012, New Astronomy, 17, 175,
  \dodoi{10.1016/j.newast.2011.07.004}

\bibitem[{Benson {et~al.}(2000)Benson, Cole, Frenk, Baugh, \&
  Lacey}]{Benson2000}
Benson, A.~J., Cole, S., Frenk, C.~S., Baugh, C.~M., \& Lacey, C.~G. 2000,
  MNRAS, 311, 793, \dodoi{10.1046/j.1365-8711.2000.03101.x}

\bibitem[{Bernardi {et~al.}(2005)Bernardi, Sheth, Nichol, Schneider, \&
  Brinkmann}]{Bernardi2005}
Bernardi, M., Sheth, R.~K., Nichol, R.~C., Schneider, D.~P., \& Brinkmann, J.
  2005, ApJ, 129, 61, \dodoi{10.1086/426336}

\bibitem[{Binney(1977)}]{Binney1977}
Binney, J. 1977, ApJ, 215, 483, \dodoi{10.1086/155378}

\bibitem[{{Blanton} \& {Moustakas}(2009)}]{Blanton2009}
{Blanton}, M.~R., \& {Moustakas}, J. 2009, \araa, 47, 159,
  \dodoi{10.1146/annurev-astro-082708-101734}

\bibitem[{Bonvin {et~al.}(2005)Bonvin, Durrer, Gasparini, Bonvin, Durrer, \&
  Gasparini}]{Bonvin2005}
Bonvin, C., Durrer, R., Gasparini, M.~A., {et~al.} 2005, PhRvD, 73, 023523,
  \dodoi{10.1103/PhysRevD.73.023523}

\bibitem[{Branchini {et~al.}(2012)Branchini, Davis, \& Nusser}]{Branchini2012}
Branchini, E., Davis, M., \& Nusser, A. 2012, MNRAS, 424, 472,
  \dodoi{10.1111/j.1365-2966.2012.21210.x}

\bibitem[{Comparat {et~al.}(2017)Comparat, Prada, Yepes, \&
  Klypin}]{Comparat2017}
Comparat, J., Prada, F., Yepes, G., \& Klypin, A. 2017, MNRAS, 469, 4157,
  \dodoi{10.1093/mnras/stx1183}

\bibitem[{{Contreras} {et~al.}(2013){Contreras}, {Blake}, {Poole}, {Marin},
  {Brough}, {Colless}, {Couch}, {Croom}, {Croton}, {Davis}, {Drinkwater},
  {Forster}, {Gilbank}, {Gladders}, {Glazebrook}, {Jelliffe}, {Jurek}, {Li},
  {Madore}, {Martin}, {Pimbblet}, {Pracy}, {Sharp}, {Wisnioski}, {Woods},
  {Wyder}, \& {Yee}}]{contreras13}
{Contreras}, C., {Blake}, C., {Poole}, G.~B., {et~al.} 2013, \mnras, 430, 924,
  \dodoi{10.1093/mnras/sts608}

\bibitem[{Cora {et~al.}(2018)Cora, Vega-Mart{\'{i}}nez, Hough, Ruiz, Orsi,
  {Mu{\~{n}}oz Arancibia}, Gargiulo, Collacchioni, Padilla, Gottl{\"{o}}ber, \&
  Yepes}]{Cora2018}
Cora, S.~A., Vega-Mart{\'{i}}nez, C.~A., Hough, T., {et~al.} 2018, MNRAS, 479,
  2, \dodoi{10.1093/mnras/sty1131}

\bibitem[{Croton {et~al.}(2016)Croton, Stevens, Tonini, Garel, Bernyk, Bibiano,
  Hodkinson, Mutch, Poole, \& Shattow}]{Croton2016}
Croton, D.~J., Stevens, A. R.~H., Tonini, C., {et~al.} 2016, The Astrophysical
  Journal Supplement Series, 222, 22, \dodoi{10.3847/0067-0049/222/2/22}

\bibitem[{Davis {et~al.}(2011{\natexlab{a}})Davis, Nusser, Masters, Springob,
  Huchra, \& Lemson}]{DN10}
Davis, M., Nusser, A., Masters, K., {et~al.} 2011{\natexlab{a}}, MNRAS, 413,
  2906, \dodoi{10.1111/j.1365-2966.2011.18362.x}

\bibitem[{Davis {et~al.}(2011{\natexlab{b}})Davis, Nusser, Masters, Springob,
  Huchra, \& Lemson}]{Davis2011}
Davis, M., Nusser, A., Masters, K.~L., {et~al.} 2011{\natexlab{b}}, MNRAS, 413,
  2906, \dodoi{10.1111/j.1365-2966.2011.18362.x}

\bibitem[{Davis {et~al.}(1996)Davis, Nusser, \& Willick}]{DNW96}
Davis, M., Nusser, A., \& Willick, J. 1996, ApJ, 473, 22,
  \dodoi{10.1086/178124}

\bibitem[{Dekel \& Lahav(1999)}]{dh99}
Dekel, A., \& Lahav, O. 1999, ApJ, 520, 24, \dodoi{10.1086/307428}

\bibitem[{Dekel \& Silk(1986)}]{DS86}
Dekel, A., \& Silk, J. 1986, ApJ, 303, 39, \dodoi{10.1086/164050}

\bibitem[{{DESI Collaboration} {et~al.}(2016){DESI Collaboration}, Aghamousa,
  Aguilar, Ahlen, Alam, Allen, Prieto, Annis, Bailey, Balland, Ballester,
  Baltay, Beaufore, Bebek, Beers, Bell, Bernal, Besuner, Beutler, Blake,
  Bleuler, Blomqvist, Blum, Bolton, Briceno, Brooks, Brownstein, Buckley-Geer,
  Burden, Burtin, Busca, Cahn, Cai, Cardiel-Sas, Carlberg, Carton, Casas,
  Castander, Cervantes-Cota, Claybaugh, Close, Coker, Cole, Comparat, Cooper,
  Cousinou, Crocce, Cuby, Cunningham, Davis, Dawson, de~la Macorra, {De
  Vicente}, Delubac, Derwent, Dey, Dhungana, Ding, Doel, Duan, Ealet,
  Edelstein, Eftekharzadeh, Eisenstein, Elliott, Escoffier, Evatt, Fagrelius,
  Fan, Fanning, Farahi, Farihi, Favole, Feng, Fernandez, Findlay, Finkbeiner,
  Fitzpatrick, Flaugher, Flender, Font-Ribera, Forero-Romero, Fosalba, Frenk,
  Fumagalli, Gaensicke, Gallo, Garcia-Bellido, Gaztanaga, Fusillo, Gerard,
  Gershkovich, Giannantonio, Gillet, Gonzalez-de Rivera, Gonzalez-Perez, Gott,
  Graur, Gutierrez, Guy, Habib, Heetderks, Heetderks, Heitmann, Hellwing,
  Herrera, Ho, Holland, Honscheid, Huff, Hutchinson, Huterer, Hwang, Laguna,
  Ishikawa, Jacobs, Jeffrey, Jelinsky, Jennings, Jiang, Jimenez, Johnson,
  Joyce, Jullo, Juneau, Kama, Karcher, Karkar, Kehoe, Kennamer, Kent,
  Kilbinger, Kim, Kirkby, Kisner, Kitanidis, Kneib, Koposov, Kovacs, Koyama,
  Kremin, Kron, Kronig, Kueter-Young, Lacey, Lafever, Lahav, Lambert, Lampton,
  Landriau, Lang, Lauer, Goff, Guillou, {Van Suu}, Lee, Lee, Leitner, Lesser,
  Levi, L'Huillier, Li, Liang, Lin, Linder, Loebman, Luki{\'{c}}, Ma, MacCrann,
  Magneville, Makarem, Manera, Manser, Marshall, Martini, Massey, Matheson,
  McCauley, McDonald, McGreer, Meisner, Metcalfe, Miller, Miquel, Moustakas,
  Myers, Naik, Newman, Nichol, Nicola, da~Costa, Nie, Niz, Norberg, Nord,
  Norman, Nugent, O'Brien, Oh, Olsen, Padilla, Padmanabhan, Padmanabhan,
  Palanque-Delabrouille, Palmese, Pappalardo, P{\^{a}}ris, Park, Patej,
  Peacock, Peiris, Peng, Percival, Perruchot, Pieri, Pogge, Pollack, Poppett,
  Prada, Prakash, Probst, Rabinowitz, Raichoor, Ree, Refregier, Regal, Reid,
  Reil, Rezaie, Rockosi, Roe, Ronayette, Roodman, Ross, Ross, Rossi, Rozo,
  Ruhlmann-Kleider, Rykoff, Sabiu, Samushia, Sanchez, Sanchez, Schlegel,
  Schneider, Schubnell, Secroun, Seljak, Seo, Serrano, Shafieloo, Shan,
  Sharples, Sholl, Shourt, Silber, Silva, Sirk, Slosar, Smith, Smoot, Som,
  Song, Sprayberry, Staten, Stefanik, Tarle, Tie, Tinker, Tojeiro, Valdes,
  Valenzuela, Valluri, Vargas-Magana, Verde, Walker, Wang, Wang, Weaver,
  Weaverdyck, Wechsler, Weinberg, White, Yang, Yeche, Zhang, Zhao, Zheng, Zhou,
  Zhou, Zhu, Zou, \& Zu}]{DESICollaboration2016b}
{DESI Collaboration}, Aghamousa, A., Aguilar, J., {et~al.} 2016, ArXiv
  e-prints, 1611.00036.
\newblock \doarXiv{1611.00036}

\bibitem[{{Disney} {et~al.}(2008){Disney}, {Romano}, {Garcia-Appadoo}, {West},
  {Dalcanton}, \& {Cortese}}]{Disney2008}
{Disney}, M.~J., {Romano}, J.~D., {Garcia-Appadoo}, D.~A., {et~al.} 2008, \nat,
  455, 1082, \dodoi{10.1038/nature07366}

\bibitem[{Dolag {et~al.}(2016)Dolag, Komatsu, \& Sunyaev}]{Dolag2016}
Dolag, K., Komatsu, E., \& Sunyaev, R. 2016, MNRAS, 463, 1797,
  \dodoi{10.1093/mnras/stw2035}

\bibitem[{{Dom{\'{i}}nguez S{\'{a}}nchez} {et~al.}(2012){Dom{\'{i}}nguez
  S{\'{a}}nchez}, Mignoli, Pozzi, Calura, Cimatti, Gruppioni, Cepa,
  {S{\'{a}}nchez Portal}, Zamorani, Berta, Elbaz, {Le Floc'h}, Granato, Lutz,
  Maiolino, Matteucci, Nair, Nordon, Pozzetti, Silva, Silverman, Wuyts,
  Carollo, Contini, Kneib, {Le F{\`{e}}vre}, Lilly, Mainieri, Renzini,
  Scodeggio, Bardelli, Bolzonella, Bongiorno, Caputi, Coppa, Cucciati, de~la
  Torre, de~Ravel, Franzetti, Garilli, Iovino, Kampczyk, Knobel, Kova{\v{c}},
  Lamareille, {Le Borgne}, {Le Brun}, Maier, Magnelli, Pell{\'{o}}, Peng,
  Perez-Montero, Ricciardelli, Riguccini, Tanaka, Tasca, Tresse, Vergani, \&
  Zucca}]{Dom2012}
{Dom{\'{i}}nguez S{\'{a}}nchez}, H., Mignoli, M., Pozzi, F., {et~al.} 2012,
  MNRAS, 426, 330, \dodoi{10.1111/j.1365-2966.2012.21710.x}

\bibitem[{Dubois {et~al.}(2016)Dubois, Peirani, Pichon, Devriendt, Gavazzi,
  Welker, \& Volonteri}]{Dubois2016}
Dubois, Y., Peirani, S., Pichon, C., {et~al.} 2016, MNRAS, 463, 3948,
  \dodoi{10.1093/mnras/stw2265}

\bibitem[{{eBOSS Collaboration} {et~al.}(2020){eBOSS Collaboration}, {Alam},
  {Aubert}, {Avila}, {Balland}, {Bautista}, {Bershady}, {Bizyaev}, {Blanton},
  {Bolton}, {Bovy}, {Brinkmann}, {Brownstein}, {Burtin}, {Chabanier},
  {Chapman}, {Choi}, {Chuang}, {Comparat}, {Cuceu}, {Dawson}, {de la Macorra},
  {de la Torre}, {de Mattia}, {de Sainte Agathe}, {du Mas des Bourboux},
  {Escoffier}, {Etourneau}, {Farr}, {Font-Ribera}, {Frinchaboy}, {Fromenteau},
  {Gil-Mar{\'\i}n}, {Gonzalez-Morales}, {Gonzalez-Perez}, {Grabowski}, {Guy},
  {Hawken}, {Hou}, {Kong}, {Klaene}, {Kneib}, {Le Goff}, {Lin}, {Long}, {Lyke},
  {Cousinou}, {Martini}, {Masters}, {Mohammad}, {Moon}, {Mueller},
  {Mun{\~o}z-Gutie{\'r}rez}, {Myers}, {Nadathur}, {Neveux}, {Newman},
  {Noterdaeme}, {Oravetz}, {Oravetz}, {Palanque-Delabrouille}, {Pan}, {Parker},
  {Paviot}, {Percival}, {Pe{\'r}ez-Rafols}, {Petitjean}, {Pieri}, {Prakash},
  {Raichoor}, {Ravoux}, {Rezaie}, {Rich}, {Ross}, {Rossi}, {Ruggeri},
  {Ruhlmann-Kleider}, {Sa{\'n}chez}, {Sa{\'n}chez}, {Sa{\'n}chez-Gallego},
  {Sayres}, {Schneider}, {Seo}, {Shafieloo}, {Slosar}, {Smith}, {Stermer},
  {Tamone}, {Tinker}, {Tojeiro}, {Vargas-Maga{\~n}a}, {Variu}, {Wang},
  {Weaver}, {Weijmans}, {Yeche}, {Zarrouk}, {Zhao}, {Zhao}, \& {Zheng}}]{eboss}
{eBOSS Collaboration}, {Alam}, S., {Aubert}, M., {et~al.} 2020, arXiv e-prints,
  arXiv:2007.08991.
\newblock \doarXiv{2007.08991}

\bibitem[{{Eisenstein} {et~al.}(2007){Eisenstein}, {Seo}, {Sirko}, \&
  {Spergel}}]{Eisenstein2007a}
{Eisenstein}, D.~J., {Seo}, H.-J., {Sirko}, E., \& {Spergel}, D.~N. 2007, \apj,
  664, 675, \dodoi{10.1086/518712}

\bibitem[{{Euclid Collaboration} {et~al.}(2019){Euclid Collaboration},
  Blanchard, Camera, Carbone, Cardone, Casas, Ili{\'{c}}, Kilbinger, Kitching,
  Kunz, Lacasa, Linder, Majerotto, Markovi{\v{c}}, Martinelli, Pettorino,
  Pourtsidou, Sakr, S{\'{a}}nchez, Sapone, Tutusaus, Yahia-Cherif, Yankelevich,
  Andreon, Aussel, Balaguera-Antol{\'{i}}nez, Baldi, Bardelli, Bender, Biviano,
  Bonino, Boucaud, Bozzo, Branchini, Brau-Nogue, Brescia, Brinchmann, Burigana,
  Cabanac, Capobianco, Cappi, Carretero, Carvalho, Casas, Castander,
  Castellano, Cavuoti, Cimatti, Cledassou, Colodro-Conde, Congedo, Conselice,
  Conversi, Copin, Corcione, Coupon, Courtois, Cropper, {Da Silva}, de~la
  Torre, {Di Ferdinando}, Dubath, Ducret, Duncan, Dupac, Dusini, Fabbian,
  Fabricius, Farrens, Fosalba, Fotopoulou, Fourmanoit, Frailis, Franceschi,
  Franzetti, Fumana, Galeotta, Gillard, Gillis, Giocoli, G{\'{o}}mez-Alvarez,
  Graci{\'{a}}-Carpio, Grupp, Guzzo, Hoekstra, Hormuth, Israel, Jahnke,
  Keihanen, Kermiche, Kirkpatrick, Kohley, Kubik, Kurki-Suonio, Ligori, Lilje,
  Lloro, Maino, Maiorano, Marggraf, Martinet, Marulli, Massey, Medinaceli, Mei,
  Mellier, Metcalf, Metge, Meylan, Moresco, Moscardini, Munari, Nichol, Niemi,
  Nucita, Padilla, Paltani, Pasian, Percival, Pires, Polenta, Poncet, Pozzetti,
  Racca, Raison, Renzi, Rhodes, Romelli, Roncarelli, Rossetti, Saglia,
  Schneider, Scottez, Secroun, Sirri, Stanco, Starck, Sureau,
  Tallada-Cresp{\'{i}}, Tavagnacco, Taylor, Tenti, Tereno, Toledo-Moreo,
  Torradeflot, Valenziano, Vassallo, Kleijn, Viel, Wang, Zacchei, Zoubian, \&
  Zucca}]{EuclidCollaboration2019}
{Euclid Collaboration}, Blanchard, A., Camera, S., {et~al.} 2019, ArXiv
  e-prints, 1910.09273.
\newblock \doarXiv{1910.09273}

\bibitem[{Feix {et~al.}(2017)Feix, Branchini, \& Nusser}]{Feix2016}
Feix, M., Branchini, E., \& Nusser, A. 2017, MNRAS, 467, 468.
\newblock \doarXiv{1612.07809}

\bibitem[{Feix {et~al.}(2014)Feix, Nusser, \& Branchini}]{Feix2014}
Feix, M., Nusser, A., \& Branchini, E. 2014, JCAP, 09, 19,
  \dodoi{10.1088/1475-7516/2014/09/019}

\bibitem[{Feix {et~al.}(2015)Feix, Nusser, \& Branchini}]{Feix2015}
---. 2015, PRL, 115, 011301, \dodoi{10.1103/PhysRevLett.115.011301}

\bibitem[{Gao \& White(2007)}]{Gao2007}
Gao, L., \& White, S. D.~M. 2007, MNRAS, 377, L5,
  \dodoi{10.1111/j.1745-3933.2007.00292.x}

\bibitem[{Gene {et~al.}(2014)Gene, Vogelsberger, Springel, Sijacki, Nelson,
  Snyder, Rodriguez-Gomez, Torrey, \& Hernquist}]{Gene2014}
Gene, S., Vogelsberger, M., Springel, V., {et~al.} 2014, MNRAS, 445, 175,
  \dodoi{10.1093/mnras/stu1654}

\bibitem[{Gottlober {et~al.}(2001)Gottlober, Klypin, \&
  Kravtsov}]{Gottlober2001}
Gottlober, S., Klypin, A., \& Kravtsov, A.~V. 2001, ApJ, 546, 223,
  \dodoi{10.1086/318248}

\bibitem[{Gruppioni {et~al.}(2015)Gruppioni, Calura, Pozzi, Delvecchio, Berta,
  {De Lucia}, Fontanot, Franceschini, Marchetti, Menci, Monaco, \&
  Vaccari}]{Gruppioni2015}
Gruppioni, C., Calura, F., Pozzi, F., {et~al.} 2015, MNRAS, 451, 3419,
  \dodoi{10.1093/mnras/stv1204}

\bibitem[{Guo {et~al.}(2011)Guo, White, Boylan-Kolchin, {De Lucia}, Kauffmann,
  Lemson, Li, Springel, \& Weinmann}]{Guo2011}
Guo, Q., White, S., Boylan-Kolchin, M., {et~al.} 2011, MNRAS, 413, 101,
  \dodoi{10.1111/j.1365-2966.2010.18114.x}

\bibitem[{{Hahn} {et~al.}(2015){Hahn}, {Blanton}, {Moustakas}, {Coil}, {Cool},
  {Eisenstein}, {Skibba}, {Wong}, \& {Zhu}}]{Hahn2015}
{Hahn}, C., {Blanton}, M.~R., {Moustakas}, J., {et~al.} 2015, \apj, 806, 162,
  \dodoi{10.1088/0004-637X/806/2/162}

\bibitem[{Hoyle {et~al.}(2005)Hoyle, Rojas, Vogeley, \& Brinkmann}]{Hoyle2005}
Hoyle, F., Rojas, R.~R., Vogeley, M.~S., \& Brinkmann, J. 2005, ApJ, 620, 618,
  \dodoi{10.1086/427176}

\bibitem[{Huchra {et~al.}(2012)Huchra, Macri, Masters, Jarrett, Berlind,
  Calkins, Crook, Cutri, Erdogdu, Falco, George, Hutcheson, Lahav, Mader, Mink,
  Martimbeau, Schneider, Skrutskie, Tokarz, \& Westover}]{2mrs2012}
Huchra, J., Macri, L., Masters, K., {et~al.} 2012, ApJS, 199, 26,
  \dodoi{10.1088/0067-0049/199/2/26}

\bibitem[{Hui \& Greene(2006)}]{2006PhRvD..73l3526H}
Hui, L., \& Greene, P. 2006, $\backslash$prd, 73, 123526,
  \dodoi{10.1103/PhysRevD.73.123526}

\bibitem[{Jing(2005)}]{Jing2005}
Jing, Y.~P. 2005, The Astrophysical Journal, 620, 559, \dodoi{10.1086/427087}

\bibitem[{Kaiser(1987)}]{k87}
Kaiser, N. 1987, MNRAS, 227, 1

\bibitem[{Kauffmann {et~al.}(1999)Kauffmann, Colberg, Diaferio, \&
  White}]{Kauffmann1999}
Kauffmann, G., Colberg, J.~M., Diaferio, A., \& White, S.~D. 1999, MNRAS, 303,
  188, \dodoi{10.1046/j.1365-8711.1999.02202.x}

\bibitem[{Kauffmann {et~al.}(1997)Kauffmann, Nusser, \& Steinmetz}]{kn97}
Kauffmann, G., Nusser, A., \& Steinmetz, M. 1997, MNRAS, 286, 795

\bibitem[{Kauffmann {et~al.}(1993)Kauffmann, White, \&
  Guiderdoni}]{Kauffmann1993}
Kauffmann, G., White, S. D.~M., \& Guiderdoni, B. 1993, MNRAS, 264, 201,
  \dodoi{10.1093/mnras/264.1.201}

\bibitem[{Keselman \& Nusser(2016)}]{KN16}
Keselman, A., \& Nusser, A. 2016, MNRAS, 467, 1915,
  \dodoi{10.1093/mnras/stx152}

\bibitem[{Khandai {et~al.}(2015)Khandai, {Di Matteo}, Croft, Wilkins, Feng,
  Tucker, DeGraf, \& Liu}]{Khandai2015}
Khandai, N., {Di Matteo}, T., Croft, R., {et~al.} 2015, MNRAS, 450, 1349,
  \dodoi{10.1093/mnras/stv627}

\bibitem[{Knebe {et~al.}(2018)Knebe, Stoppacher, Prada, Behrens, Benson, Cora,
  Croton, Padilla, Ruiz, Sinha, Stevens, Vega-Mart{\'{i}}nez, Behroozi,
  Gonzalez-Perez, Gottl{\"{o}}ber, Klypin, Yepes, Enke, Libeskind, Riebe, \&
  Steinmetz}]{Knebe2018}
Knebe, A., Stoppacher, D., Prada, F., {et~al.} 2018, MNRAS, 474, 5206,
  \dodoi{10.1093/mnras/stx2662}

\bibitem[{Krumholz {et~al.}(2009)Krumholz, McKee, \& Tumlinson}]{Krumholz2009}
Krumholz, M.~R., McKee, C.~F., \& Tumlinson, J. 2009, ApJ, 699, 850,
  \dodoi{10.1088/0004-637X/699/1/850}

\bibitem[{Lacey {et~al.}(1993)Lacey, Guiderdoni, Rocca-Volmerange, \&
  Silk}]{Lacey1993}
Lacey, C., Guiderdoni, B., Rocca-Volmerange, B., \& Silk, J. 1993, ApJ, 402,
  15, \dodoi{10.1086/172109}

\bibitem[{Larson(1974)}]{Larson74}
Larson, R.~B. 1974, MNRAS, 169, 229, \dodoi{10.1093/mnras/169.2.229}

\bibitem[{Linder(2005)}]{Lind05}
Linder, E.~V. 2005, Physical Review D, 72, 043529,
  \dodoi{10.1103/PhysRevD.72.043529}

\bibitem[{Mo {et~al.}(2004)Mo, Yang, van~den Bosch, \& Jing}]{moyang04}
Mo, H., Yang, X., van~den Bosch, F., \& Jing, Y. 2004, MNRAS, 349, 205,
  \dodoi{10.1111/j.1365-2966.2004.07485.x}

\bibitem[{Moustakas {et~al.}(2013)Moustakas, Coil, Aird, Blanton, Cool,
  Eisenstein, Mendez, Wong, Zhu, \& Arnouts}]{Moustakas2013}
Moustakas, J., Coil, A.~L., Aird, J., {et~al.} 2013, ApJ, 767, 50,
  \dodoi{10.1088/0004-637X/767/1/50}

\bibitem[{{Nair} {et~al.}(2010){Nair}, {van den Bergh}, \&
  {Abraham}}]{Nair2010}
{Nair}, P.~B., {van den Bergh}, S., \& {Abraham}, R.~G. 2010, \apj, 715, 606,
  \dodoi{10.1088/0004-637X/715/1/606}

\bibitem[{{Nusser} \& {Branchini}(2000)}]{nusser2000least}
{Nusser}, A., \& {Branchini}, E. 2000, \mnras, 313, 587,
  \dodoi{10.1046/j.1365-8711.2000.03261.x}

\bibitem[{Nusser {et~al.}(2011)Nusser, Branchini, \& Davis}]{Nusser2011a}
Nusser, A., Branchini, E., \& Davis, M. 2011, $\backslash$apj, 735,
  \dodoi{10.1088/0004-637X/735/2/77}

\bibitem[{{Nusser} {et~al.}(2012){Nusser}, {Branchini}, \&
  {Davis}}]{Nusser2012}
{Nusser}, A., {Branchini}, E., \& {Davis}, M. 2012, \apj, 744, 193,
  \dodoi{10.1088/0004-637X/744/2/193}

\bibitem[{Nusser {et~al.}(2013)Nusser, Branchini, \& Feix}]{Nusser2013}
Nusser, A., Branchini, E., \& Feix, M. 2013, Journal of Cosmology and
  Astroparticle Physics, 2013, 018

\bibitem[{Nusser \& Davis(1994)}]{ND94}
Nusser, A., \& Davis, M. 1994, $\backslash$apjl, 421, L1,
  \dodoi{10.1086/187172}

\bibitem[{Nusser \& Davis(1995)}]{Nusser1995}
---. 1995, MNRAS, 276, 1391

\bibitem[{{Padmanabhan} {et~al.}(2012){Padmanabhan}, {Xu}, {Eisenstein},
  {Scalzo}, {Cuesta}, {Mehta}, \& {Kazin}}]{Padmanabhan+2012}
{Padmanabhan}, N., {Xu}, X., {Eisenstein}, D.~J., {et~al.} 2012, \mnras, 427,
  2132, \dodoi{10.1111/j.1365-2966.2012.21888.x}

\bibitem[{Park {et~al.}(2007)Park, Choi, Vogeley, {Gott III}, \&
  Blanton}]{Park2007}
Park, C., Choi, Y., Vogeley, M.~S., {Gott III}, J.~R., \& Blanton, M.~R. 2007,
  ApJ, 658, 898, \dodoi{10.1086/511059}

\bibitem[{Peebles(1980)}]{Peeb80}
Peebles, P. J.~E. 1980, {The large-scale structure of the universe} (Princeton
  University Press, NJ).
\newblock \url{http://adsabs.harvard.edu/abs/1980lssu.book.....P}

\bibitem[{{Peebles}(1989)}]{peebles1989tracing}
{Peebles}, P.~J.~E. 1989, \apjl, 344, L53, \dodoi{10.1086/185529}

\bibitem[{{Pezzotta} {et~al.}(2017){Pezzotta}, {de la Torre}, {Bel}, {Granett},
  {Guzzo}, {Peacock}, {Garilli}, {Scodeggio}, {Bolzonella}, {Abbas}, {Adami},
  {Bottini}, {Cappi}, {Cucciati}, {Davidzon}, {Franzetti}, {Fritz}, {Iovino},
  {Krywult}, {Le Brun}, {Le F{\`e}vre}, {Maccagni}, {Ma{\l}ek}, {Marulli},
  {Polletta}, {Pollo}, {Tasca}, {Tojeiro}, {Vergani}, {Zanichelli}, {Arnouts},
  {Branchini}, {Coupon}, {De Lucia}, {Koda}, {Ilbert}, {Mohammad}, {Moutard},
  \& {Moscardini}}]{pezzotta17}
{Pezzotta}, A., {de la Torre}, S., {Bel}, J., {et~al.} 2017, \aap, 604, A33,
  \dodoi{10.1051/0004-6361/201630295}

\bibitem[{Pike \& Hudson(2005)}]{p05}
Pike, R., \& Hudson, M. 2005, $\backslash$apj, 635, 11, \dodoi{10.1086/497359}

\bibitem[{Pozzetti {et~al.}(2016)Pozzetti, Hirata, Geach, Cimatti, Baugh,
  Cucciati, Merson, Norberg, \& Shi}]{Pozzetti2016}
Pozzetti, L., Hirata, C.~M., Geach, J.~E., {et~al.} 2016, A$\backslash${\&}A,
  590, A3, \dodoi{10.1051/0004-6361/201527081}

\bibitem[{Rees \& Ostriker(1977)}]{Rees1977}
Rees, M.~J., \& Ostriker, J.~P. 1977, MNRAS, 179, 541,
  \dodoi{10.1093/mnras/179.4.541}

\bibitem[{Sachs \& Wolfe(1967)}]{SW}
Sachs, R., \& Wolfe, A. 1967, ApJ, 147, 73, \dodoi{10.1086/148982}

\bibitem[{{Sargent} \& {Turner}(1977)}]{Sargent1977}
{Sargent}, W.~L.~W., \& {Turner}, E.~L. 1977, \apjl, 212, L3,
  \dodoi{10.1086/182362}

\bibitem[{{Sarpa} {et~al.}(2019){Sarpa}, {Schimd}, {Branchini}, \&
  {Matarrese}}]{Sarpa:2018ucb}
{Sarpa}, E., {Schimd}, C., {Branchini}, E., \& {Matarrese}, S. 2019, \mnras,
  484, 3818, \dodoi{10.1093/mnras/stz278}

\bibitem[{Schaye {et~al.}(2015)Schaye, Crain, Bower, Furlong, Schaller, Theuns,
  {Dalla Vecchia}, Frenk, Mccarthy, Helly, Jenkins, Rosas-Guevara, White, Baes,
  Booth, Camps, Navarro, Qu, Rahmati, Sawala, Thomas, \& Trayford}]{Schaye2015}
Schaye, J., Crain, R.~A., Bower, R.~G., {et~al.} 2015, Monthly Notices of the
  Royal Astronomical Society, 446, 521, \dodoi{10.1093/mnras/stu2058}

\bibitem[{Sheth \& Tormen(2002)}]{Sheth2002}
Sheth, R.~K., \& Tormen, G. 2002, MNRAS, 329, 61,
  \dodoi{10.1046/j.1365-8711.2002.04950.x}

\bibitem[{Silk(1977)}]{Silk1977}
Silk, J. 1977, ApJ, 211, 638, \dodoi{10.1086/154972}

\bibitem[{Silk \& Rees(1998)}]{Silk1998}
Silk, J., \& Rees, M.~J. 1998, A{\&}A, 331, L1

\bibitem[{Somerville \& Primack(1999)}]{Somerville1999}
Somerville, R.~S., \& Primack, J.~R. 1999, MNRAS, 310, 1087,
  \dodoi{10.1046/j.1365-8711.1999.03032.x}

\bibitem[{{Springel} {et~al.}(2005){Springel}, {White}, {Jenkins}, {Frenk},
  {Yoshida}, {Gao}, {Navarro}, {Thacker}, {Croton}, {Helly}, {Peacock}, {Cole},
  {Thomas}, {Couchman}, {Evrard}, {Colberg}, \& {Pearce}}]{Springel2005}
{Springel}, V., {White}, S. D.~M., {Jenkins}, A., {et~al.} 2005, \nat, 435,
  629, \dodoi{10.1038/nature03597}

\bibitem[{Springel {et~al.}(2018)Springel, Pakmor, Pillepich, Weinberger,
  Nelson, Hernquist, Vogelsberger, Genel, Torrey, Marinacci, \&
  Naiman}]{Springel2018}
Springel, V., Pakmor, R., Pillepich, A., {et~al.} 2018, MNRAS, 475, 676,
  \dodoi{10.1093/mnras/stx3304}

\bibitem[{Strauss {et~al.}(1992)Strauss, Yahil, Davis, Huchra, \&
  Fisher}]{Strauss1992}
Strauss, M.~A., Yahil, A., Davis, M., Huchra, J.~P., \& Fisher, K. 1992, ApJ,
  397, 395, \dodoi{10.1086/171796}

\bibitem[{Tammann {et~al.}(1979)Tammann, Yahil, \& Sandage}]{TYS}
Tammann, G., Yahil, A., \& Sandage, A. 1979, ApJ, 234, 775,
  \dodoi{10.1086/157556}

\bibitem[{{Tamone} {et~al.}(2020){Tamone}, {Raichoor}, {Zhao}, {de Mattia},
  {Gorgoni}, {Burtin}, {Ruhlmann-Kleider}, {Ross}, {Alam}, {Percival}, {Avila},
  {Chapman}, {Chuang}, {Comparat}, {Dawson}, {de la Torre}, {des Mas du
  Bourboux}, {Escoffier}, {Gonzalez-Perez}, {Hou}, {Kneib}, {Mohammad},
  {Mueller}, {Paviot}, {Rossi}, {Schneider}, {Wang}, \& {Zhao}}]{Tamone2020}
{Tamone}, A., {Raichoor}, A., {Zhao}, C., {et~al.} 2020, arXiv e-prints,
  arXiv:2007.09009.
\newblock \doarXiv{2007.09009}

\bibitem[{{Tegmark, M. Peebles}(1998)}]{Tegmark1998}
{Tegmark, M. Peebles}, P. 1998, ApJL, 500

\bibitem[{Tempel {et~al.}(2011)Tempel, Saar, Liivam{\"{a}}gi, Tamm, Einasto,
  Einasto, \& M{\"{u}}ller}]{Tempel2011a}
Tempel, E., Saar, E., Liivam{\"{a}}gi, L.~J., {et~al.} 2011, A{\&}A, 529, A53,
  \dodoi{10.1051/0004-6361/201016196}

\bibitem[{White \& Frenk(1991)}]{White1991}
White, S. D.~M., \& Frenk, C.~S. 1991, ApJ, 379, 52, \dodoi{10.1086/170483}

\bibitem[{White \& Rees(1978)}]{White1978a}
White, S. D.~M., \& Rees, M.~J. 1978, MNRAS, 183, 341,
  \dodoi{10.1093/mnras/183.3.341}

\bibitem[{Xu {et~al.}(2020)Xu, Zehavi, \& Contreras}]{Xu2020}
Xu, X., Zehavi, I., \& Contreras, S. 2020, arXiv e-prints, 2007.05545.
\newblock \doarXiv{2007.05545}

\end{thebibliography}

\appendix

\section{shot noise}
\label{sec:SN}
The variance of the shot noise in  a TH smoothed galaxy density field $\dgal$ is 
\begin{equation}
\sigma^2_\textrm{SN}=\frac{1}{\bar n V_\mathrm{s}} 
\end{equation}
where $V_\mathrm{s}=4\pi R_\mathrm{s}^3/3$.  For the \sag\ mocks with the Euclid cut,  $\bar n= 3.29 \times 10^{-3} [ \hmpc ]^{-3}$
giving $\sigma^2_\textrm{SN}=9\times 10^{-3}$ and    $7.2\times 10^{-5}$ for $\rs=20$ and $100\hmpc$, respectively. 

The shot noise in the velocity can be derived as follows \citep[c.f.][]{Strauss1992}
For simplicity, we choose to calculate the shot noise variance for the velocity recovered at the origin, $\br =0$. By homogeneity,  the result will be valid  at any other point. 
The linear theory relation gives 
\begin{equation}
\bv(0)=-\frac{a H \beta}{4 \pi \bar n} \sum_{r_i>\rs}\frac{\br_i}{r_i^3}-\frac{a H f}{4 \pi \bar n}\sum_{r_i\le \rs} \frac{4\pi}{3V_\mathrm{s}}\br_i\; .
\end{equation}
where in the TH smoothing, each galaxy is represented as a sphere of radius $\rs$. 
To estimate the shot noise in this quantity, we consider boot-strap realizations in which each galaxy is replaced 
by $N_i $ particles where $N_i$ is a random integer drawn from 
 a Poisson distribution with a mean of unity.  The difference between the velocity reconstructed from  one of these realizations and $\bv(0)$ is 
\begin{equation}
\delta \bv=-\frac{a H \beta}{4 \pi \bar n}\left[ \sum_{r_i>\rs}\frac{(N_i-1)\br_i}{r_i^3}+\frac{1}{\rs^3}  \sum_{r_i\le \rs } (N_i-1)\br_i\right] \; ,
\end{equation}
leading to a 1D variance 
\begin{equation}
\sigma^2_{V,\textrm{SN}}=\av{\frac{|\delta \bv|^2}{3}}=\frac{1}{3}\left(\frac{a H \beta}{4 \pi \bar n}\right)^2\left[ \sum_{r_i>\rs}\frac{1}{r_i^4}+\frac{1}{\rs^6} \sum_{r_i\le \rs} r_i^2\right]\; .
\end{equation}
Transforming the summation into a volume integral using  $\sum_i \rightarrow4\pi  \bar n\int r^2 \dd r $ yields 
\begin{equation}
\sigma^2_{V,\textrm{SN}}=\frac{(a H \beta)^2}{10 \pi \bar n\rs}\; .
\end{equation}

\section{Regressions}
\label{sec:slopes}
The aim here is to clarify the difference between various regressions.  
 We seek a parameter (slope) $p$ which renders a minimum in the expression
\begin{equation}
\tilde \chi^2=\sum_i (y_i -p x_i)^2 \; ,
\end{equation}
where $x $ and $y$, respectively,  have  vanishing mean values.
This yields the regression slope of  $y_{i}$ on $x_i$ as
\begin{equation}
p=\frac{\sum x_i y_i}{\sum x_i^2} \;  , 
\end{equation}
with a $1\sigma$ uncertainty 
\begin{equation}
\label{eq:sigmaA}
\sigma^2_p=\frac{\sum(y_i-p x_i)^2}{\sum x_i^2}\; .
\end{equation}
This implies the slope of the regression of $x$ on $y$ 
is 
\begin{equation}
q=\frac{\sigma_x^2}{\sigma_y^2} p \;  .
\end{equation}

\begin{enumerate}[label=\Roman*.] 

\item Statistically  unbiased estimate of $b$ via density-density regressions:
Let $x=\ds $  the mass density contrast (smoothed or otherwise) and $y=\dgs $ to be  the smoothed number  density contrast of galaxies.
 In this case, the regression of $y$ on $x$  yields
\begin{equation}
p = \frac{\langle  \ds \dgs\rangle}{\sigs^2} 
=  b\; .
\end{equation}
where in the last line  we have assumed linear biasing $\dgs=\ds+\epsilon^s$ and that the smoothed 
random noise term $\epsilon^s$ satisfied $\av{\epsilon^s \ds}$. This regression yields  a statistically unbiased estimate for $b$.

Now, take $x=\ds$  and $y=\dgs $. This is the inverse regression to the above and it gives.
\begin{equation}
p= \frac{\langle  \ds \dgs\rangle}{\siggs^2 }=b \frac{\sigs^2}{\siggs^2}=\frac{b}{1+\sigeps^2/\sigs^2}\; ,
\end{equation}
which equals $b$ only in the limit of vanishing $\sigeps$.

\item Let  $x=\delta $  be the un-smoothed the mass density contrast and, as before, $y=\dgs $.
The regression of $y$ on $x$  yields
\begin{equation}
\label{eq:dgonds}
p = \frac{\langle  \delta \dgs \rangle}{\langle \delta^2\rangle} 
=  b \frac{\langle \delta \ds\rangle}{\langle\delta^2\rangle}\; . 
\end{equation}

Note that for a sharp k-cutoff smoothing  $\langle \delta \ds\rangle=\langle  (\ds)^2\rangle $ since $\delta -\ds$ are composed of Fourier modes that are entirely independent from $\ds$. 

\item Consider the regression of  $y=\delta$  on $ x=\dgs $.
Then the slope of this regression is
\begin{equation}
q = \frac{\langle \delta \dgs\rangle}{\langle(\dgs)^2\rangle} 
=  b^{-1} \frac{\langle \delta \ds\rangle}{\langle (\ds)^2\rangle +\langle (\epsilon^s)^2\rangle/b^2}\; .
\end{equation}
Only for $\epsilon^s=0$ and a sharp $k$-cutoff smoothing, this regression yields $1/b$. 

\item The form of the \meth\ is basically a regression of true on predicted velocies.
 We identify $x$ with the radial peculiar velocities, $\vpr$,   predicted from the distribution of galaxies with $\beta=f$.
 We write 
\begin{equation}
\label{eq:xvrd}
x=\vpr =  \vg^{s} +\epsilon_V \;  ,
\end{equation}
where we take the  predicted velocity  $\vpr$ as obtained from the smoothed galaxy distribution using linear theory using the true value of $\beta$.
Further, we have assumed that the smoothing is on sufficiently large scales (cf. Fig.~\ref{fig:gvDMDM})  so that  $\vpr$ differs from the true 
smoothed velocities $\vg^s$ only due to shot noise and scatter in the biasing relation as represented by the term $\epsilon_V$

As for   $y$ we take 
\begin{equation}
\label{eq:y}
y=\vg +\epsilon_\textrm{SFR}\; ,
\end{equation}
where $\epsilon_\textrm{SFR}$ represented the scatter due to the spread of the SFR.  In this case, the slope  is 
\begin{equation}
p = \frac{\langle \vg \vg^s\rangle} {\langle (\vg^s)^2\rangle+ \langle (\epsilon_V)^2\rangle} \; .
\end{equation} 
Therefore, only if  $\epsilon_V=0$. i.e. no scatter  and  if $\vpr$ is obtained with $k$-cutoff smoothing (or without any smoothing), we find
the $ p=1/b$.   In the application to real catalogs $k$-cutoff smoothing is unrealistic and  it is impossible to recover 
the galaxy velocities without smoothing, but it is rather easy to model the expectation values of velocity products. 
In linear theory. Therefore, one needs to carefully calibrate the results in order to obtain a statistically estimate of $\beta$. Fortunately, this is easy to do. 

\end{enumerate}

\section{Luminosity modulation}
\label{sec:dl}

Neglecting terms proportional to the gravitational potential, we have 
\begin{equation}
\label{eq:zzc}
\frac{z-z_t}{1+z}\approx \frac{v}{c}\; , 
\end{equation}
where $c$ is the speed of light. 
A galaxy with  measured redshift $z$ and  an observed flux, $F$, in units of 
$\textrm{energy}\;  (\textrm{time})^{-1}  \; (\textrm{area})^{-1}$ (e.g. $\mathrm{ergs} \; s^{-1} \mathrm{cm}^{-2} $, is assigned a luminosity, 
$\lobs$, 
according 
\begin{equation}
\label{eq:dtoL}
\lobs=4\pi \dl^2(z) F \; , 
\end{equation}
where $\dl(z) $ is the luminosity distance evaluated at redshift $z$.

Let us explore now how $\lobs$ is related to the true intrinsic luminosity, $\lint$, of the galaxy.
Let the galaxy cover an area $A$ perpendicular to the los and let 
$I_\nu$ be the specific  intensity of light emitted by this area in units of  $\textrm{energy}\ $ $ (\textrm{second})^{-1} $
 $(\textrm{area})^{-1} $ $(\textrm{frequency})^{-1} $ $(\textrm{solid \; angle})^{-1}$. We assume a uniform $I_\nu$ across $A$.
 We assume that the galaxy emits at a single frequency with a very narrow line such that 
\begin{equation}
\label{eq:nuint}
I_\nu=I \delta^\textrm{D}(\nu-\nuint)\; ,
\end{equation}
where $\delta^\textrm{D}$ is the Dirac-$\delta $  function.
Now, $ I_\nu \dd A \dd \Omega $ is the energy  emitted per second per frequency from a small patch $\dd A$ into a solid angle $\dd \Omega$. Therefore, we  have the 
\begin{equation}
\lint= \int I_\nu \dd \nu \dd A \dd \Omega=4\pi I A\; .
\end{equation}

The observer at redshift zero measures a specific intensity $\IO$ which is related to $I_\nu$ by 
invariant 
\begin{equation}
\label{eq:adia}
\frac{\IO}{\nu_0^3}=\frac{I_\nu}{\nu^3}\; ,
\end{equation}
where 
\begin{equation}
\label{eq:nunu}
\nu_0=\frac{\nu}{1+z}
\end{equation}
 is the observed frequency. 
The flux measured by the observer's  detector is
\begin{equation}
F=\Omega_A \int \IO \dd\nu_0 \; 
\end{equation}
where $\Omega_A=A/\da^2(\zc)$ is the solid angle subtended by the area $A$.  The angular diameter is 
evaluated at $\zc$ since the area  is perpendicular to the los and thus, to first order,  it is unaffected by the peculiar velocity of the galaxy. 
Now, using the relations  \ref{eq:nuint}, \ref{eq:adia} and \ref{eq:nunu}, we obtain
\begin{equation}
\int \IO \dd\nu_0 ={I_\nu}\left(\frac{\nu_0}{\nu}\right)^3 \frac{\dd \nu_0}{\dd \nu} \dd\nu=\frac{I}{(1+z)^4}\; . 
\end{equation}
This is Tolman's surface brightness law.
Therefore, 
\begin{equation}
\label{eq:flux}
F=\frac{I A}{\da^2(\zc)(1+z)^4}=\frac{\lint}{4\pi \da^2 (\zc)(1+z)^4}
\end{equation}
Using \eq{eq:dtoL} and remembering that $\dl(z)=(1+z)^2\da(z)$ we find 
\begin{equation}
\label{eq:ratio}
\lobs=\left[\frac{\da(z)}{\da(\zc)}\right]^{2}\lint \; .
\end{equation}
We have arrived at a peculiarity result that the modulation of the luminosity is actually 
via the angular diameter distance rather that the luminosity distance. 

Let us expand the distance ratio to first order in $V$. We  start with 
\begin{equation}
\label{eq:dldc}
\da=\frac{\dc}{1+z}
\end{equation}
where
\begin{equation}
\label{eq:dc}
\dc=c\int_0^z\frac{\dd z}{H(z)}\; ,
\end{equation}
is the comoving distance.

First order Taylor expansion of  $\da(z)$ in the vicinity of $z\approx \zc$ is 
\begin{eqnarray}
\da(z)&=&\da(\zc)-\frac{z-\zc}{1+\zc} \frac{\dc(\zc)}{1+\zc}+\frac{z-\zc}{1+\zc} \frac{\dd \dc(z_c)}{\dd z}\\
&=& \da(\zc)\left[1-\frac{V}{c}  +\frac{V}{H(\zc)\da(\zc)}\right]\; .
\end{eqnarray}
The last line is consistent with previous findings in the literature \citep{Bonvin2005,2006PhRvD..73l3526H}. In the square brackets, The $V/c$  reflects relativistic beaming which goes in the opposite direction of the second term arising from associating the distance with redshift.
To first order, we can replace $\zc$ by $z$ in the term in brackets. Therefore, given the velocities (via a model)  $\lint$ can be estimated as
\begin{equation}
\label{eq:lintmodel}
\lint^\textrm{e}=\lobs\left[1+\frac{2V}{c}  -\frac{2V}{H(z)\da(z)}\right]\; .
\end{equation}
 
 We can easily include all other relativistic effected related to the gravitational potential. 
 Over the scales considered here, gravitational lensing is dominant. The modification 
to the  equations \eq{eq:flux} and \eq{eq:ratio} due to lensing is simple: $\da(z)$ remains the same and given by the expressions \eq{eq:dldc} \& \eq{eq:dc}, while 
 \begin{equation}
 \da(\zc) \rightarrow \tda(\zc,\kappa)=\da(\zc)(1-\kappa)\; ,
 \end{equation}
 where $\kappa$ is the convergence given in terms of a los integral over he density contrast.
 This reflects the fact that objects appear to occupy larger solid angles for positive $\kappa$. 
 Therefore, 
 \begin{equation}
 \da(z)= \da(\zc)\left[1-\frac{V}{c}  +\frac{V}{H(\zc)\da(\zc)}-\kappa\right]\; .
 \end{equation}

 Let us consider a small patch of the survey at a given redshift and where $\kappa$ and $V$   can be 
 assumed constant. 
 Let $F_{l}$ be the limiting flux of the survey. 
 The minimum threshold observed  luminosity for this patch is 
 \begin{equation}
{\lobsl}=4\pi(1+z)^2\da^2(z) F_l \; . 
\end{equation}
The threshold  intrinsic luminosity is  (cf. \eq{eq:ratio})
\begin{equation}
\lintl= 4\pi (1+z)^2\tda^2(\zc)F_l\; .
\end{equation}
Thus the actual threshold intrinsic luminosity  of galaxies  in a given range  of  observed redshift $z$ in the patch will depend on $\kappa$ and the los velocity, $V$.
The number density will therefore chance and this will 
affect the  variance of $\log \lint^\textrm{e} $ estimated from $\lobs$ using \eq{eq:lintmodel}, in addition to the modulation  of $\lint^\textrm{e}$ by $V$. 
In the tests provided in the paper, we do not include the effect related to the change in the threshold luminosity. 

In the above, it may seem that there is a degeneracy between $\kappa$ and the velocity corrections.  The convergence  $\kappa$ is a los integrated quantity and can actually be estimated from the distribution of galaxies assuming a biasing relation. 

The observer's
 Since we are considering line luminosity, there are no issues related to k-correction present in the case of measuring magnitudes in a given band.

\section{Gravitational lensing}
Gravitational lensing by foreground mass distribution  modifies the  observed luminosity/SFR  of a galaxy  by  a multiplicative factor $1+2\kappa$ where \citep{BS01}
\begin{equation}
\kappa=\frac{3 H_0^2 \Omega_m}{2c^2}\int_0^r\dd r' g(r',r)\delta(r\hat{\br})\; ,
\end{equation}
where the galaxy is at $\br$ and $g(r',r)= (r'-r'^2/r)/a(r')$. Following the derivation in \cite{Nusser2013}, we obtain
\begin{equation}
C_l=\frac{2}{\pi}\left(\frac{3 H_0^2 \Omega_m}{2c^2}\right)^2\int \dd k k^2 P(k) |\int_0^r \dd r' g(r',r)j_l(kr')|^2\; .
\end{equation}
Working with the Limber's approximation for the spherical Bessel functions $j_l(kr)\sim \sqrt{\frac{\pi}{l+1/2}}\delta^\textrm{D}(l+1/2-kr)$, the relation becomes 
\begin{equation}
C_l=(2l+1)\left(\frac{3 H_0^2 \Omega_m}{2c^2}\right)^2\int_{k=(l+1/2)/r}^\infty\dd k P(k)\left(1-\frac{2l+1}{2kr}\right)^2
\end{equation}
The variance of $\kappa$ is then 
\begin{equation}
\sigma^2_\kappa=\sum_{l=0}^\infty W_l^2 C_l\approx \sum_{l=0}^{l_\textrm{max}}\frac{2l+1}{4\pi}C_l\; ,
\end{equation}
where $W_l$ represents averaging over a solid angle $\pi \theta^2_\textrm{s}$ where  $\theta_\textrm{s}=\rs/r$.
In the second part of this equation, the effect of this  window function is approximated in terms the sharp $l-$cutoff at  $l_\textrm{lmax}=2r/\rs$. For $\rs=20$ and $100\hmpc$ the expression yields, respectively, $\sigma_\kappa\approx 2\times 10^{-3}$ and $3\times 10^{-4}$.
This translates into $\lsfr$  shifts of $\log(1+2
\sigma_\kappa)\approx 1.6\times 10^{-3}$ and $2.5\times 10^{-4}$ for the smaller and larger $\rs$ respectively.

 \end{document}